\documentclass[aip,apl,reprint,final]{revtex4-2}

\pdfoutput=1

\usepackage[colorlinks=true,allcolors=blue]{hyperref} 
\usepackage{graphicx}
\usepackage{amsmath,amssymb,mathrsfs,array,longtable,delarray}
\usepackage{latexsym}
\usepackage{dcolumn}
\usepackage{bm}
\usepackage{textcomp} 
\usepackage{siunitx}

\usepackage{verbatim}

\begin{document}

\title{High transparency induced superconductivity in field effect two-dimensional electron gases in undoped InAs/AlGaSb surface quantum wells}

\author{E. Annelise Bergeron}
\affiliation{Institute for Quantum Computing, University of Waterloo, Waterloo N2L 3G1, Canada}
\affiliation{Department of Physics, University of Waterloo, Waterloo N2L 3G1, Canada}

\author{F. Sfigakis}
\altaffiliation{corresponding author: francois.sfigakis@uwaterloo.ca}
\affiliation{Institute for Quantum Computing, University of Waterloo, Waterloo N2L 3G1, Canada}
\affiliation{Northern Quantum Lights inc., Waterloo N2B 1N5, Canada}
\affiliation{Department of Chemistry, University of Waterloo, Waterloo N2L 3G1, Canada}

\author{A. Elbaroudy}
\affiliation{Department of Electrical and Computer Engineering, University of Waterloo, Waterloo N2L 3G1, Canada}
\affiliation{Department of Physics, University of Waterloo, Waterloo N2L 3G1, Canada}

\author{A. W. M. Jordan}
\affiliation{Institute for Quantum Computing, University of Waterloo, Waterloo N2L 3G1, Canada}
\affiliation{Department of Chemistry, University of Waterloo, Waterloo N2L 3G1, Canada}

\author{F. Thompson}
\affiliation{Institute for Quantum Computing, University of Waterloo, Waterloo N2L 3G1, Canada}

\author{\\George Nichols}
\affiliation{Institute for Quantum Computing, University of Waterloo, Waterloo N2L 3G1, Canada}
\affiliation{Department of Physics, University of Waterloo, Waterloo N2L 3G1, Canada}

\author{Y. Shi}
\affiliation{Department of Electrical and Computer Engineering, University of Waterloo, Waterloo N2L 3G1, Canada}
\affiliation{Department of Physics, University of Waterloo, Waterloo N2L 3G1, Canada}
\affiliation{Waterloo Institute for Nanotechnology, University of Waterloo, Waterloo N2L 3G1, Canada}

\author{Man Chun Tam}
\affiliation{Department of Electrical and Computer Engineering, University of Waterloo, Waterloo N2L 3G1, Canada}
\affiliation{Waterloo Institute for Nanotechnology, University of Waterloo, Waterloo N2L 3G1, Canada}

\author{Z. R. Wasilewski}
\affiliation{Institute for Quantum Computing, University of Waterloo, Waterloo N2L 3G1, Canada}
\affiliation{Department of Physics, University of Waterloo, Waterloo N2L 3G1, Canada}
\affiliation{Northern Quantum Lights inc., Waterloo N2B 1N5, Canada}
\affiliation{Department of Electrical and Computer Engineering, University of Waterloo, Waterloo N2L 3G1, Canada}
\affiliation{Waterloo Institute for Nanotechnology, University of Waterloo, Waterloo N2L 3G1, Canada}

\author{J. Baugh}
\altaffiliation{baugh@uwaterloo.ca}
\affiliation{Institute for Quantum Computing, University of Waterloo, Waterloo N2L 3G1, Canada}
\affiliation{Department of Physics, University of Waterloo, Waterloo N2L 3G1, Canada}
\affiliation{Northern Quantum Lights inc., Waterloo N2B 1N5, Canada}
\affiliation{Department of Chemistry, University of Waterloo, Waterloo N2L 3G1, Canada}
\affiliation{Waterloo Institute for Nanotechnology, University of Waterloo, Waterloo N2L 3G1, Canada}


\begin{abstract}
We report on transport characteristics of field effect two-dimensional electron gases (2DEG) in 24~nm wide indium arsenide surface quantum wells. High quality single-subband magnetotransport with clear quantized integer quantum Hall plateaus are observed to filling factor $\nu=2$ in magnetic fields of up to $B=18$~T, at electron densities up to 8$\times 10^{11}$ /cm$^2$.
Peak mobility is 11,000 cm$^2$/Vs at 2$\times 10^{12}$ /cm$^2$. Large Rashba spin-orbit coefficients up to 124 meV$\cdot${\AA} are obtained through weak anti-localization (WAL) measurements. Proximitized superconductivity is demonstrated in Nb-based superconductor-normal-superconductor (SNS) junctions, yielding 78$-$99\,\% interface transparencies  from superconducting contacts fabricated ex-situ (post-growth), using
two commonly-used experimental techniques for measuring transparencies. These transparencies are on a par with those reported for epitaxially-grown superconductors. These SNS junctions show characteristic voltages $I_c R_{\textsc{n}}$ up to 870~$\mu$V and critical current densities up to 9.6~$\mu$A/$\mu$m, among the largest values reported for Nb-InAs SNS devices.
\end{abstract}

\maketitle

The last decade has seen spectacular progress in InAs/AlGaSb two-dimensional electron gases (2DEGs).\cite{Shojaei15,Shojaei16-A,Shojaei16-B,Hatke17,LeeJS19,Mittag21} This material system has joined the small club where the fractional quantum Hall effect can be routinely observed,\cite{MaMK17,KomatsuS22} in addition to 2DEGs in GaAs/AlGaAs,\cite{Pan08,Kleinbaum20} in AlAs/AlGaAs,\cite{ChungYJ18-B} in graphene,\cite{Bolotin09,Dean11} in Si/SiGe,\cite{LaiK04,LuTM12}, in Ge/SiGe,\cite{Mironov16} in CdTe,\cite{Piot10,Betthausen14} and in ZnO/MgZnO.\cite{Tsukazaki10,Falson18} The highest mobility reported in InAs/AlGaSb is 2.4$\times$10$^6$ cm$^2$/Vs,\cite{Tschirky17,ThomasC18} only exceeded by GaAs/AlGaAs,\cite{Umansky09,ChungYJ21} Ge/SiGe,\cite{Myronov23} and AlAs/AlGaAs\cite{ChungYJ18-B} material systems. The combination of high mobilities, strong spin orbit interactions (SOI), pinning of the Fermi level in the conduction band, small effective mass, and large Land\'e g-factor could make InAs/AlGaSb a strong candidate material system for topological quantum computing with Majorana zero modes.\cite{Shabani16,Karzig17,KeCT19}

In the last decade, most efforts towards realizing Majorana fermions in a scalable platform have focused on surface quantum wells in the In(Ga)As/In$_{0.8}$Al$_{0.2}$As material system, where mobilities have significantly improved from 1$\times 10^4$~cm$^2$/Vs initially\cite{Shabani16} to more than 1$\times 10^5$~cm$^2$/Vs recently.\cite{ZhangT23} However, in the context of topological quantum computing, the InAs/Al$_{0.8}$Ga$_{0.2}$Sb material system could offer possible advantages over the In(Ga)As/In$_{0.8}$Al$_{0.2}$As system, including better strain engineering,\footnote{The critical thickness of an InAs quantum well (QW) grown on Al$_{0.8}$Ga$_{0.2}$Sb is much larger ($>$24 nm; tensile strain) than on In$_{0.8}$Al$_{0.2}$As (7 nm; compressive strain) despite similar differences in lattice constant mismatch ($\sim$~8 pm) between InAs and either barrier material.} higher electron densities,\footnote{Al$_{0.8}$Ga$_{0.2}$Sb has a larger conduction band offset ($\sim$\,1.9~eV) relative to InAs than In$_{0.8}$Al$_{0.2}$As does ($\sim$\,0.3~eV), thus providing a higher barrier next to the quantum well and allowing higher carrier densities to be achieved within a single 2D subband. Higher electron densities in turn can enable higher mobilities and stronger SOI than at lower electron densities.} and higher mobilities.\footnote{A 2DEG hosted in a binary alloy quantum well instead of a ternary alloy quantum well would not suffer from alloy scattering, which only occurs in ternary alloys and is a significant mechanism limiting mobilities.}

Furthermore, most efforts in this field have also centered on semiconductor-superconductor hybrid devices proximitized by ``epitaxial aluminum,'' grown directly on In(Ga)As/InAlAs heterostructures in the same growth chamber. This approach appeared to be the only way to reliably generate strong, ``hard'' superconducting gaps in the semiconductor,\cite{ChangW15} as opposed to smaller, ``soft'' gaps, typically produced by superconductor contacts deposited ex-situ, post-growth.

In this Letter, we demonstrate gated 2DEGs in InAs/Al$_{0.8}$Ga$_{0.2}$Sb surface quantum wells, without parallel conduction from another conducting layer in magnetic fields up to 18~T and at electron densities up to $3 \times 10^{12}$~cm$^{-2}$. Single-subband operation is demonstrated at lower 2DEG densities. Using SiO$_2$ as dielectric yielded stable and reproducible gating operations all the way down to pinch-off. Rashba spin-orbit coefficients up to 124~meV$\cdot${\AA} are obtained through weak anti-localization (WAL) measurements. Proximitized superconductivity is demonstrated in Nb-based superconductor-normal-superconductor (SNS) Josephson junctions, yielding deep  gaps with up to unity transparencies from superconducting contacts fabricated entirely ex-situ (post-growth).

Two nominally identical heterostructures (G743 and G782) were grown by molecular beam epitaxy (MBE), with the following sequence of layers, starting from a GaSb (001) substrate: a 25~nm GaSb nucleation layer, a 800~nm Al$_{0.80}$Ga$_{0.20}$Sb$_{0.93}$As$_{0.07}$ quaternary buffer, a 20~nm Al$_{0.8}$Ga$_{0.2}$Sb bottom barrier, a 24~nm InAs quantum well, and
a 6~nm In$_{0.75}$Ga$_{0.25}$As cap layer. There is no intentional doping anywhere in the heterostructure. Section~I in the supplementary material provides additional details about MBE growth. Figure~S1 from Section~II in the supplementary material shows nextnano\texttrademark ~self-consistent simulations\cite{NextNano-A,NextNano-B} of bandstructure profiles, showing the extent of the 2DEG wavefunction within the InAs quantum well, and compares it to an In(Ga)As quantum well profile.

Hall bars were fabricated using standard optical lithography and wet-etching techniques, keeping all processes at or below a temperature of 150$^\circ$C to prevent the deterioration of device characteristics.\cite{Uddin13,YiW15,Kulesh20} Ti/Au Ohmic contacts were deposited directly on the InGaAs cap layer, with typical resistances of 400$-$500~$\Omega$ at magnetic field $B=0$, and 10~k$\Omega$ at $B=18$ T. Finally, 60 nm thick silicon dioxide (SiO$_2$) or hafnium dioxide (HfO$_2$) was deposited by atomic layer deposition (ALD) at 150$^\circ$C, followed by the deposition of a Ti/Au global top-gate. See Section III.A of the supplementary material for more details on Hall bar fabrication. Using standard four-terminal ac lock-in measurement techniques, all transport experiments were performed in either a pumped-$^4$He cryostat or $^3$He/$^4$He dilution refrigerator, with a base temperature of 1.6 K and 11 mK respectively.

\begin{figure}[t]
  \includegraphics[width=1.0\columnwidth]{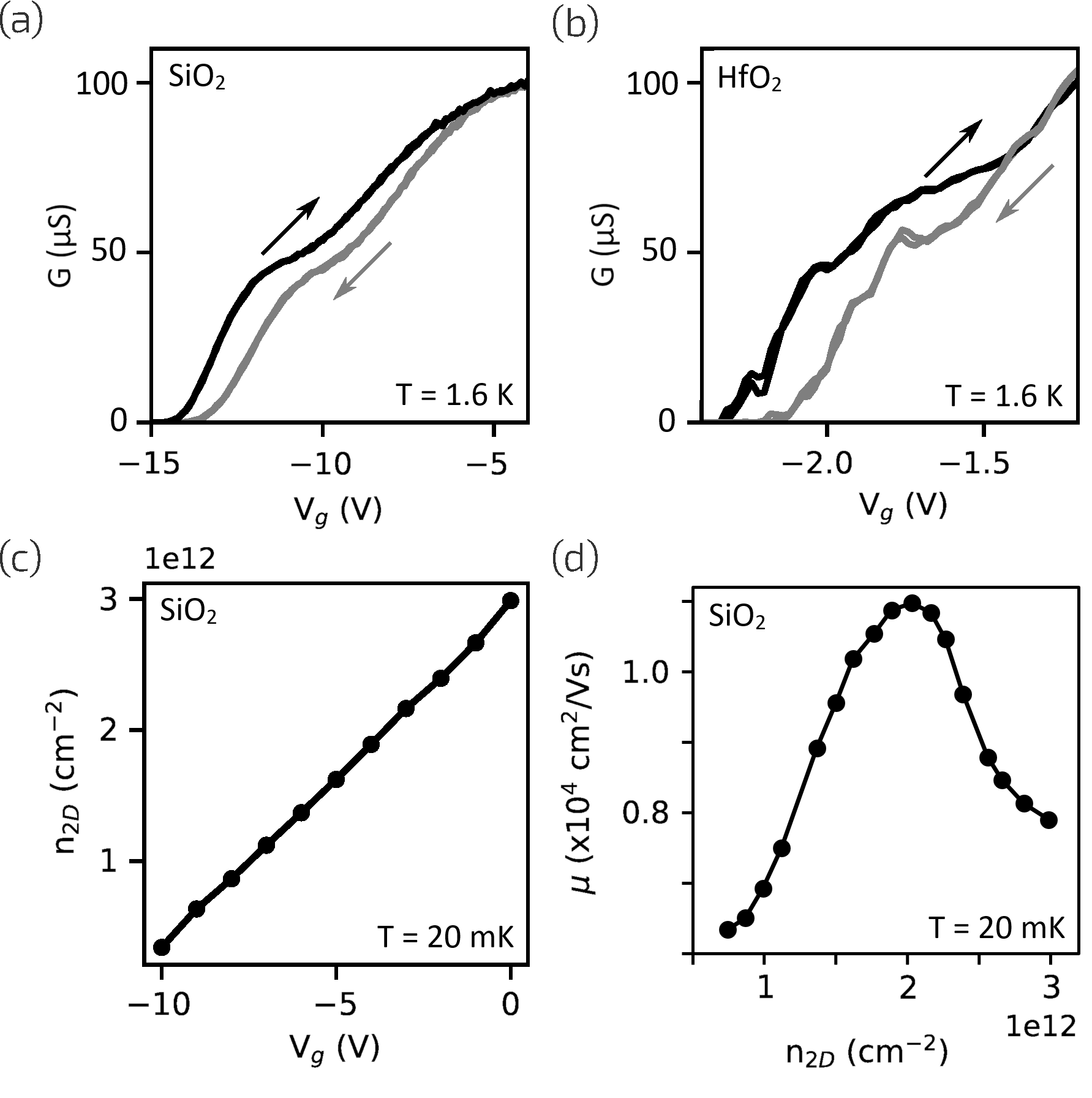}
  \caption{Typical differential conductance $G(V_g)=dI/dV_{sd}$ (using 100~$\mu$V ac excitation) showing the turn-on voltage of a gated Hall bar with gate dielectric (a) SiO2 and (b) HfO2. In panel (a), eight traces are shown, four while increasing $V_g$ (black) and four while decreasing $V_g$ (gray). Only four traces are shown in panel (b), two going up and two going down in $V_g$. (c) Typical $n_{\text{2D}}(V_g)$ of a Hall bar with SiO2. The 2DEG density increases linearly with $V_g$ in all samples, and is reproducible along the linear traces. (d) Electron mobility in the same device as in (c).}
  \label{fig:gating}
\end{figure}

The electron density at top-gate voltage V$_g=0$ in all gated Hall bars was significantly larger than the as-grown electron densities in ungated Hall bars. Figure \ref{fig:gating} shows the pinch-off characteristics of Hall bars with different gate dielectrics. With SiO$_2$, the pinch-off curves are stable and reproducible, overlapping perfectly when $V_g$ is swept in the same direction, but showing some hysteresis when $V_g$ is swept in opposite directions [see Fig.~\ref{fig:gating}(a)]. After pinch-off, the 2DEG does not turn itself back on with time.\cite{ThomasC18} With HfO$_2$ however, gating characteristics become unstable and non-reproducible near pinch-off [see Fig.~\ref{fig:gating}(b)].

At high conductances ($G > 100$~$\mu$S), both SiO$_2$ and HfO$_2$ produce stable and reproducible gating behavior. Figure \ref{fig:gating}(c) shows a typical electron density function $n_{\text{2D}}(V_g)$ with SiO$_2$. Section IV in the supplementary material shows representative gating and mobility characteristics from four gated Hall bars with SiO$_2$ and HfO$_2$. Figure \ref{fig:gating}(d) shows a peak in the transport mobility $\mu = 11 \times 10^{3}$ cm$^{2}$/Vs near 2DEG density $n_{\text{2D}} = 2.0 \times 10^{12}$ cm$^{-2}$. At lower densities, ionized impurity scattering is most likely limiting mobility. The much higher 2DEG carrier density at $V_g=0$ (by a factor of 2$-$3) in gated Hall bars (i.e., after the dielectric deposition) relative to the as-grown 2DEG densities in ungated Hall bars strongly hints at a high density of charge traps forming at the semiconductor-oxide interface. To eliminate the risk of parallel conduction, often observed in similar structures when electrons are supplied by remote doping, the AlGaSb/InAs interface was engineered to have an AlAs character.  Such interfaces are known to have high concentrations of As antisites, which act as double donors, supplying carriers to the 2DEG in the InAs quantum well.\cite{Tuttle90} The resulting high density of As$^{2+}$ ions at the interface subjects the 2DEG to strong Coulomb scattering, reducing its mobility. However, with the top quantum well barrier being InGaAs $–$ a necessary element of our design $–$ such a tradeoff is well justified. Indeed, Lee \textit{et al.}\cite{LeeJS19} demonstrated a severe reduction of mobility, from \num{650e3} cm$^2$/Vs down to \num{24e3} cm$^2$/Vs, a value comparable with our results, by only replacing the AlGaSb top barrier layer with InGaAs in otherwise identical heterostructures in near-surface InAs quantum wells.  Such reduction of mobility is likely due to the strong extent of the 2DEG wavefunction into the InGaAs barrier, with a lower conduction band offset than AlGaSb. At carrier densities above the mobility peak in Fig.~\ref{fig:gating}(d), the observed decline in mobility with increasing 2DEG density can be attributed to interface roughness scattering and/or alloy scattering. This is likely because the electron wavefunction is drawn closer to the surface by the stronger electric field from the top-gate.\cite{Ando82-A,Arjun22}

\begin{figure*}[t]
  \includegraphics[width=2.0\columnwidth]{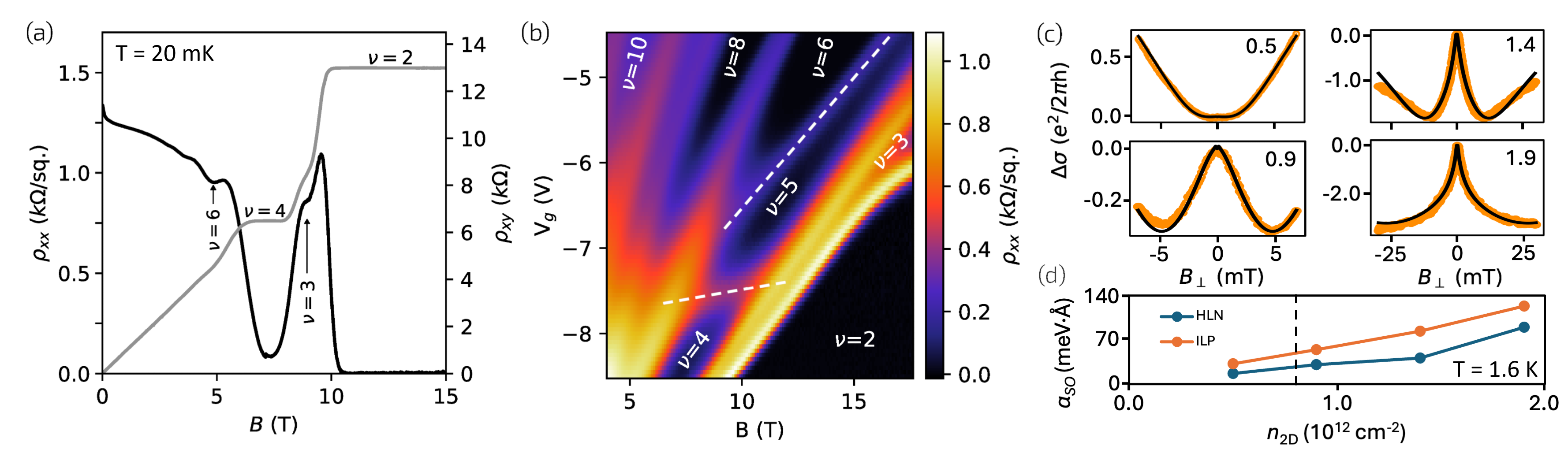}
  \caption{(a) SdH and Hall traces for $n_{\text{2D}} = 8.57 \times 10^{11}$~cm$^{-2}$ ($V_g=-8.0$~V). (b) Landau fan of SdH oscillations with densities ranging from $n_{\text{2D}} = 7.3 \times 10^{11}$ cm$^{-2}$ to $n_{\text{2D}} = 1.8 \times 10^{12}$ cm$^{-2}$. White labels identify corresponding effective filling factors (directly obtained from the Hall resistance with $\nu = h/e^2 R_H$, regardless of subband occupation), and dashed white lines identify populating Landau levels from the second 2D subband. (c) Weak anti-localization peak at $n_{\text{2D}} = (0.5; 0.9; 1.4; 1.9)\times 10^{12}$ cm$^{-2}$. Colored circles are experimental points, black lines are fits to the ILP model, and $n_{\text{2D}}$ is indicated in the upper right corner of each plot. (d) Spin orbit coefficient $\alpha_{so}$ as a function of $n_{\text{2D}}$, extracted from fits of the WAL data to the ILP and HLN models. The dashed vertical line indicates where the 2$^{\text{nd}}$ subband populates.}
  \label{fig:LandauFan}
\end{figure*}

In a Hall bar with SiO$_2$, Figure \ref{fig:LandauFan}(a) shows Shubnikov-de-Haas (SdH) oscillations in the longitudinal resistivity $\rho_{xx}$ and well-defined quantized quantum Hall plateaus in the Hall resistance $R_{\textsc{h}}$. Quantum Hall plateaus occur at specific resistance values  $R_{\textsc{h}}=h/\nu e^2$ at filling factors $\nu = hn_{2D}/eB$, where $h$ is the Planck constant and $e$ is the single electron charge. The presence of quantized Hall plateaus at $\nu=2,4$ confirms the formation of a 2DEG. We note SdH oscillations are not, by themselves, proof of the presence of a 2DEG, since they are also observed in 3D conductors,\cite{Shubnikov-de-Haas1930} albeit with much smaller amplitudes. Despite a large Land\'{e} $g$-factor ($g^*\sim$\,15), the spin-split $\nu=3$ Hall plateau is only starting to be resolved at $B \approx 10$~T, because of disorder. The visibility of spin splitting is dictated by $(g^*\mu_{\textsc{b}}B-\Gamma)>k_{\textsc{b}} T$, where $\mu_{\textsc{b}}$ is the Bohr magneton, $\Gamma$ is disorder associated with Landau level broadening, and $k_{\textsc{b}}$ is the Boltzmann constant. Thus the late onset in field of spin splitting implies $\Gamma\sim 9$~meV in our samples. This same disorder $\Gamma \sim 9$~meV is also responsible for the very late onset of SdH oscillations ($B\approx 3$~T), whose visibility is determined by $(\hbar e B/m^*-\Gamma)>k_{\textsc{b}} T$.

For $B>10$~T, the SdH oscillation minimum at $\nu=2$ reaches $\rho_{xx}=0$ in Fig.~\ref{fig:LandauFan}(a), implying no parallel conduction from another conductive layer. This remains true until at least $n_{\text{2D}} = 2.8 \times 10^{12}$ cm$^{-2}$ ($V_g=-0.5$~V). However, this does not exclude the possibility of a second subband populating the 2DEG. Indeed, at $n_{\text{2D}} = 7.3 \times 10^{11}$ cm$^{-2}$ ($V_g=-8.5$~V), there is a small mismatch ($<$3\%) between the classical Hall density $n_{total}=B/eR_{\textsc{h}}$ and the 2DEG density determined from the periodicity of Shubnikov-de-Haas oscillations versus inverse field, given by $n_{SdH} = \frac{2e}{h}\left( \frac{1}{B_{\nu+1}} - \frac{1}{B_{\nu}}\right)^{-1}$. This mismatch grows significantly as the 2DEG density increases (until $n_{total}\approx 2n_{SdH}$ near $V_g=0$). Energy level crossings in the Landau fan from Figure~\ref{fig:LandauFan}(b) confirm the presence of another 2D subband, most likely corresponding to the two spin-split branches of the lowest Landau level. We estimate the second subband populates near $n_{\text{2D}} \approx 8 \times 10^{11}$~cm$^{-2}$, which is consistent with similar reports of populating second subbands in 24 nm wide InAs/AlGaSb quantum wells.\cite{Tschirky17,ThomasC18} The ``knee'' observed in the pinch-off characteristics in Figure~\ref{fig:gating}(a) is consistent with the population of a second subband in the 2DEG. We note that the second subband's branch separating the regions labeled $\nu=4$ and $\nu=5$ at $B\sim10$~T in Fig.~\ref{fig:LandauFan}(b) does not appear to cross the $\nu=3$ branch and does not extend into the $\nu=2$ region. This is reminiscent of similar occurrences in the ring-like structures from Landau fans with Landau level crossings between the first and second subbands of GaAs/AlGaAs 2DEGs\cite{Muraki01,ZhangXC05,Ellenberger06} and InAs/AlGaSb 2DEGs.\cite{Tschirky17} Nonetheless, our Landau fan, obtained by sweeping the top-gate at magnetic field increments, showcases the reproducibility and stability of gating characteristics with dielectric SiO$_2$.

\begin{figure*}[t]
  \includegraphics[width=2.0\columnwidth]{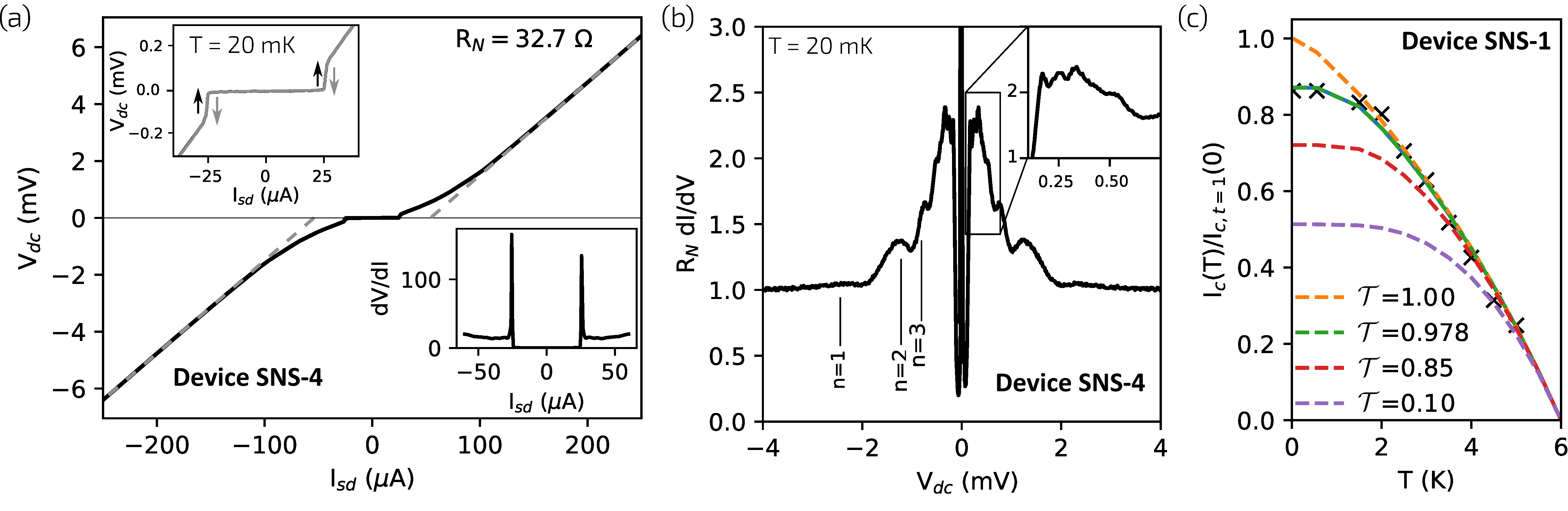}
  \caption{(a) Two overlapped I-V traces (solid) in a SNS device, where $I_{sd}$ is the source-drain dc current and $V_{dc}$ is the four-terminal dc voltage drop across the SNS junction. The dashed lines are the extrapolated normal resistance $R_{\textsc{n}}$ and their intercepts at $V_{dc}=0$ are the excess currents $I_{ex}$. (upper inset) Zoom-in of the superconducting transition, showing critical current $I_c$. Two overlapping traces are shown (black, grey), sweeping $I_{sd}$ in opposite directions. (lower inset) Sharp, deep gap in the four-terminal differential resistance of device SNS-4. (b) Normalized four-terminal differential conductance $dI/dV$, identifying MAR peaks in a SNS device. (c) Temperature dependence of $I_c$ of a SNS device, where crosses are experimental data and dashed lines are fits to Eqn.~(\ref{eq:Ic-Tdep}) for different values of $\mathcal{T}$.}
  \label{fig:SC}
\end{figure*}

The curvature of the energy level associated with $\nu=2$ near $B=17$~T at high densities in Fig.~\ref{fig:LandauFan}(b) is consistent with strong spin orbit interactions in the system. To quantify the strength of spin-orbit interactions, the Rashba coefficient $\alpha_{so}$ was determined from fits to the weak antilocalization (WAL) conductivity peak $\Delta\sigma_{xx}$ [see Fig.~\ref{fig:LandauFan}(c)] using the Hikami-Larkin-Nagaoka (HLN) and Iordanskii-Lyanda-Geller-Pikus (ILP) models,\cite{Hikami80,Iordanskii94} where $\Delta\sigma_{xx}$ = $\sigma_{xx}(B)-\sigma_{xx}(0)$, $\sigma_{xx}(B)$ is the field-dependent conductivity, and $\sigma_{xx}(0)$ is a constant background conductivity. Figure~\ref{fig:LandauFan}(d) summarizes the resulting fit values of $\alpha_{so}$ with both models, ranging from 16 to 124~meV$\cdot${\AA}, roughly linear with $n_{2D}$. These are consistent with published values for 2DEGs in InAs/AlGaSb\cite{Shojaei16-A,Hatke17,ThomasC18} and In(Ga)As/InAlAs.\cite{Wickramasinghe18,ZhangT23,Witt23,Farzaneh24} Some of these literature values (and ours) span both the single subband and multi-subband regimes. Figure~S6 from Section~IV in the supplementary material presents a detailed comparison of our $\alpha_{so}(n_{\text{2D}})$ with literature. Both the HLN and ILP models are only applicable at magnetic fields where the mean free path $\ell_e$ is smaller than the magnetic length $\ell_m$, leading to the condition $B < B_{tr}$ where $B_{tr} = \hbar/2e\ell_e^2$ is the transport field. In our sample, one of the distinctive features of the conductivity used for fitting (the minima on either side of the WAL peak) occurs well beyond $B_{tr}$ for the data point at $n_{2D} = \num{1.9e12}$~cm$^{-2}$, and the latter should therefore not be considered reliable.

\begin{table}[b]
    \begin{ruledtabular}
    \begin{tabular}{cccccccc}
    Device & width & gap & $I_c$ & $I_{ex}$ & $R_{\textsc{n}}$
    & $\Delta_{\textsc{mar}}$ & $\mathcal{T}$\\
    ID & ($\mu$m) & (nm) & ($\mu$A) & ($\mu$A) & ($\Omega$)
    & (meV) & (\%) \vspace{0.5 mm}\\
    \hline \vspace{-2.5 mm} \\
    SNS-1 & 10 & 200 & 82 & 158 & 8.8 & 1.13 & 78\,$\pm$\,5 \\ 
    SNS-2 & 10 & 200 & 89 & 183 & 8.6 & 1.16 & 81\,$\pm$\,4 \\ 
    SNS-3 & 3 & 120 & 29 & 58 & 30 & 1.13 & 85\,$\pm$\,2 \\  
    SNS-4 & 3 & 120 & 25 & 54 & 33 & 1.05 & 87\,$\pm$\,3 \\  
    SNS-5 & 3 & 400 & 5 & 24 & 53 & 0.91 & 82\,$\pm$\,2 \\   
    SNS-6 & 3 & 400 & 2 & 26 & 55 & 0.94 & 84\,$\pm$\,3 \\   
    \end{tabular}
    \end{ruledtabular}
    \caption{List of SNS devices.}
    \label{tab:SNS}
\end{table}

Three nominally identical pairs of superconductor-normal-superconductor (SNS) devices [see Table~\ref{tab:SNS}] were fabricated from wafers G743 and G782, using standard optical lithography, electron beam lithography, and wet-etching techniques, keeping all processes at or below a temperature of 180$^\circ$C to prevent the deterioration of device characteristics.\cite{Uddin13,YiW15,Kulesh20} Ti/Nb (2/80 nm) ohmic contacts were deposited directly on the InGaAs cap layer by sputtering, after Ar ion milling the contact areas for 6.5 minutes at 50 Watts. Immediately prior to loading samples in the Nb deposition system, the contact areas were treated with sulfur passivation.\cite{Bergeron23} The latter is designed to etch away native oxides, prevent further surface oxidation during transfer in air ($\sim$\,30~s) from the wetbench to the deposition chamber, and possibly dope the surface.\cite{Tajik12,LebedevMV20,Bessolov98} Four devices (SNS-3, ...~, SNS-6) were 3~$\mu$m wide, of which two (two) were fabricated with a gap $L=120$~nm ($L=400$~nm) between the Ti/Nb contacts. Devices  SNS-1 and SNS-2 were fabricated on wafer G743 rather than G782, with a width $W=10$~$\mu$m and a gap of $L=200$~nm. Note the gap region is not gated in any device. See Section III.B of the supplementary material for more details on sample fabrication.

The sputter-deposited Nb had a critical temperature $T_c=8.1$~K, yielding the superconducting energy gap $\Delta_{\text{Nb}} = 1.76\,k_{\textsc{b}} T_c = 1.23$~meV. Using this value, the coherence length in the proximitized semiconductor of our SNS junctions is $\xi_b = \hbar v_{\textsc{f}} / 2 \Delta_{\text{Nb}} = 220$~nm (246~nm) in the ballistic regime for wafer G782 (G743). Using the as-grown 2DEG density and mobility, the mean free path is $\ell_e = \hbar \mu \sqrt{2\pi n_{\text{2D}}}/e = 244$~nm (175~nm) for wafer G782 (G743). Devices SNS-1 and SNS-2 are in the diffusive, short junction regime, since $\ell_e < L < \xi_b$. Devices SNS-3 and SNS-4 are in the quasi-ballistic, short junction regime, since $L < \ell_e, \xi_b$. Devices SNS-5 and SNS-6 are in the diffusive, long junction regime, since $L > \ell_e, \xi_b$. All devices are in the dirty regime, since $\ell_e \lessapprox \xi_b$.

All six SNS devices demonstrated a supercurrent. Figure~\ref{fig:SC}(a) shows typical four-terminal dc I-V traces in a quasi-ballistic device. Remarkably, the I-V traces do not display any hysterectic behavior, as emphasized in the upper inset of Fig.~\ref{fig:SC}(a); this is true for all six SNS devices reported here. All devices are well thermalized and do not adversely suffer from local Joule heating effects,\cite{Hazra10,Vodolazov11} due to the dilution refrigerator (Kelvinox TLM from Oxford Instruments) used in the experiments, where samples and wiring are completely immersed in the $^3$He/$^4$He mixture. The lower inset in Fig.~\ref{fig:SC}(a) demonstrates a deep, sharply-defined superconductivity gap.

Figure~\ref{fig:SC}(b) plots the four-terminal differential conductance $G=dI/dV$ normalized by the conductance in the normal regime $G_{\textsc{n}}=1/R_{\textsc{n}}$, with periodic peaks arising from multiple Andreev reflections (MAR). The peak periodicity in $V_{dc}$ is described by $eV_n = 2\Delta_{\textsc{mar}}/n$ where $n$ is an integer. Up to six peaks are observed in Fig.~\ref{fig:SC}(b), with the first three ($n=1, ... , 3$) following the MAR sequence with $\Delta_{\textsc{mar}}=1.05$~meV. This value is only slightly smaller than $\Delta_{\text{Nb}}=1.23$~meV, and is consistent with a high-quality SNS junction. Section V in the supplementary material shows I-V traces and MAR analysis for all six SNS devices reported here (Figs.~S8$-$S10).

The ratio $e I_c R_{\textsc{n}}/\Delta_{\textsc{mar}}$ and critical current density $I_c/W$ are often used as a figure of merit for SNS junctions, with typical ranges 0.02$-$0.15 and 0.3$-$1.6 $\mu$A/$\mu$m respectively for planar In(Ga)As quantum wells proximitized to Nb.\cite{Nitta92,Takayanagi95,Heida98,Giazotto04} For the four devices in the short junction regime (SNS\nobreakdash-1, ..., SNS\nobreakdash-4 in Table~\ref{tab:SNS}), the ratio $e I_c R_{\textsc{n}}/\Delta_{\textsc{mar}}$ ranges from 0.64 to 0.79, and the critical current density $I_c/W$ ranges from 8.2 to 9.6 $\mu$A/$\mu$m. Both set of numbers indicate a very strong proximity effect from the Nb parent superconductor to the proximitized InAs 2DEG in the gap region.

A high-quality SNS junction is characterized by a high interface transparency $\mathcal{T}$ at the SN interfaces, which corresponds to a high probability of Andreev reflection.\cite{Blonder82,Octavio83,Flensberg88,Cuevas96} It can be analytically calculated with the generalized Octavion-Tinkham–Blonder-Klapwijk (OTBK) model:\cite{Niebler09}
\begin{eqnarray}
\frac{eI_{ex}R_{\textsc{n}}}{\Delta_{\textsc{mar}}}&=&
\frac{2(1+2Z^2)\tanh^{-1}(2\tilde{Z})}{Z\sqrt{(1+Z^2)(1+6Z^2+4Z^4)}}-\frac{4}{3}\qquad \label{eq:Z}\\
\text{and~~}\mathcal{T}&=&(1+Z^2)^{-1} \label{eq:transparency}
\end{eqnarray}
where $\tilde{Z}=Z\sqrt{(1+Z^2)/(1+6Z^2+4Z^4)}$, $I_{ex}$ is the excess current obtained from an I-V trace [see Fig.~\ref{fig:SC}(a)], $\Delta_{\textsc{mar}}$ is the superconducting gap determined from the MAR periodicity [see Fig.~\ref{fig:SC}(b)], and $Z$ is a dimensionless scattering parameter related to the barrier height at the SN interface. The formalism above explicitly assumes that both SN interfaces in the SNS junction are symmetric. To calculate the SN transparency, $Z$ is first used as a free variable in  Eq.~(\ref{eq:Z}) to fit the experimental data ($I_{ex}$, $R_{\textsc{n}}$, $\Delta_{\textsc{mar}}$). Once a value for $Z$ is found, it is then used in Eq.~(\ref{eq:transparency}) to calculate $\mathcal{T}$.

Using only the data at base temperature and Eqns.~(\ref{eq:Z})$-$(\ref{eq:transparency}), all six SNS devices in Table~\ref{tab:SNS} show consistently high transparencies $\mathcal{T}$ ranging in values from 78\% to 87\%. For context, reported interface transparencies of epitaxial aluminum to In(Ga)As surface quantum wells range from 75\% to 97\%,\cite{Kjaergaard17,Mayer20-A,Hertel21} when using the same experimental measurement method.

Another method for measuring $\mathcal{T}$ involves the temperature dependence of $I_c$ to be fit to the generalized Kulik-Omelyanchuk (KO) model:\cite{Haberkorn78}
\begin{eqnarray}
I_c(T,\phi,\mathcal{T})&=&\frac{\pi \Delta(T)\sin(\phi)}{2eR_{\textsc{n}}\sqrt{1-\mathcal{T}\sin^2(\phi/2)}}
\qquad\qquad\qquad\qquad\nonumber\\
&&\times\tanh\left(\frac{\Delta(T)\sqrt{1-\mathcal{T}\sin^2(\phi/2)}}
{2k_{\textsc{b}}T}\right) \label{eq:Ic-Tdep}
\end{eqnarray}
where $\phi$ is the superconducting phase picked up from Andreev reflections at the semiconductor-superconductor interface, and $\Delta(T)$ is the temperature-dependent  superconducting gap, which we model with the BCS theory relation $\Delta(T)=\Delta_{\textsc{mar}}\sqrt{1-(T/T_c)^2}$ with $T_c = 6.0$~K. The conventional fitting procedure involves finding the value of $\phi$ that maximizes $I_c$ for each temperature.\cite{Mayer19,LiT18,LeeGH15,Borzenets16}

Figure \ref{fig:SC}(c) shows the temperature dependence and fit of $I_c$ on device SNS-1, where the experimental data is normalized by the maximum current for $\mathcal{T}=100$\% predicted by Eqn.~(\ref{eq:Ic-Tdep}). The experiment yielded a transparency of $\mathcal{T}=97.8$\%. The same experiment was also performed on device SNS-3 (shown in Fig.~S11 of the supplementary material), which yielded $\mathcal{T}=99.5$\%. Both values are significantly larger than the transparencies obtained by the MAR analysis ($\mathcal{T} = 78 \pm 5$\,\%, $85 \pm 2$\%) in the same sample [see Table~\ref{tab:SNS}].

Regardless of which experimental method is used for measuring $\mathcal{T}$, our main result is that semiconductor-superconductor interface transparencies achieved in samples with a superconductor deposited post-growth can be comparable to those from ``epitaxial'' superconductors. This could dramatically expand the repertoire of possible superconductors available for realizing semiconductor-superconductor hybrid devices, potentially impacting the fields of topological quantum computing, superconducting qubits (via the gatemon design), and superconducting logic circuits.

In conclusion, we presented Josephson SNS junctions fabricated with ex-situ sputtered Nb contacts to 2DEGs hosted in InAs/AlGaSb surface quantum wells. We observed consistent and highly transparent interfaces with values of $\mathcal{T}$ ranging 78$-$99\%. Post-growth superconducting contacts to InAs quantum wells can be a viable method on a par with epitaxial aluminum systems, and do not depend on unreasonably stringent fabrication parameters in the InAs material system.

\section*{supplementary material}

The supplementary material contains additional information on MBE growth, bandstructure profiles, sample fabrication, characterization of Hall bars, and I-V/MAR traces of SNS junctions.

E.A.B., F.S., and A.E. contributed equally to this paper. The authors thank Kaveh Gharavi, Sean Walker, and Christine Nicoll for useful discussions. E.A.B. acknowledges support from a Mike and Ophelia Lazaridis Fellowship. This research was undertaken thanks in part to funding from the Canada First Research Excellence Fund (Transformative Quantum Technologies) and the Natural Sciences and Engineering Research Council (NSERC) of Canada. The University of Waterloo's QNFCF facility was used for this work. This infrastructure would not be possible without the significant contributions of CFREF-TQT, CFI, ISED, the Ontario Ministry of Research and Innovation, and Mike and Ophelia Lazaridis. Their support is gratefully acknowledged.


\begin{thebibliography}{73}%
\makeatletter
\providecommand \@ifxundefined [1]{%
 \@ifx{#1\undefined}
}%
\providecommand \@ifnum [1]{%
 \ifnum #1\expandafter \@firstoftwo
 \else \expandafter \@secondoftwo
 \fi
}%
\providecommand \@ifx [1]{%
 \ifx #1\expandafter \@firstoftwo
 \else \expandafter \@secondoftwo
 \fi
}%
\providecommand \natexlab [1]{#1}%
\providecommand \enquote  [1]{``#1''}%
\providecommand \bibnamefont  [1]{#1}%
\providecommand \bibfnamefont [1]{#1}%
\providecommand \citenamefont [1]{#1}%
\providecommand \href@noop [0]{\@secondoftwo}%
\providecommand \href [0]{\begingroup \@sanitize@url \@href}%
\providecommand \@href[1]{\@@startlink{#1}\@@href}%
\providecommand \@@href[1]{\endgroup#1\@@endlink}%
\providecommand \@sanitize@url [0]{\catcode `\\12\catcode `\$12\catcode
  `\&12\catcode `\#12\catcode `\^12\catcode `\_12\catcode `\%12\relax}%
\providecommand \@@startlink[1]{}%
\providecommand \@@endlink[0]{}%
\providecommand \url  [0]{\begingroup\@sanitize@url \@url }%
\providecommand \@url [1]{\endgroup\@href {#1}{\urlprefix }}%
\providecommand \urlprefix  [0]{URL }%
\providecommand \Eprint [0]{\href }%
\providecommand \doibase [0]{https://doi.org/}%
\providecommand \selectlanguage [0]{\@gobble}%
\providecommand \bibinfo  [0]{\@secondoftwo}%
\providecommand \bibfield  [0]{\@secondoftwo}%
\providecommand \translation [1]{[#1]}%
\providecommand \BibitemOpen [0]{}%
\providecommand \bibitemStop [0]{}%
\providecommand \bibitemNoStop [0]{.\EOS\space}%
\providecommand \EOS [0]{\spacefactor3000\relax}%
\providecommand \BibitemShut  [1]{\csname bibitem#1\endcsname}%
\let\auto@bib@innerbib\@empty
\bibitem [{\citenamefont {Shojaei}\ \emph {et~al.}(2015)\citenamefont
  {Shojaei}, \citenamefont {McFadden}, \citenamefont {Shabani}, \citenamefont
  {Schultz},\ and\ \citenamefont {Palmstr{\o}m}}]{Shojaei15}%
  \BibitemOpen
  \bibfield  {author} {\bibinfo {author} {\bibfnamefont {B.}~\bibnamefont
  {Shojaei}}, \bibinfo {author} {\bibfnamefont {A.}~\bibnamefont {McFadden}},
  \bibinfo {author} {\bibfnamefont {J.}~\bibnamefont {Shabani}}, \bibinfo
  {author} {\bibfnamefont {B.~D.}\ \bibnamefont {Schultz}},\ and\ \bibinfo
  {author} {\bibfnamefont {C.~J.}\ \bibnamefont {Palmstr{\o}m}},\ }\bibfield
  {title} {\enquote {\bibinfo {title} {Studies of scattering mechanisms in gate
  tunable {InAs/(Al,Ga)Sb} two dimensional electron gases},}\ }\href@noop {}
  {\bibfield  {journal} {\bibinfo  {journal} {Appl. Phys. Lett.}\ }\textbf
  {\bibinfo {volume} {106}},\ \bibinfo {pages} {222101} (\bibinfo {year}
  {2015})}\BibitemShut {NoStop}%
\bibitem [{\citenamefont {Shojaei}\ \emph
  {et~al.}(2016{\natexlab{a}})\citenamefont {Shojaei}, \citenamefont
  {O’Malley}, \citenamefont {Shabani}, \citenamefont {Roushan}, \citenamefont
  {Schultz}, \citenamefont {Lutchyn}, \citenamefont {Nayak}, \citenamefont
  {Martinis},\ and\ \citenamefont {Palmstr{\o}m}}]{Shojaei16-A}%
  \BibitemOpen
  \bibfield  {author} {\bibinfo {author} {\bibfnamefont {B.}~\bibnamefont
  {Shojaei}}, \bibinfo {author} {\bibfnamefont {P.~J.~J.}\ \bibnamefont
  {O’Malley}}, \bibinfo {author} {\bibfnamefont {J.}~\bibnamefont {Shabani}},
  \bibinfo {author} {\bibfnamefont {P.}~\bibnamefont {Roushan}}, \bibinfo
  {author} {\bibfnamefont {B.~D.}\ \bibnamefont {Schultz}}, \bibinfo {author}
  {\bibfnamefont {R.~M.}\ \bibnamefont {Lutchyn}}, \bibinfo {author}
  {\bibfnamefont {C.}~\bibnamefont {Nayak}}, \bibinfo {author} {\bibfnamefont
  {J.~M.}\ \bibnamefont {Martinis}},\ and\ \bibinfo {author} {\bibfnamefont
  {C.~J.}\ \bibnamefont {Palmstr{\o}m}},\ }\bibfield  {title} {\enquote
  {\bibinfo {title} {Demonstration of gate control of spin splitting in a
  high-mobility {InAs/AlSb} two-dimensional electron gas},}\ }\href@noop {}
  {\bibfield  {journal} {\bibinfo  {journal} {Phys. Rev. B}\ }\textbf {\bibinfo
  {volume} {93}},\ \bibinfo {pages} {075302} (\bibinfo {year}
  {2016}{\natexlab{a}})}\BibitemShut {NoStop}%
\bibitem [{\citenamefont {Shojaei}\ \emph
  {et~al.}(2016{\natexlab{b}})\citenamefont {Shojaei}, \citenamefont
  {Drachmann}, \citenamefont {Pendharkar}, \citenamefont {Pennachio},
  \citenamefont {Echlin}, \citenamefont {Callahan}, \citenamefont {Kraemer},
  \citenamefont {Pollock}, \citenamefont {Marcus},\ and\ \citenamefont
  {Palmstr{\o}m}}]{Shojaei16-B}%
  \BibitemOpen
  \bibfield  {author} {\bibinfo {author} {\bibfnamefont {B.}~\bibnamefont
  {Shojaei}}, \bibinfo {author} {\bibfnamefont {A.~C.~C.}\ \bibnamefont
  {Drachmann}}, \bibinfo {author} {\bibfnamefont {M.}~\bibnamefont
  {Pendharkar}}, \bibinfo {author} {\bibfnamefont {D.~J.}\ \bibnamefont
  {Pennachio}}, \bibinfo {author} {\bibfnamefont {M.~P.}\ \bibnamefont
  {Echlin}}, \bibinfo {author} {\bibfnamefont {P.~G.}\ \bibnamefont
  {Callahan}}, \bibinfo {author} {\bibfnamefont {S.}~\bibnamefont {Kraemer}},
  \bibinfo {author} {\bibfnamefont {T.~M.}\ \bibnamefont {Pollock}}, \bibinfo
  {author} {\bibfnamefont {C.~M.}\ \bibnamefont {Marcus}},\ and\ \bibinfo
  {author} {\bibfnamefont {C.~J.}\ \bibnamefont {Palmstr{\o}m}},\ }\bibfield
  {title} {\enquote {\bibinfo {title} {Limits to mobility in {InAs} quantum
  wells with nearly lattice-matched barriers},}\ }\href@noop {} {\bibfield
  {journal} {\bibinfo  {journal} {Phys. Rev. B}\ }\textbf {\bibinfo {volume}
  {94}},\ \bibinfo {pages} {245306} (\bibinfo {year}
  {2016}{\natexlab{b}})}\BibitemShut {NoStop}%
\bibitem [{\citenamefont {Hatke}\ \emph {et~al.}(2017)\citenamefont {Hatke},
  \citenamefont {Wang}, \citenamefont {Thomas}, \citenamefont {Gardner},\ and\
  \citenamefont {Manfra}}]{Hatke17}%
  \BibitemOpen
  \bibfield  {author} {\bibinfo {author} {\bibfnamefont {A.~T.}\ \bibnamefont
  {Hatke}}, \bibinfo {author} {\bibfnamefont {T.}~\bibnamefont {Wang}},
  \bibinfo {author} {\bibfnamefont {C.}~\bibnamefont {Thomas}}, \bibinfo
  {author} {\bibfnamefont {G.~C.}\ \bibnamefont {Gardner}},\ and\ \bibinfo
  {author} {\bibfnamefont {M.~J.}\ \bibnamefont {Manfra}},\ }\bibfield  {title}
  {\enquote {\bibinfo {title} {Mobility in excess of {10$^6$} {cm$^2$/Vs} in
  {InAs} quantum wells grown on lattice mismatched {InP} substrates},}\
  }\href@noop {} {\bibfield  {journal} {\bibinfo  {journal} {Appl. Phys.
  Lett.}\ }\textbf {\bibinfo {volume} {111}},\ \bibinfo {pages} {142106}
  (\bibinfo {year} {2017})}\BibitemShut {NoStop}%
\bibitem [{\citenamefont {Lee}\ \emph {et~al.}(2019)\citenamefont {Lee},
  \citenamefont {Shojaei}, \citenamefont {Pendharkar}, \citenamefont {Feldman},
  \citenamefont {Mukherjee},\ and\ \citenamefont {Palmstr{\o}m}}]{LeeJS19}%
  \BibitemOpen
  \bibfield  {author} {\bibinfo {author} {\bibfnamefont {J.~S.}\ \bibnamefont
  {Lee}}, \bibinfo {author} {\bibfnamefont {B.}~\bibnamefont {Shojaei}},
  \bibinfo {author} {\bibfnamefont {M.}~\bibnamefont {Pendharkar}}, \bibinfo
  {author} {\bibfnamefont {M.}~\bibnamefont {Feldman}}, \bibinfo {author}
  {\bibfnamefont {K.}~\bibnamefont {Mukherjee}},\ and\ \bibinfo {author}
  {\bibfnamefont {C.~J.}\ \bibnamefont {Palmstr{\o}m}},\ }\bibfield  {title}
  {\enquote {\bibinfo {title} {Contribution of top barrier materials to high
  mobility in near-surface {InAs} quantum wells grown on {GaSb(001)}},}\
  }\href@noop {} {\bibfield  {journal} {\bibinfo  {journal} {Phys. Rev.
  Mater.}\ }\textbf {\bibinfo {volume} {3}},\ \bibinfo {pages} {014603}
  (\bibinfo {year} {2019})}\BibitemShut {NoStop}%
\bibitem [{\citenamefont {Mittag}\ \emph {et~al.}(2021)\citenamefont {Mittag},
  \citenamefont {Koski}, \citenamefont {Karalic}, \citenamefont {Thomas},
  \citenamefont {Tuaz}, \citenamefont {Hatke}, \citenamefont {Gardner},
  \citenamefont {Manfra}, \citenamefont {Danon}, \citenamefont {Ihn},\ and\
  \citenamefont {Ensslin}}]{Mittag21}%
  \BibitemOpen
  \bibfield  {author} {\bibinfo {author} {\bibfnamefont {C.}~\bibnamefont
  {Mittag}}, \bibinfo {author} {\bibfnamefont {J.~V.}\ \bibnamefont {Koski}},
  \bibinfo {author} {\bibfnamefont {M.}~\bibnamefont {Karalic}}, \bibinfo
  {author} {\bibfnamefont {C.}~\bibnamefont {Thomas}}, \bibinfo {author}
  {\bibfnamefont {A.}~\bibnamefont {Tuaz}}, \bibinfo {author} {\bibfnamefont
  {A.~T.}\ \bibnamefont {Hatke}}, \bibinfo {author} {\bibfnamefont {G.~C.}\
  \bibnamefont {Gardner}}, \bibinfo {author} {\bibfnamefont {M.~J.}\
  \bibnamefont {Manfra}}, \bibinfo {author} {\bibfnamefont {J.}~\bibnamefont
  {Danon}}, \bibinfo {author} {\bibfnamefont {T.}~\bibnamefont {Ihn}},\ and\
  \bibinfo {author} {\bibfnamefont {K.}~\bibnamefont {Ensslin}},\ }\bibfield
  {title} {\enquote {\bibinfo {title} {Few-electron single and double quantum
  dots in an {InAs} two-dimensional electron gas},}\ }\href@noop {} {\bibfield
  {journal} {\bibinfo  {journal} {PRX Quantum}\ }\textbf {\bibinfo {volume}
  {2}},\ \bibinfo {pages} {010321} (\bibinfo {year} {2021})}\BibitemShut
  {NoStop}%
\bibitem [{\citenamefont {Ma}\ \emph {et~al.}(2017)\citenamefont {Ma},
  \citenamefont {Hossain}, \citenamefont {Rosales}, \citenamefont {Deng},
  \citenamefont {Tschirky}, \citenamefont {Wegscheider},\ and\ \citenamefont
  {Shayegan}}]{MaMK17}%
  \BibitemOpen
  \bibfield  {author} {\bibinfo {author} {\bibfnamefont {M.~K.}\ \bibnamefont
  {Ma}}, \bibinfo {author} {\bibfnamefont {M.~S.}\ \bibnamefont {Hossain}},
  \bibinfo {author} {\bibfnamefont {K.~A.~V.}\ \bibnamefont {Rosales}},
  \bibinfo {author} {\bibfnamefont {H.}~\bibnamefont {Deng}}, \bibinfo {author}
  {\bibfnamefont {T.}~\bibnamefont {Tschirky}}, \bibinfo {author}
  {\bibfnamefont {W.}~\bibnamefont {Wegscheider}},\ and\ \bibinfo {author}
  {\bibfnamefont {M.}~\bibnamefont {Shayegan}},\ }\bibfield  {title} {\enquote
  {\bibinfo {title} {Observation of fractional quantum hall effect in an {InAs}
  quantum well},}\ }\href@noop {} {\bibfield  {journal} {\bibinfo  {journal}
  {Phys. Rev. B}\ }\textbf {\bibinfo {volume} {96}},\ \bibinfo {pages}
  {241301(R)} (\bibinfo {year} {2017})}\BibitemShut {NoStop}%
\bibitem [{\citenamefont {Komatsu}\ \emph {et~al.}(2022)\citenamefont
  {Komatsu}, \citenamefont {Irie}, \citenamefont {Akiho}, \citenamefont
  {Nojima}, \citenamefont {Akazaki},\ and\ \citenamefont
  {Muraki}}]{KomatsuS22}%
  \BibitemOpen
  \bibfield  {author} {\bibinfo {author} {\bibfnamefont {S.}~\bibnamefont
  {Komatsu}}, \bibinfo {author} {\bibfnamefont {H.}~\bibnamefont {Irie}},
  \bibinfo {author} {\bibfnamefont {T.}~\bibnamefont {Akiho}}, \bibinfo
  {author} {\bibfnamefont {T.}~\bibnamefont {Nojima}}, \bibinfo {author}
  {\bibfnamefont {T.}~\bibnamefont {Akazaki}},\ and\ \bibinfo {author}
  {\bibfnamefont {K.}~\bibnamefont {Muraki}},\ }\bibfield  {title} {\enquote
  {\bibinfo {title} {Gate tuning of fractional quantum {Hall} states in an
  {InAs} two-dimensional electron gas},}\ }\href@noop {} {\bibfield  {journal}
  {\bibinfo  {journal} {Phys. Rev. B}\ }\textbf {\bibinfo {volume} {105}},\
  \bibinfo {pages} {075305} (\bibinfo {year} {2022})}\BibitemShut {NoStop}%
\bibitem [{\citenamefont {Pan}\ \emph {et~al.}(2008)\citenamefont {Pan},
  \citenamefont {Xia}, \citenamefont {{St\"{o}rmer}}, \citenamefont {Tsui},
  \citenamefont {Vicente}, \citenamefont {Adams}, \citenamefont {Sullivan},
  \citenamefont {Pfeiffer}, \citenamefont {Baldwin},\ and\ \citenamefont
  {West}}]{Pan08}%
  \BibitemOpen
  \bibfield  {author} {\bibinfo {author} {\bibfnamefont {W.}~\bibnamefont
  {Pan}}, \bibinfo {author} {\bibfnamefont {J.~S.}\ \bibnamefont {Xia}},
  \bibinfo {author} {\bibfnamefont {H.~L.}\ \bibnamefont {{St\"{o}rmer}}},
  \bibinfo {author} {\bibfnamefont {D.~C.}\ \bibnamefont {Tsui}}, \bibinfo
  {author} {\bibfnamefont {C.}~\bibnamefont {Vicente}}, \bibinfo {author}
  {\bibfnamefont {E.~D.}\ \bibnamefont {Adams}}, \bibinfo {author}
  {\bibfnamefont {N.~S.}\ \bibnamefont {Sullivan}}, \bibinfo {author}
  {\bibfnamefont {L.~N.}\ \bibnamefont {Pfeiffer}}, \bibinfo {author}
  {\bibfnamefont {K.~W.}\ \bibnamefont {Baldwin}},\ and\ \bibinfo {author}
  {\bibfnamefont {K.~W.}\ \bibnamefont {West}},\ }\bibfield  {title} {\enquote
  {\bibinfo {title} {Experimental studies of the fractional quantum hall effect
  in the first excited {Landau} level},}\ }\href@noop {} {\bibfield  {journal}
  {\bibinfo  {journal} {Phys. Rev. B}\ }\textbf {\bibinfo {volume} {77}},\
  \bibinfo {pages} {075307} (\bibinfo {year} {2008})}\BibitemShut {NoStop}%
\bibitem [{\citenamefont {Kleinbaum}\ \emph {et~al.}(2020)\citenamefont
  {Kleinbaum}, \citenamefont {Li}, \citenamefont {Deng}, \citenamefont
  {Gardner}, \citenamefont {Manfra},\ and\ \citenamefont
  {{Cs\'{a}thy}}}]{Kleinbaum20}%
  \BibitemOpen
  \bibfield  {author} {\bibinfo {author} {\bibfnamefont {E.}~\bibnamefont
  {Kleinbaum}}, \bibinfo {author} {\bibfnamefont {H.}~\bibnamefont {Li}},
  \bibinfo {author} {\bibfnamefont {N.}~\bibnamefont {Deng}}, \bibinfo {author}
  {\bibfnamefont {G.~C.}\ \bibnamefont {Gardner}}, \bibinfo {author}
  {\bibfnamefont {M.~J.}\ \bibnamefont {Manfra}},\ and\ \bibinfo {author}
  {\bibfnamefont {G.~A.}\ \bibnamefont {{Cs\'{a}thy}}},\ }\bibfield  {title}
  {\enquote {\bibinfo {title} {Disorder broadening of even-denominator
  fractional quantum {Hall} states in the presence of a short-range alloy
  potential},}\ }\href@noop {} {\bibfield  {journal} {\bibinfo  {journal}
  {Phys. Rev. B}\ }\textbf {\bibinfo {volume} {102}},\ \bibinfo {pages}
  {035140} (\bibinfo {year} {2020})}\BibitemShut {NoStop}%
\bibitem [{\citenamefont {Chung}\ \emph {et~al.}(2018)\citenamefont {Chung},
  \citenamefont {Rosales}, \citenamefont {Deng}, \citenamefont {Baldwin},
  \citenamefont {West}, \citenamefont {Shayegan},\ and\ \citenamefont
  {Pfeiffer}}]{ChungYJ18-B}%
  \BibitemOpen
  \bibfield  {author} {\bibinfo {author} {\bibfnamefont {Y.~J.}\ \bibnamefont
  {Chung}}, \bibinfo {author} {\bibfnamefont {K.~A.~V.}\ \bibnamefont
  {Rosales}}, \bibinfo {author} {\bibfnamefont {H.}~\bibnamefont {Deng}},
  \bibinfo {author} {\bibfnamefont {K.~W.}\ \bibnamefont {Baldwin}}, \bibinfo
  {author} {\bibfnamefont {K.~W.}\ \bibnamefont {West}}, \bibinfo {author}
  {\bibfnamefont {M.}~\bibnamefont {Shayegan}},\ and\ \bibinfo {author}
  {\bibfnamefont {L.~N.}\ \bibnamefont {Pfeiffer}},\ }\bibfield  {title}
  {\enquote {\bibinfo {title} {Multivalley two-dimensional electron system in
  an {AlAs} quantum well with mobility exceeding {2$\times$10$^6$
  cm$^2$/Vs}},}\ }\href@noop {} {\ \textbf {\bibinfo {volume} {2}},\ \bibinfo
  {pages} {071001(R)} (\bibinfo {year} {2018})}\BibitemShut {NoStop}%
\bibitem [{\citenamefont {Bolotin}\ \emph {et~al.}(2009)\citenamefont
  {Bolotin}, \citenamefont {Ghahari}, \citenamefont {Shulman}, \citenamefont
  {Stormer},\ and\ \citenamefont {Kim}}]{Bolotin09}%
  \BibitemOpen
  \bibfield  {author} {\bibinfo {author} {\bibfnamefont {K.~I.}\ \bibnamefont
  {Bolotin}}, \bibinfo {author} {\bibfnamefont {F.}~\bibnamefont {Ghahari}},
  \bibinfo {author} {\bibfnamefont {M.~D.}\ \bibnamefont {Shulman}}, \bibinfo
  {author} {\bibfnamefont {H.~L.}\ \bibnamefont {Stormer}},\ and\ \bibinfo
  {author} {\bibfnamefont {P.}~\bibnamefont {Kim}},\ }\bibfield  {title}
  {\enquote {\bibinfo {title} {Observation of the fractional quantum {Hall}
  effect in graphene},}\ }\href@noop {} {\bibfield  {journal} {\bibinfo
  {journal} {Nature}\ }\textbf {\bibinfo {volume} {462}},\ \bibinfo {pages}
  {196} (\bibinfo {year} {2009})}\BibitemShut {NoStop}%
\bibitem [{\citenamefont {Dean}\ \emph {et~al.}(2011)\citenamefont {Dean},
  \citenamefont {Young}, \citenamefont {Cadden-Zimansky}, \citenamefont {Wang},
  \citenamefont {Ren}, \citenamefont {Watanabe}, \citenamefont {Taniguchi},
  \citenamefont {Kim}, \citenamefont {Hone},\ and\ \citenamefont
  {Shepard}}]{Dean11}%
  \BibitemOpen
  \bibfield  {author} {\bibinfo {author} {\bibfnamefont {C.~R.}\ \bibnamefont
  {Dean}}, \bibinfo {author} {\bibfnamefont {A.~F.}\ \bibnamefont {Young}},
  \bibinfo {author} {\bibfnamefont {P.}~\bibnamefont {Cadden-Zimansky}},
  \bibinfo {author} {\bibfnamefont {L.}~\bibnamefont {Wang}}, \bibinfo {author}
  {\bibfnamefont {H.}~\bibnamefont {Ren}}, \bibinfo {author} {\bibfnamefont
  {K.}~\bibnamefont {Watanabe}}, \bibinfo {author} {\bibfnamefont
  {T.}~\bibnamefont {Taniguchi}}, \bibinfo {author} {\bibfnamefont
  {P.}~\bibnamefont {Kim}}, \bibinfo {author} {\bibfnamefont {J.}~\bibnamefont
  {Hone}},\ and\ \bibinfo {author} {\bibfnamefont {K.~L.}\ \bibnamefont
  {Shepard}},\ }\bibfield  {title} {\enquote {\bibinfo {title} {Multicomponent
  fractional quantum hall effect in graphene},}\ }\href@noop {} {\bibfield
  {journal} {\bibinfo  {journal} {Nat. Phys.}\ }\textbf {\bibinfo {volume}
  {7}},\ \bibinfo {pages} {693} (\bibinfo {year} {2011})}\BibitemShut {NoStop}%
\bibitem [{\citenamefont {Lai}\ \emph {et~al.}(2004)\citenamefont {Lai},
  \citenamefont {Pan}, \citenamefont {Tsui}, \citenamefont {Lyon},
  \citenamefont {M{\"{u}}hlberger},\ and\ \citenamefont
  {Sch{\"{a}}ffler}}]{LaiK04}%
  \BibitemOpen
  \bibfield  {author} {\bibinfo {author} {\bibfnamefont {K.}~\bibnamefont
  {Lai}}, \bibinfo {author} {\bibfnamefont {W.}~\bibnamefont {Pan}}, \bibinfo
  {author} {\bibfnamefont {D.~C.}\ \bibnamefont {Tsui}}, \bibinfo {author}
  {\bibfnamefont {S.}~\bibnamefont {Lyon}}, \bibinfo {author} {\bibfnamefont
  {M.}~\bibnamefont {M{\"{u}}hlberger}},\ and\ \bibinfo {author} {\bibfnamefont
  {F.}~\bibnamefont {Sch{\"{a}}ffler}},\ }\bibfield  {title} {\enquote
  {\bibinfo {title} {Two-flux composite fermion series of the fractional
  quantum {Hall} states in strained {Si}},}\ }\href@noop {} {\bibfield
  {journal} {\bibinfo  {journal} {Phys. Rev. Lett.}\ }\textbf {\bibinfo
  {volume} {93}},\ \bibinfo {pages} {156805} (\bibinfo {year}
  {2004})}\BibitemShut {NoStop}%
\bibitem [{\citenamefont {Lu}\ \emph {et~al.}(2012)\citenamefont {Lu},
  \citenamefont {Pan}, \citenamefont {Tsui}, \citenamefont {Lee},\ and\
  \citenamefont {Liu}}]{LuTM12}%
  \BibitemOpen
  \bibfield  {author} {\bibinfo {author} {\bibfnamefont {T.~M.}\ \bibnamefont
  {Lu}}, \bibinfo {author} {\bibfnamefont {W.}~\bibnamefont {Pan}}, \bibinfo
  {author} {\bibfnamefont {D.~C.}\ \bibnamefont {Tsui}}, \bibinfo {author}
  {\bibfnamefont {C.-H.}\ \bibnamefont {Lee}},\ and\ \bibinfo {author}
  {\bibfnamefont {C.~W.}\ \bibnamefont {Liu}},\ }\bibfield  {title} {\enquote
  {\bibinfo {title} {Fractional quantum {Hall} effect of two-dimensional
  electrons in high-mobility {Si/SiGe} field-effect transistors},}\ }\href@noop
  {} {\bibfield  {journal} {\bibinfo  {journal} {Phys. Rev. B}\ }\textbf
  {\bibinfo {volume} {85}},\ \bibinfo {pages} {121307(R)} (\bibinfo {year}
  {2012})}\BibitemShut {NoStop}%
\bibitem [{\citenamefont {Mironov}\ \emph {et~al.}(2016)\citenamefont
  {Mironov}, \citenamefont {{d'Ambrumenil}}, \citenamefont {Dobbie},
  \citenamefont {Leadley}, \citenamefont {Suslov},\ and\ \citenamefont
  {Green}}]{Mironov16}%
  \BibitemOpen
  \bibfield  {author} {\bibinfo {author} {\bibfnamefont {O.~A.}\ \bibnamefont
  {Mironov}}, \bibinfo {author} {\bibfnamefont {N.}~\bibnamefont
  {{d'Ambrumenil}}}, \bibinfo {author} {\bibfnamefont {A.}~\bibnamefont
  {Dobbie}}, \bibinfo {author} {\bibfnamefont {D.~R.}\ \bibnamefont {Leadley}},
  \bibinfo {author} {\bibfnamefont {A.~V.}\ \bibnamefont {Suslov}},\ and\
  \bibinfo {author} {\bibfnamefont {E.}~\bibnamefont {Green}},\ }\bibfield
  {title} {\enquote {\bibinfo {title} {Fractional quantum {Hall} states in a
  {Ge} quantum well},}\ }\href@noop {} {\bibfield  {journal} {\bibinfo
  {journal} {Phys. Rev. Lett.}\ }\textbf {\bibinfo {volume} {116}},\ \bibinfo
  {pages} {176802} (\bibinfo {year} {2016})}\BibitemShut {NoStop}%
\bibitem [{\citenamefont {Piot}\ \emph {et~al.}(2010)\citenamefont {Piot},
  \citenamefont {Kunc}, \citenamefont {Potemski}, \citenamefont {Maude},
  \citenamefont {Betthausen}, \citenamefont {Vogl}, \citenamefont {Weiss},
  \citenamefont {Karczewski},\ and\ \citenamefont {Wojtowicz}}]{Piot10}%
  \BibitemOpen
  \bibfield  {author} {\bibinfo {author} {\bibfnamefont {B.~A.}\ \bibnamefont
  {Piot}}, \bibinfo {author} {\bibfnamefont {J.}~\bibnamefont {Kunc}}, \bibinfo
  {author} {\bibfnamefont {M.}~\bibnamefont {Potemski}}, \bibinfo {author}
  {\bibfnamefont {D.~K.}\ \bibnamefont {Maude}}, \bibinfo {author}
  {\bibfnamefont {C.}~\bibnamefont {Betthausen}}, \bibinfo {author}
  {\bibfnamefont {A.}~\bibnamefont {Vogl}}, \bibinfo {author} {\bibfnamefont
  {D.}~\bibnamefont {Weiss}}, \bibinfo {author} {\bibfnamefont
  {G.}~\bibnamefont {Karczewski}},\ and\ \bibinfo {author} {\bibfnamefont
  {T.}~\bibnamefont {Wojtowicz}},\ }\bibfield  {title} {\enquote {\bibinfo
  {title} {Fractional quantum {Hall} effect in {CdTe}},}\ }\href@noop {}
  {\bibfield  {journal} {\bibinfo  {journal} {Phys. Rev. B}\ }\textbf {\bibinfo
  {volume} {82}},\ \bibinfo {pages} {081307(R)} (\bibinfo {year}
  {2010})}\BibitemShut {NoStop}%
\bibitem [{\citenamefont {Betthausen}\ \emph {et~al.}(2014)\citenamefont
  {Betthausen}, \citenamefont {Giudici}, \citenamefont {Iankilevitch},
  \citenamefont {Preis}, \citenamefont {Kolkovsky}, \citenamefont {Wiater},
  \citenamefont {Karczewski}, \citenamefont {Piot}, \citenamefont {Kunc},
  \citenamefont {Potemski}, \citenamefont {Wojtowicz},\ and\ \citenamefont
  {Weiss}}]{Betthausen14}%
  \BibitemOpen
  \bibfield  {author} {\bibinfo {author} {\bibfnamefont {C.}~\bibnamefont
  {Betthausen}}, \bibinfo {author} {\bibfnamefont {P.}~\bibnamefont {Giudici}},
  \bibinfo {author} {\bibfnamefont {A.}~\bibnamefont {Iankilevitch}}, \bibinfo
  {author} {\bibfnamefont {C.}~\bibnamefont {Preis}}, \bibinfo {author}
  {\bibfnamefont {V.}~\bibnamefont {Kolkovsky}}, \bibinfo {author}
  {\bibfnamefont {M.}~\bibnamefont {Wiater}}, \bibinfo {author} {\bibfnamefont
  {G.}~\bibnamefont {Karczewski}}, \bibinfo {author} {\bibfnamefont {B.~A.}\
  \bibnamefont {Piot}}, \bibinfo {author} {\bibfnamefont {J.}~\bibnamefont
  {Kunc}}, \bibinfo {author} {\bibfnamefont {M.}~\bibnamefont {Potemski}},
  \bibinfo {author} {\bibfnamefont {T.}~\bibnamefont {Wojtowicz}},\ and\
  \bibinfo {author} {\bibfnamefont {D.}~\bibnamefont {Weiss}},\ }\bibfield
  {title} {\enquote {\bibinfo {title} {Fractional quantum {Hall} effect in a
  dilute magnetic semiconductor},}\ }\href@noop {} {\bibfield  {journal}
  {\bibinfo  {journal} {Phys. Rev. B}\ }\textbf {\bibinfo {volume} {90}},\
  \bibinfo {pages} {115302} (\bibinfo {year} {2014})}\BibitemShut {NoStop}%
\bibitem [{\citenamefont {Tsukazaki1}\ \emph {et~al.}(2010)\citenamefont
  {Tsukazaki1}, \citenamefont {Akasaka}, \citenamefont {Nakahara},
  \citenamefont {Ohno}, \citenamefont {Ohno}, \citenamefont {Maryenko},
  \citenamefont {Ohtomo},\ and\ \citenamefont {Kawasaki}}]{Tsukazaki10}%
  \BibitemOpen
  \bibfield  {author} {\bibinfo {author} {\bibfnamefont {A.}~\bibnamefont
  {Tsukazaki1}}, \bibinfo {author} {\bibfnamefont {S.}~\bibnamefont {Akasaka}},
  \bibinfo {author} {\bibfnamefont {K.}~\bibnamefont {Nakahara}}, \bibinfo
  {author} {\bibfnamefont {Y.}~\bibnamefont {Ohno}}, \bibinfo {author}
  {\bibfnamefont {H.}~\bibnamefont {Ohno}}, \bibinfo {author} {\bibfnamefont
  {D.}~\bibnamefont {Maryenko}}, \bibinfo {author} {\bibfnamefont
  {A.}~\bibnamefont {Ohtomo}},\ and\ \bibinfo {author} {\bibfnamefont
  {M.}~\bibnamefont {Kawasaki}},\ }\bibfield  {title} {\enquote {\bibinfo
  {title} {Observation of the fractional quantum {Hall} effect in an oxide},}\
  }\href@noop {} {\bibfield  {journal} {\bibinfo  {journal} {Nat. Mater.}\
  }\textbf {\bibinfo {volume} {9}},\ \bibinfo {pages} {889} (\bibinfo {year}
  {2010})}\BibitemShut {NoStop}%
\bibitem [{\citenamefont {Falson}\ and\ \citenamefont
  {Kawasaki}(2018)}]{Falson18}%
  \BibitemOpen
  \bibfield  {author} {\bibinfo {author} {\bibfnamefont {J.}~\bibnamefont
  {Falson}}\ and\ \bibinfo {author} {\bibfnamefont {M.}~\bibnamefont
  {Kawasaki}},\ }\bibfield  {title} {\enquote {\bibinfo {title} {A review of
  the quantum {Hall} effects in {MgZnO/ZnO} heterostructures},}\ }\href@noop {}
  {\bibfield  {journal} {\bibinfo  {journal} {Rep. Prog. Phys.}\ }\textbf
  {\bibinfo {volume} {81}},\ \bibinfo {pages} {056501} (\bibinfo {year}
  {2018})}\BibitemShut {NoStop}%
\bibitem [{\citenamefont {Tschirky}\ \emph {et~al.}(2017)\citenamefont
  {Tschirky}, \citenamefont {Mueller}, \citenamefont {Lehner}, \citenamefont
  {{F\"{a}lt}}, \citenamefont {Ihn}, \citenamefont {Ensslin},\ and\
  \citenamefont {Wegscheider}}]{Tschirky17}%
  \BibitemOpen
  \bibfield  {author} {\bibinfo {author} {\bibfnamefont {T.}~\bibnamefont
  {Tschirky}}, \bibinfo {author} {\bibfnamefont {S.}~\bibnamefont {Mueller}},
  \bibinfo {author} {\bibfnamefont {C.~A.}\ \bibnamefont {Lehner}}, \bibinfo
  {author} {\bibfnamefont {S.}~\bibnamefont {{F\"{a}lt}}}, \bibinfo {author}
  {\bibfnamefont {T.}~\bibnamefont {Ihn}}, \bibinfo {author} {\bibfnamefont
  {K.}~\bibnamefont {Ensslin}},\ and\ \bibinfo {author} {\bibfnamefont
  {W.}~\bibnamefont {Wegscheider}},\ }\bibfield  {title} {\enquote {\bibinfo
  {title} {Scattering mechanisms of highest-mobility
  {{InAs/Al$_x$Ga$_{1-x}$Sb}} quantum wells},}\ }\href@noop {} {\bibfield
  {journal} {\bibinfo  {journal} {Phys. Rev. B}\ }\textbf {\bibinfo {volume}
  {95}},\ \bibinfo {pages} {115304} (\bibinfo {year} {2017})}\BibitemShut
  {NoStop}%
\bibitem [{\citenamefont {Thomas}\ \emph {et~al.}(2018)\citenamefont {Thomas},
  \citenamefont {Hatke}, \citenamefont {Tuaz}, \citenamefont {Kallaher},
  \citenamefont {Wu}, \citenamefont {Wang}, \citenamefont {Diaz}, \citenamefont
  {Gardner}, \citenamefont {Capano},\ and\ \citenamefont {Manfra}}]{ThomasC18}%
  \BibitemOpen
  \bibfield  {author} {\bibinfo {author} {\bibfnamefont {C.}~\bibnamefont
  {Thomas}}, \bibinfo {author} {\bibfnamefont {A.~T.}\ \bibnamefont {Hatke}},
  \bibinfo {author} {\bibfnamefont {A.}~\bibnamefont {Tuaz}}, \bibinfo {author}
  {\bibfnamefont {R.}~\bibnamefont {Kallaher}}, \bibinfo {author}
  {\bibfnamefont {T.}~\bibnamefont {Wu}}, \bibinfo {author} {\bibfnamefont
  {T.}~\bibnamefont {Wang}}, \bibinfo {author} {\bibfnamefont {R.~E.}\
  \bibnamefont {Diaz}}, \bibinfo {author} {\bibfnamefont {G.~C.}\ \bibnamefont
  {Gardner}}, \bibinfo {author} {\bibfnamefont {M.~A.}\ \bibnamefont
  {Capano}},\ and\ \bibinfo {author} {\bibfnamefont {M.~J.}\ \bibnamefont
  {Manfra}},\ }\bibfield  {title} {\enquote {\bibinfo {title} {High-mobility
  {InAs} {2DEGs} on {GaSb} substrates: {A} platform for mesoscopic quantum
  transport},}\ }\href@noop {} {\bibfield  {journal} {\bibinfo  {journal}
  {Phys. Rev. Mater.}\ }\textbf {\bibinfo {volume} {2}},\ \bibinfo {pages}
  {104602} (\bibinfo {year} {2018})}\BibitemShut {NoStop}%
\bibitem [{\citenamefont {Umansky}\ \emph {et~al.}(2009)\citenamefont
  {Umansky}, \citenamefont {Heiblum}, \citenamefont {Levinson}, \citenamefont
  {Smet}, \citenamefont {N{\"{u}}bler},\ and\ \citenamefont
  {Dolev}}]{Umansky09}%
  \BibitemOpen
  \bibfield  {author} {\bibinfo {author} {\bibfnamefont {V.}~\bibnamefont
  {Umansky}}, \bibinfo {author} {\bibfnamefont {M.}~\bibnamefont {Heiblum}},
  \bibinfo {author} {\bibfnamefont {Y.}~\bibnamefont {Levinson}}, \bibinfo
  {author} {\bibfnamefont {J.}~\bibnamefont {Smet}}, \bibinfo {author}
  {\bibfnamefont {J.}~\bibnamefont {N{\"{u}}bler}},\ and\ \bibinfo {author}
  {\bibfnamefont {M.}~\bibnamefont {Dolev}},\ }\bibfield  {title} {\enquote
  {\bibinfo {title} {{MBE} growth of ultra-low disorder {2DEG} with mobility
  exceeding {$35\times 10^6$}cm{$^2$/Vs}},}\ }\href@noop {} {\bibfield
  {journal} {\bibinfo  {journal} {J. Cryst. Growth}\ }\textbf {\bibinfo
  {volume} {311}},\ \bibinfo {pages} {1658} (\bibinfo {year}
  {2009})}\BibitemShut {NoStop}%
\bibitem [{\citenamefont {Chung}\ \emph {et~al.}(2021)\citenamefont {Chung},
  \citenamefont {Rosales}, \citenamefont {Baldwin}, \citenamefont {Madathil},
  \citenamefont {West}, \citenamefont {Shayegan},\ and\ \citenamefont
  {Pfeiffer}}]{ChungYJ21}%
  \BibitemOpen
  \bibfield  {author} {\bibinfo {author} {\bibfnamefont {Y.~J.}\ \bibnamefont
  {Chung}}, \bibinfo {author} {\bibfnamefont {K.~A.~V.}\ \bibnamefont
  {Rosales}}, \bibinfo {author} {\bibfnamefont {K.~W.}\ \bibnamefont
  {Baldwin}}, \bibinfo {author} {\bibfnamefont {P.~T.}\ \bibnamefont
  {Madathil}}, \bibinfo {author} {\bibfnamefont {K.~W.}\ \bibnamefont {West}},
  \bibinfo {author} {\bibfnamefont {M.}~\bibnamefont {Shayegan}},\ and\
  \bibinfo {author} {\bibfnamefont {L.~N.}\ \bibnamefont {Pfeiffer}},\
  }\bibfield  {title} {\enquote {\bibinfo {title} {Ultra-high-quality
  two-dimensional electron systems},}\ }\href@noop {} {\bibfield  {journal}
  {\bibinfo  {journal} {Nat. Mater.}\ }\textbf {\bibinfo {volume} {20}},\
  \bibinfo {pages} {632} (\bibinfo {year} {2021})}\BibitemShut {NoStop}%
\bibitem [{\citenamefont {Myronov}\ \emph {et~al.}(2023)\citenamefont
  {Myronov}, \citenamefont {Kycia}, \citenamefont {Waldron}, \citenamefont
  {Jiang}, \citenamefont {Barrios}, \citenamefont {Bogan}, \citenamefont
  {Coleridge},\ and\ \citenamefont {Studenikin}}]{Myronov23}%
  \BibitemOpen
  \bibfield  {author} {\bibinfo {author} {\bibfnamefont {M.}~\bibnamefont
  {Myronov}}, \bibinfo {author} {\bibfnamefont {J.}~\bibnamefont {Kycia}},
  \bibinfo {author} {\bibfnamefont {P.}~\bibnamefont {Waldron}}, \bibinfo
  {author} {\bibfnamefont {W.}~\bibnamefont {Jiang}}, \bibinfo {author}
  {\bibfnamefont {P.}~\bibnamefont {Barrios}}, \bibinfo {author} {\bibfnamefont
  {A.}~\bibnamefont {Bogan}}, \bibinfo {author} {\bibfnamefont
  {P.}~\bibnamefont {Coleridge}},\ and\ \bibinfo {author} {\bibfnamefont
  {S.}~\bibnamefont {Studenikin}},\ }\bibfield  {title} {\enquote {\bibinfo
  {title} {Holes outperform electrons in group {IV} semiconductor materials},}\
  }\href@noop {} {\bibfield  {journal} {\bibinfo  {journal} {Small Sci.}\
  }\textbf {\bibinfo {volume} {3}},\ \bibinfo {pages} {2200094} (\bibinfo
  {year} {2023})}\BibitemShut {NoStop}%
\bibitem [{\citenamefont {Shabani}\ \emph {et~al.}(2016)\citenamefont
  {Shabani}, \citenamefont {Kj{\ae}rgaard}, \citenamefont {Suominen},
  \citenamefont {Kim}, \citenamefont {Nichele}, \citenamefont {Pakrouski},
  \citenamefont {Stankevic}, \citenamefont {Lutchyn}, \citenamefont
  {Krogstrup}, \citenamefont {Feidenhans}, \citenamefont {Kraemer},
  \citenamefont {Nayak}, \citenamefont {Troyer}, \citenamefont {Marcus},\ and\
  \citenamefont {Palmstr{\o}m}}]{Shabani16}%
  \BibitemOpen
  \bibfield  {author} {\bibinfo {author} {\bibfnamefont {J.}~\bibnamefont
  {Shabani}}, \bibinfo {author} {\bibfnamefont {M.}~\bibnamefont
  {Kj{\ae}rgaard}}, \bibinfo {author} {\bibfnamefont {H.~J.}\ \bibnamefont
  {Suominen}}, \bibinfo {author} {\bibfnamefont {Y.}~\bibnamefont {Kim}},
  \bibinfo {author} {\bibfnamefont {F.}~\bibnamefont {Nichele}}, \bibinfo
  {author} {\bibfnamefont {K.}~\bibnamefont {Pakrouski}}, \bibinfo {author}
  {\bibfnamefont {T.}~\bibnamefont {Stankevic}}, \bibinfo {author}
  {\bibfnamefont {R.~M.}\ \bibnamefont {Lutchyn}}, \bibinfo {author}
  {\bibfnamefont {P.}~\bibnamefont {Krogstrup}}, \bibinfo {author}
  {\bibfnamefont {R.}~\bibnamefont {Feidenhans}}, \bibinfo {author}
  {\bibfnamefont {S.}~\bibnamefont {Kraemer}}, \bibinfo {author} {\bibfnamefont
  {C.}~\bibnamefont {Nayak}}, \bibinfo {author} {\bibfnamefont
  {M.}~\bibnamefont {Troyer}}, \bibinfo {author} {\bibfnamefont {C.~M.}\
  \bibnamefont {Marcus}},\ and\ \bibinfo {author} {\bibfnamefont {C.~J.}\
  \bibnamefont {Palmstr{\o}m}},\ }\bibfield  {title} {\enquote {\bibinfo
  {title} {Two-dimensional epitaxial superconductor-semiconductor
  heterostructures: A platform for topological superconducting networks},}\
  }\href@noop {} {\bibfield  {journal} {\bibinfo  {journal} {Phys. Rev. B}\
  }\textbf {\bibinfo {volume} {93}},\ \bibinfo {pages} {155402} (\bibinfo
  {year} {2016})}\BibitemShut {NoStop}%
\bibitem [{\citenamefont {Karzig}\ \emph {et~al.}(2017)\citenamefont {Karzig},
  \citenamefont {Knapp}, \citenamefont {Lutchyn}, \citenamefont {Bonderson},
  \citenamefont {Hastings}, \citenamefont {Nayak}, \citenamefont {Alicea},
  \citenamefont {Flensberg}, \citenamefont {Plugge}, \citenamefont {Oreg},
  \citenamefont {Marcus},\ and\ \citenamefont {Freedman}}]{Karzig17}%
  \BibitemOpen
  \bibfield  {author} {\bibinfo {author} {\bibfnamefont {T.}~\bibnamefont
  {Karzig}}, \bibinfo {author} {\bibfnamefont {C.}~\bibnamefont {Knapp}},
  \bibinfo {author} {\bibfnamefont {R.~M.}\ \bibnamefont {Lutchyn}}, \bibinfo
  {author} {\bibfnamefont {P.}~\bibnamefont {Bonderson}}, \bibinfo {author}
  {\bibfnamefont {M.~B.}\ \bibnamefont {Hastings}}, \bibinfo {author}
  {\bibfnamefont {C.}~\bibnamefont {Nayak}}, \bibinfo {author} {\bibfnamefont
  {J.}~\bibnamefont {Alicea}}, \bibinfo {author} {\bibfnamefont
  {K.}~\bibnamefont {Flensberg}}, \bibinfo {author} {\bibfnamefont
  {S.}~\bibnamefont {Plugge}}, \bibinfo {author} {\bibfnamefont
  {Y.}~\bibnamefont {Oreg}}, \bibinfo {author} {\bibfnamefont {C.~M.}\
  \bibnamefont {Marcus}},\ and\ \bibinfo {author} {\bibfnamefont {M.~H.}\
  \bibnamefont {Freedman}},\ }\bibfield  {title} {\enquote {\bibinfo {title}
  {Scalable designs for quasiparticle-poisoning-protected topological quantum
  computation with {{Majorana}} zero modes},}\ }\href@noop {} {\bibfield
  {journal} {\bibinfo  {journal} {Phys. Rev. B}\ }\textbf {\bibinfo {volume}
  {95}},\ \bibinfo {pages} {235305} (\bibinfo {year} {2017})}\BibitemShut
  {NoStop}%
\bibitem [{\citenamefont {Ke}\ \emph {et~al.}(2019)\citenamefont {Ke},
  \citenamefont {Moehle}, \citenamefont {de~Vries}, \citenamefont {Thomas},
  \citenamefont {Metti}, \citenamefont {Guinn}, \citenamefont {Kallaher},
  \citenamefont {Lodari}, \citenamefont {Scappucci}, \citenamefont {Wang},
  \citenamefont {Diaz}, \citenamefont {Gardner}, \citenamefont {Manfra},\ and\
  \citenamefont {Goswami}}]{KeCT19}%
  \BibitemOpen
  \bibfield  {author} {\bibinfo {author} {\bibfnamefont {C.~T.}\ \bibnamefont
  {Ke}}, \bibinfo {author} {\bibfnamefont {C.~M.}\ \bibnamefont {Moehle}},
  \bibinfo {author} {\bibfnamefont {F.~K.}\ \bibnamefont {de~Vries}}, \bibinfo
  {author} {\bibfnamefont {C.}~\bibnamefont {Thomas}}, \bibinfo {author}
  {\bibfnamefont {S.}~\bibnamefont {Metti}}, \bibinfo {author} {\bibfnamefont
  {C.~R.}\ \bibnamefont {Guinn}}, \bibinfo {author} {\bibfnamefont
  {R.}~\bibnamefont {Kallaher}}, \bibinfo {author} {\bibfnamefont
  {M.}~\bibnamefont {Lodari}}, \bibinfo {author} {\bibfnamefont
  {G.}~\bibnamefont {Scappucci}}, \bibinfo {author} {\bibfnamefont
  {T.}~\bibnamefont {Wang}}, \bibinfo {author} {\bibfnamefont {R.~E.}\
  \bibnamefont {Diaz}}, \bibinfo {author} {\bibfnamefont {G.~C.}\ \bibnamefont
  {Gardner}}, \bibinfo {author} {\bibfnamefont {M.~J.}\ \bibnamefont
  {Manfra}},\ and\ \bibinfo {author} {\bibfnamefont {S.}~\bibnamefont
  {Goswami}},\ }\bibfield  {title} {\enquote {\bibinfo {title} {Ballistic
  superconductivity and tunable {$\pi$–junctions} in {InSb} quantum wells},}\
  }\href@noop {} {\bibfield  {journal} {\bibinfo  {journal} {Nat. Commun.}\
  }\textbf {\bibinfo {volume} {10}},\ \bibinfo {pages} {3764} (\bibinfo {year}
  {2019})}\BibitemShut {NoStop}%
\bibitem [{\citenamefont {Zhang}\ \emph {et~al.}(2023)\citenamefont {Zhang},
  \citenamefont {Lindemann}, \citenamefont {Gardner}, \citenamefont {Gronin},
  \citenamefont {Wu},\ and\ \citenamefont {Manfra}}]{ZhangT23}%
  \BibitemOpen
  \bibfield  {author} {\bibinfo {author} {\bibfnamefont {T.}~\bibnamefont
  {Zhang}}, \bibinfo {author} {\bibfnamefont {T.}~\bibnamefont {Lindemann}},
  \bibinfo {author} {\bibfnamefont {G.~C.}\ \bibnamefont {Gardner}}, \bibinfo
  {author} {\bibfnamefont {S.}~\bibnamefont {Gronin}}, \bibinfo {author}
  {\bibfnamefont {T.}~\bibnamefont {Wu}},\ and\ \bibinfo {author}
  {\bibfnamefont {M.~J.}\ \bibnamefont {Manfra}},\ }\bibfield  {title}
  {\enquote {\bibinfo {title} {Mobility exceeding 100,000 {cm$^2$/Vs} in
  modulation-doped shallow {InAs} quantum wells coupled to epitaxial
  aluminium},}\ }\href@noop {} {\bibfield  {journal} {\bibinfo  {journal}
  {Phys. Rev. Mater.}\ }\textbf {\bibinfo {volume} {7}},\ \bibinfo {pages}
  {056201} (\bibinfo {year} {2023})}\BibitemShut {NoStop}%
\bibitem [{Note1()}]{Note1}%
  \BibitemOpen
  \bibinfo {note} {The critical thickness of an InAs quantum well (QW) grown on
  Al$_{0.8}$Ga$_{0.2}$Sb is much larger ($>$24 nm; tensile strain) than on
  In$_{0.8}$Al$_{0.2}$As (7 nm; compressive strain) despite similar differences
  in lattice constant mismatch ($\sim $~8 pm) between InAs and either barrier
  material.}\BibitemShut {Stop}%
\bibitem [{Note2()}]{Note2}%
  \BibitemOpen
  \bibinfo {note} {Al$_{0.8}$Ga$_{0.2}$Sb has a larger conduction band offset
  ($\sim $\protect \,1.9~eV) relative to InAs than In$_{0.8}$Al$_{0.2}$As does
  ($\sim $\protect \,0.3~eV), thus providing a higher barrier next to the
  quantum well and allowing higher carrier densities to be achieved within a
  single 2D subband. Higher electron densities in turn can enable higher
  mobilities and stronger SOI than at lower electron densities.}\BibitemShut
  {Stop}%
\bibitem [{Note3()}]{Note3}%
  \BibitemOpen
  \bibinfo {note} {A 2DEG hosted in a binary alloy quantum well instead of a
  ternary alloy quantum well would not suffer from alloy scattering, which only
  occurs in ternary alloys and is a significant mechanism limiting
  mobilities.}\BibitemShut {Stop}%
\bibitem [{\citenamefont {Chang}\ \emph {et~al.}(2015)\citenamefont {Chang},
  \citenamefont {Albrecht}, \citenamefont {Jespersen}, \citenamefont
  {Kuemmeth}, \citenamefont {Krogstrup}, \citenamefont {Nyg{\aa}rd},\ and\
  \citenamefont {Marcus}}]{ChangW15}%
  \BibitemOpen
  \bibfield  {author} {\bibinfo {author} {\bibfnamefont {W.}~\bibnamefont
  {Chang}}, \bibinfo {author} {\bibfnamefont {S.~M.}\ \bibnamefont {Albrecht}},
  \bibinfo {author} {\bibfnamefont {T.~S.}\ \bibnamefont {Jespersen}}, \bibinfo
  {author} {\bibfnamefont {F.}~\bibnamefont {Kuemmeth}}, \bibinfo {author}
  {\bibfnamefont {P.}~\bibnamefont {Krogstrup}}, \bibinfo {author}
  {\bibfnamefont {J.}~\bibnamefont {Nyg{\aa}rd}},\ and\ \bibinfo {author}
  {\bibfnamefont {C.~M.}\ \bibnamefont {Marcus}},\ }\bibfield  {title}
  {\enquote {\bibinfo {title} {Hard gap in epitaxial
  semiconductor–superconductor nanowires},}\ }\href@noop {} {\bibfield
  {journal} {\bibinfo  {journal} {Nat. Nanotechnol.}\ }\textbf {\bibinfo
  {volume} {10}},\ \bibinfo {pages} {232} (\bibinfo {year} {2015})}\BibitemShut
  {NoStop}%
\bibitem [{\citenamefont {Birner}\ \emph {et~al.}(2007)\citenamefont {Birner},
  \citenamefont {Zibold}, \citenamefont {Kubis}, \citenamefont {Sabathil},
  \citenamefont {Trellakis},\ and\ \citenamefont {Vogl}}]{NextNano-A}%
  \BibitemOpen
  \bibfield  {author} {\bibinfo {author} {\bibfnamefont {S.}~\bibnamefont
  {Birner}}, \bibinfo {author} {\bibfnamefont {T.}~\bibnamefont {Zibold}},
  \bibinfo {author} {\bibfnamefont {T.}~\bibnamefont {Kubis}}, \bibinfo
  {author} {\bibfnamefont {M.}~\bibnamefont {Sabathil}}, \bibinfo {author}
  {\bibfnamefont {A.}~\bibnamefont {Trellakis}},\ and\ \bibinfo {author}
  {\bibfnamefont {P.}~\bibnamefont {Vogl}},\ }\bibfield  {title} {\enquote
  {\bibinfo {title} {{nextnano: General Purpose 3-D Simulations}},}\
  }\href@noop {} {\bibfield  {journal} {\bibinfo  {journal} {IEEE Trans.
  Electron Dev.}\ }\textbf {\bibinfo {volume} {54}},\ \bibinfo {pages} {2137}
  (\bibinfo {year} {2007})}\BibitemShut {NoStop}%
\bibitem [{\citenamefont {Trellakis}\ \emph {et~al.}(2006)\citenamefont
  {Trellakis}, \citenamefont {Zibold}, \citenamefont {Andlauer}, \citenamefont
  {Birner}, \citenamefont {Smith}, \citenamefont {Morschl},\ and\ \citenamefont
  {Vogl}}]{NextNano-B}%
  \BibitemOpen
  \bibfield  {author} {\bibinfo {author} {\bibfnamefont {A.}~\bibnamefont
  {Trellakis}}, \bibinfo {author} {\bibfnamefont {T.}~\bibnamefont {Zibold}},
  \bibinfo {author} {\bibfnamefont {T.}~\bibnamefont {Andlauer}}, \bibinfo
  {author} {\bibfnamefont {S.}~\bibnamefont {Birner}}, \bibinfo {author}
  {\bibfnamefont {R.~K.}\ \bibnamefont {Smith}}, \bibinfo {author}
  {\bibfnamefont {R.}~\bibnamefont {Morschl}},\ and\ \bibinfo {author}
  {\bibfnamefont {P.}~\bibnamefont {Vogl}},\ }\bibfield  {title} {\enquote
  {\bibinfo {title} {{The 3D nanometer device project nextnano: Concepts,
  methods, results}},}\ }\href@noop {} {\bibfield  {journal} {\bibinfo
  {journal} {J. Comput. Electron.}\ }\textbf {\bibinfo {volume} {5}},\ \bibinfo
  {pages} {285} (\bibinfo {year} {2006})}\BibitemShut {NoStop}%
\bibitem [{\citenamefont {Uddin}\ \emph {et~al.}(2013)\citenamefont {Uddin},
  \citenamefont {Liu}, \citenamefont {Yang}, \citenamefont {Nagase},
  \citenamefont {Sekine}, \citenamefont {Gaspe}, \citenamefont {Mishima},
  \citenamefont {Santos},\ and\ \citenamefont {Hirayama}}]{Uddin13}%
  \BibitemOpen
  \bibfield  {author} {\bibinfo {author} {\bibfnamefont {M.~M.}\ \bibnamefont
  {Uddin}}, \bibinfo {author} {\bibfnamefont {H.~W.}\ \bibnamefont {Liu}},
  \bibinfo {author} {\bibfnamefont {K.~F.}\ \bibnamefont {Yang}}, \bibinfo
  {author} {\bibfnamefont {K.}~\bibnamefont {Nagase}}, \bibinfo {author}
  {\bibfnamefont {K.}~\bibnamefont {Sekine}}, \bibinfo {author} {\bibfnamefont
  {C.~K.}\ \bibnamefont {Gaspe}}, \bibinfo {author} {\bibfnamefont {T.~D.}\
  \bibnamefont {Mishima}}, \bibinfo {author} {\bibfnamefont {M.~B.}\
  \bibnamefont {Santos}},\ and\ \bibinfo {author} {\bibfnamefont
  {Y.}~\bibnamefont {Hirayama}},\ }\bibfield  {title} {\enquote {\bibinfo
  {title} {Gate depletion of an {InSb} two-dimensional electron gas},}\
  }\href@noop {} {\bibfield  {journal} {\bibinfo  {journal} {Appl. Phys.
  Lett.}\ }\textbf {\bibinfo {volume} {103}},\ \bibinfo {pages} {123502}
  (\bibinfo {year} {2013})}\BibitemShut {NoStop}%
\bibitem [{\citenamefont {Yi}\ \emph {et~al.}(2015)\citenamefont {Yi},
  \citenamefont {Kiselev}, \citenamefont {Thorp}, \citenamefont {Noah},
  \citenamefont {Nguyen}, \citenamefont {Bui}, \citenamefont {Rajavel},
  \citenamefont {Hussain}, \citenamefont {Gyure}, \citenamefont {Kratz},
  \citenamefont {Qian}, \citenamefont {Manfra}, \citenamefont {Pribiag},
  \citenamefont {Kouwenhoven}, \citenamefont {Marcus},\ and\ \citenamefont
  {Sokolich}}]{YiW15}%
  \BibitemOpen
  \bibfield  {author} {\bibinfo {author} {\bibfnamefont {W.}~\bibnamefont
  {Yi}}, \bibinfo {author} {\bibfnamefont {A.~A.}\ \bibnamefont {Kiselev}},
  \bibinfo {author} {\bibfnamefont {J.}~\bibnamefont {Thorp}}, \bibinfo
  {author} {\bibfnamefont {R.}~\bibnamefont {Noah}}, \bibinfo {author}
  {\bibfnamefont {B.-M.}\ \bibnamefont {Nguyen}}, \bibinfo {author}
  {\bibfnamefont {S.}~\bibnamefont {Bui}}, \bibinfo {author} {\bibfnamefont
  {R.~D.}\ \bibnamefont {Rajavel}}, \bibinfo {author} {\bibfnamefont
  {T.}~\bibnamefont {Hussain}}, \bibinfo {author} {\bibfnamefont {M.~F.}\
  \bibnamefont {Gyure}}, \bibinfo {author} {\bibfnamefont {P.}~\bibnamefont
  {Kratz}}, \bibinfo {author} {\bibfnamefont {Q.}~\bibnamefont {Qian}},
  \bibinfo {author} {\bibfnamefont {M.~J.}\ \bibnamefont {Manfra}}, \bibinfo
  {author} {\bibfnamefont {V.~S.}\ \bibnamefont {Pribiag}}, \bibinfo {author}
  {\bibfnamefont {L.~P.}\ \bibnamefont {Kouwenhoven}}, \bibinfo {author}
  {\bibfnamefont {C.~M.}\ \bibnamefont {Marcus}},\ and\ \bibinfo {author}
  {\bibfnamefont {M.}~\bibnamefont {Sokolich}},\ }\bibfield  {title} {\enquote
  {\bibinfo {title} {Gate-tunable high mobility remote-doped
  {InSb}/{In$_{1-x}$Al$_{x}$Sb} quantum well heterostructures},}\ }\href@noop
  {} {\bibfield  {journal} {\bibinfo  {journal} {Appl. Phys. Lett.}\ }\textbf
  {\bibinfo {volume} {106}},\ \bibinfo {pages} {142103} (\bibinfo {year}
  {2015})}\BibitemShut {NoStop}%
\bibitem [{\citenamefont {Kulesh}\ \emph {et~al.}(2020)\citenamefont {Kulesh},
  \citenamefont {Ke}, \citenamefont {Thomas}, \citenamefont {Karwal},
  \citenamefont {Moehle}, \citenamefont {Metti}, \citenamefont {Kallaher},
  \citenamefont {Gardner}, \citenamefont {Manfra},\ and\ \citenamefont
  {Goswami}}]{Kulesh20}%
  \BibitemOpen
  \bibfield  {author} {\bibinfo {author} {\bibfnamefont {I.}~\bibnamefont
  {Kulesh}}, \bibinfo {author} {\bibfnamefont {C.~K.}\ \bibnamefont {Ke}},
  \bibinfo {author} {\bibfnamefont {C.}~\bibnamefont {Thomas}}, \bibinfo
  {author} {\bibfnamefont {S.}~\bibnamefont {Karwal}}, \bibinfo {author}
  {\bibfnamefont {M.~C.}\ \bibnamefont {Moehle}}, \bibinfo {author}
  {\bibfnamefont {S.}~\bibnamefont {Metti}}, \bibinfo {author} {\bibfnamefont
  {R.}~\bibnamefont {Kallaher}}, \bibinfo {author} {\bibfnamefont {C.~G.}\
  \bibnamefont {Gardner}}, \bibinfo {author} {\bibfnamefont {M.~J.}\
  \bibnamefont {Manfra}},\ and\ \bibinfo {author} {\bibfnamefont
  {S.}~\bibnamefont {Goswami}},\ }\bibfield  {title} {\enquote {\bibinfo
  {title} {Quantum dots in an {InSb} two-dimensional electron gas},}\
  }\href@noop {} {\bibfield  {journal} {\bibinfo  {journal} {Phys. Rev. Appl.}\
  }\textbf {\bibinfo {volume} {13}},\ \bibinfo {pages} {041003} (\bibinfo
  {year} {2020})}\BibitemShut {NoStop}%
\bibitem [{\citenamefont {Tuttle}, \citenamefont {Kroemer},\ and\ \citenamefont
  {English}(1990)}]{Tuttle90}%
  \BibitemOpen
  \bibfield  {author} {\bibinfo {author} {\bibfnamefont {G.}~\bibnamefont
  {Tuttle}}, \bibinfo {author} {\bibfnamefont {H.}~\bibnamefont {Kroemer}},\
  and\ \bibinfo {author} {\bibfnamefont {J.~H.}\ \bibnamefont {English}},\
  }\bibfield  {title} {\enquote {\bibinfo {title} {Effects of interface layer
  sequencing on the transport properties of {InAs/AlSb} quantum wells:
  {Evidence} for antisite donors at the {InAs/AlSb} interface},}\ }\href@noop
  {} {\bibfield  {journal} {\bibinfo  {journal} {J. Appl. Phys.}\ }\textbf
  {\bibinfo {volume} {67}},\ \bibinfo {pages} {3032} (\bibinfo {year}
  {1990})}\BibitemShut {NoStop}%
\bibitem [{\citenamefont {Ando}, \citenamefont {Fowler},\ and\ \citenamefont
  {Stern}(1982)}]{Ando82-A}%
  \BibitemOpen
  \bibfield  {author} {\bibinfo {author} {\bibfnamefont {T.}~\bibnamefont
  {Ando}}, \bibinfo {author} {\bibfnamefont {A.~B.}\ \bibnamefont {Fowler}},\
  and\ \bibinfo {author} {\bibfnamefont {F.}~\bibnamefont {Stern}},\ }\bibfield
   {title} {\enquote {\bibinfo {title} {{Electronic properties of
  two-dimensional systems}},}\ }\href@noop {} {\bibfield  {journal} {\bibinfo
  {journal} {Rev. Mod. Phys.}\ }\textbf {\bibinfo {volume} {54}},\ \bibinfo
  {pages} {437} (\bibinfo {year} {1982})}\BibitemShut {NoStop}%
\bibitem [{\citenamefont {Shetty}\ \emph {et~al.}(2022)\citenamefont {Shetty},
  \citenamefont {Sfigakis}, \citenamefont {Mak}, \citenamefont {Gupta},
  \citenamefont {Buonacorsi}, \citenamefont {Tam}, \citenamefont {Kim},
  \citenamefont {Farrer}, \citenamefont {Croxall}, \citenamefont {Beere},
  \citenamefont {Hamilton}, \citenamefont {Pepper}, \citenamefont {Austing},
  \citenamefont {Studenikin}, \citenamefont {Sachrajda}, \citenamefont
  {Reimer}, \citenamefont {Wasilewski}, \citenamefont {Ritchie},\ and\
  \citenamefont {Baugh}}]{Arjun22}%
  \BibitemOpen
  \bibfield  {author} {\bibinfo {author} {\bibfnamefont {A.}~\bibnamefont
  {Shetty}}, \bibinfo {author} {\bibfnamefont {F.}~\bibnamefont {Sfigakis}},
  \bibinfo {author} {\bibfnamefont {W.~Y.}\ \bibnamefont {Mak}}, \bibinfo
  {author} {\bibfnamefont {K.~D.}\ \bibnamefont {Gupta}}, \bibinfo {author}
  {\bibfnamefont {B.}~\bibnamefont {Buonacorsi}}, \bibinfo {author}
  {\bibfnamefont {M.~C.}\ \bibnamefont {Tam}}, \bibinfo {author} {\bibfnamefont
  {H.~S.}\ \bibnamefont {Kim}}, \bibinfo {author} {\bibfnamefont
  {I.}~\bibnamefont {Farrer}}, \bibinfo {author} {\bibfnamefont {A.~F.}\
  \bibnamefont {Croxall}}, \bibinfo {author} {\bibfnamefont {H.~E.}\
  \bibnamefont {Beere}}, \bibinfo {author} {\bibfnamefont {A.~R.}\ \bibnamefont
  {Hamilton}}, \bibinfo {author} {\bibfnamefont {M.}~\bibnamefont {Pepper}},
  \bibinfo {author} {\bibfnamefont {D.~G.}\ \bibnamefont {Austing}}, \bibinfo
  {author} {\bibfnamefont {S.~A.}\ \bibnamefont {Studenikin}}, \bibinfo
  {author} {\bibfnamefont {A.}~\bibnamefont {Sachrajda}}, \bibinfo {author}
  {\bibfnamefont {M.~E.}\ \bibnamefont {Reimer}}, \bibinfo {author}
  {\bibfnamefont {Z.~R.}\ \bibnamefont {Wasilewski}}, \bibinfo {author}
  {\bibfnamefont {D.~A.}\ \bibnamefont {Ritchie}},\ and\ \bibinfo {author}
  {\bibfnamefont {J.}~\bibnamefont {Baugh}},\ }\bibfield  {title} {\enquote
  {\bibinfo {title} {Effects of biased and unbiased illuminations on
  two-dimensional electron gases in dopant-free {GaAs/AlGaAs}},}\ }\href@noop
  {} {\bibfield  {journal} {\bibinfo  {journal} {Phys. Rev. B}\ }\textbf
  {\bibinfo {volume} {105}},\ \bibinfo {pages} {075302} (\bibinfo {year}
  {2022})}\BibitemShut {NoStop}%
\bibitem [{\citenamefont {Schubnikow}\ and\ \citenamefont
  {{de~Haas}}(1930)}]{Shubnikov-de-Haas1930}%
  \BibitemOpen
  \bibfield  {author} {\bibinfo {author} {\bibfnamefont {L.}~\bibnamefont
  {Schubnikow}}\ and\ \bibinfo {author} {\bibfnamefont {W.~J.}\ \bibnamefont
  {{de~Haas}}},\ }\bibfield  {title} {\enquote {\bibinfo {title} {A new
  phenomenon in the change of resistance in a magnetic field of single crystals
  of bismuth},}\ }\href@noop {} {\bibfield  {journal} {\bibinfo  {journal}
  {Nature}\ }\textbf {\bibinfo {volume} {126}},\ \bibinfo {pages} {500}
  (\bibinfo {year} {1930})}\BibitemShut {NoStop}%
\bibitem [{\citenamefont {Muraki}, \citenamefont {Saku},\ and\ \citenamefont
  {Hirayama}(2001)}]{Muraki01}%
  \BibitemOpen
  \bibfield  {author} {\bibinfo {author} {\bibfnamefont {K.}~\bibnamefont
  {Muraki}}, \bibinfo {author} {\bibfnamefont {T.}~\bibnamefont {Saku}},\ and\
  \bibinfo {author} {\bibfnamefont {Y.}~\bibnamefont {Hirayama}},\ }\bibfield
  {title} {\enquote {\bibinfo {title} {Charge excitations in easy-axis and
  easy-plane quantum hall ferromagnets},}\ }\href@noop {} {\bibfield  {journal}
  {\bibinfo  {journal} {Phys. Rev. Lett.}\ }\textbf {\bibinfo {volume} {87}},\
  \bibinfo {pages} {196801} (\bibinfo {year} {2001})}\BibitemShut {NoStop}%
\bibitem [{\citenamefont {Zhang}, \citenamefont {Faulhaber},\ and\
  \citenamefont {Jiang}(2005)}]{ZhangXC05}%
  \BibitemOpen
  \bibfield  {author} {\bibinfo {author} {\bibfnamefont {X.~C.}\ \bibnamefont
  {Zhang}}, \bibinfo {author} {\bibfnamefont {D.~R.}\ \bibnamefont
  {Faulhaber}},\ and\ \bibinfo {author} {\bibfnamefont {H.~W.}\ \bibnamefont
  {Jiang}},\ }\bibfield  {title} {\enquote {\bibinfo {title} {Multiple phases
  with the same quantized {Hall} conductance in a two-subband system},}\
  }\href@noop {} {\bibfield  {journal} {\bibinfo  {journal} {Phys. Rev. Lett.}\
  }\textbf {\bibinfo {volume} {95}},\ \bibinfo {pages} {216801} (\bibinfo
  {year} {2005})}\BibitemShut {NoStop}%
\bibitem [{\citenamefont {Ellenberger}\ \emph {et~al.}(2006)\citenamefont
  {Ellenberger}, \citenamefont {Simovi{\u{c}}}, \citenamefont {Leturcq},
  \citenamefont {Ihn}, \citenamefont {Ulloa}, \citenamefont {Ensslin},
  \citenamefont {Driscoll},\ and\ \citenamefont {Gossard}}]{Ellenberger06}%
  \BibitemOpen
  \bibfield  {author} {\bibinfo {author} {\bibfnamefont {C.}~\bibnamefont
  {Ellenberger}}, \bibinfo {author} {\bibfnamefont {B.}~\bibnamefont
  {Simovi{\u{c}}}}, \bibinfo {author} {\bibfnamefont {R.}~\bibnamefont
  {Leturcq}}, \bibinfo {author} {\bibfnamefont {T.}~\bibnamefont {Ihn}},
  \bibinfo {author} {\bibfnamefont {S.~E.}\ \bibnamefont {Ulloa}}, \bibinfo
  {author} {\bibfnamefont {K.}~\bibnamefont {Ensslin}}, \bibinfo {author}
  {\bibfnamefont {D.~C.}\ \bibnamefont {Driscoll}},\ and\ \bibinfo {author}
  {\bibfnamefont {A.~C.}\ \bibnamefont {Gossard}},\ }\bibfield  {title}
  {\enquote {\bibinfo {title} {Two-subband quantum hall effect in parabolic
  quantum wells},}\ }\href@noop {} {\bibfield  {journal} {\bibinfo  {journal}
  {Phys. Rev. B}\ }\textbf {\bibinfo {volume} {74}},\ \bibinfo {pages} {195313}
  (\bibinfo {year} {2006})}\BibitemShut {NoStop}%
\bibitem [{\citenamefont {Hikami}, \citenamefont {Larkin},\ and\ \citenamefont
  {Nagaoka}(1980)}]{Hikami80}%
  \BibitemOpen
  \bibfield  {author} {\bibinfo {author} {\bibfnamefont {S.}~\bibnamefont
  {Hikami}}, \bibinfo {author} {\bibfnamefont {A.~I.}\ \bibnamefont {Larkin}},\
  and\ \bibinfo {author} {\bibfnamefont {Y.}~\bibnamefont {Nagaoka}},\
  }\bibfield  {title} {\enquote {\bibinfo {title} {Spin-orbit interaction and
  magnetoresistance in the two dimensional random system},}\ }\href@noop {}
  {\bibfield  {journal} {\bibinfo  {journal} {Prog. Theoret. Phys.}\ }\textbf
  {\bibinfo {volume} {63}},\ \bibinfo {pages} {707} (\bibinfo {year}
  {1980})}\BibitemShut {NoStop}%
\bibitem [{\citenamefont {Iordanskii}\ \emph {et~al.}(1994)\citenamefont
  {Iordanskii}, \citenamefont {Lyanda-Geller}, ,\ and\ \citenamefont
  {Pikus}}]{Iordanskii94}%
  \BibitemOpen
  \bibfield  {author} {\bibinfo {author} {\bibfnamefont {S.}~\bibnamefont
  {Iordanskii}}, \bibinfo {author} {\bibfnamefont {Y.~B.}\ \bibnamefont
  {Lyanda-Geller}}, ,\ and\ \bibinfo {author} {\bibfnamefont {G.}~\bibnamefont
  {Pikus}},\ }\bibfield  {title} {\enquote {\bibinfo {title} {Weak localization
  in quantum wells with spin-orbit interaction},}\ }\href@noop {} {\bibfield
  {journal} {\bibinfo  {journal} {JETP Letters}\ }\textbf {\bibinfo {volume}
  {60}},\ \bibinfo {pages} {206} (\bibinfo {year} {1994})}\BibitemShut
  {NoStop}%
\bibitem [{\citenamefont {Wickramasinghe}\ \emph {et~al.}(2018)\citenamefont
  {Wickramasinghe}, \citenamefont {Mayer}, \citenamefont {Yuan}, \citenamefont
  {Nguyen}, \citenamefont {Jiao}, \citenamefont {Manucharyan},\ and\
  \citenamefont {Shabani}}]{Wickramasinghe18}%
  \BibitemOpen
  \bibfield  {author} {\bibinfo {author} {\bibfnamefont {K.~S.}\ \bibnamefont
  {Wickramasinghe}}, \bibinfo {author} {\bibfnamefont {W.}~\bibnamefont
  {Mayer}}, \bibinfo {author} {\bibfnamefont {J.}~\bibnamefont {Yuan}},
  \bibinfo {author} {\bibfnamefont {T.}~\bibnamefont {Nguyen}}, \bibinfo
  {author} {\bibfnamefont {L.}~\bibnamefont {Jiao}}, \bibinfo {author}
  {\bibfnamefont {V.}~\bibnamefont {Manucharyan}},\ and\ \bibinfo {author}
  {\bibfnamefont {J.}~\bibnamefont {Shabani}},\ }\bibfield  {title} {\enquote
  {\bibinfo {title} {Transport properties of near surface {InAs}
  two-dimensional heterostructures},}\ }\href@noop {} {\bibfield  {journal}
  {\bibinfo  {journal} {Appl. Phys. Lett.}\ }\textbf {\bibinfo {volume}
  {113}},\ \bibinfo {pages} {262104} (\bibinfo {year} {2018})}\BibitemShut
  {NoStop}%
\bibitem [{\citenamefont {Witt}\ \emph {et~al.}(2023)\citenamefont {Witt},
  \citenamefont {Pauka}, \citenamefont {Gardner}, \citenamefont {Gronin},
  \citenamefont {Wang}, \citenamefont {Thomas}, \citenamefont {Manfra},
  \citenamefont {Reilly},\ and\ \citenamefont {Cassidy}}]{Witt23}%
  \BibitemOpen
  \bibfield  {author} {\bibinfo {author} {\bibfnamefont {J.~D.~S.}\
  \bibnamefont {Witt}}, \bibinfo {author} {\bibfnamefont {S.~J.}\ \bibnamefont
  {Pauka}}, \bibinfo {author} {\bibfnamefont {G.~C.}\ \bibnamefont {Gardner}},
  \bibinfo {author} {\bibfnamefont {S.}~\bibnamefont {Gronin}}, \bibinfo
  {author} {\bibfnamefont {T.}~\bibnamefont {Wang}}, \bibinfo {author}
  {\bibfnamefont {C.}~\bibnamefont {Thomas}}, \bibinfo {author} {\bibfnamefont
  {M.~J.}\ \bibnamefont {Manfra}}, \bibinfo {author} {\bibfnamefont {D.~J.}\
  \bibnamefont {Reilly}},\ and\ \bibinfo {author} {\bibfnamefont {M.~C.}\
  \bibnamefont {Cassidy}},\ }\bibfield  {title} {\enquote {\bibinfo {title}
  {Spin-relaxation mechanisms in {InAs} quantum well heterostructures},}\
  }\href@noop {} {\bibfield  {journal} {\bibinfo  {journal} {Appl. Phys.
  Lett.}\ }\textbf {\bibinfo {volume} {122}},\ \bibinfo {pages} {083101}
  (\bibinfo {year} {2023})}\BibitemShut {NoStop}%
\bibitem [{\citenamefont {Farzaneh}\ \emph {et~al.}(2024)\citenamefont
  {Farzaneh}, \citenamefont {Hatefipour}, \citenamefont {Schiela},
  \citenamefont {Lotfizadeh}, \citenamefont {Yu}, \citenamefont {Elfeky},
  \citenamefont {Strickland}, \citenamefont {Matos-Abiague},\ and\
  \citenamefont {Shabani}}]{Farzaneh24}%
  \BibitemOpen
  \bibfield  {author} {\bibinfo {author} {\bibfnamefont {S.~M.}\ \bibnamefont
  {Farzaneh}}, \bibinfo {author} {\bibfnamefont {M.}~\bibnamefont
  {Hatefipour}}, \bibinfo {author} {\bibfnamefont {W.~F.}\ \bibnamefont
  {Schiela}}, \bibinfo {author} {\bibfnamefont {N.}~\bibnamefont {Lotfizadeh}},
  \bibinfo {author} {\bibfnamefont {P.}~\bibnamefont {Yu}}, \bibinfo {author}
  {\bibfnamefont {B.~H.}\ \bibnamefont {Elfeky}}, \bibinfo {author}
  {\bibfnamefont {W.~M.}\ \bibnamefont {Strickland}}, \bibinfo {author}
  {\bibfnamefont {A.}~\bibnamefont {Matos-Abiague}},\ and\ \bibinfo {author}
  {\bibfnamefont {J.}~\bibnamefont {Shabani}},\ }\bibfield  {title} {\enquote
  {\bibinfo {title} {Observing magnetoanisotropic weak antilocalization in
  near-surface quantum wells},}\ }\href@noop {} {\bibfield  {journal} {\bibinfo
   {journal} {Phys. Rev. Res.}\ }\textbf {\bibinfo {volume} {6}},\ \bibinfo
  {pages} {013039} (\bibinfo {year} {2024})}\BibitemShut {NoStop}%
\bibitem [{\citenamefont {Bergeron}\ \emph {et~al.}(2023)\citenamefont
  {Bergeron}, \citenamefont {Sfigakis}, \citenamefont {Shi}, \citenamefont
  {Nichols}, \citenamefont {Klipstein}, \citenamefont {Elbaroudy},
  \citenamefont {Walker}, \citenamefont {Wasilewski},\ and\ \citenamefont
  {Baugh}}]{Bergeron23}%
  \BibitemOpen
  \bibfield  {author} {\bibinfo {author} {\bibfnamefont {E.~A.}\ \bibnamefont
  {Bergeron}}, \bibinfo {author} {\bibfnamefont {F.}~\bibnamefont {Sfigakis}},
  \bibinfo {author} {\bibfnamefont {Y.}~\bibnamefont {Shi}}, \bibinfo {author}
  {\bibfnamefont {G.}~\bibnamefont {Nichols}}, \bibinfo {author} {\bibfnamefont
  {P.~C.}\ \bibnamefont {Klipstein}}, \bibinfo {author} {\bibfnamefont
  {A.}~\bibnamefont {Elbaroudy}}, \bibinfo {author} {\bibfnamefont {S.~M.}\
  \bibnamefont {Walker}}, \bibinfo {author} {\bibfnamefont {Z.~R.}\
  \bibnamefont {Wasilewski}},\ and\ \bibinfo {author} {\bibfnamefont
  {J.}~\bibnamefont {Baugh}},\ }\bibfield  {title} {\enquote {\bibinfo {title}
  {Field effect two-dimensional electron gases in modulation-doped {InSb}
  surface quantum wells},}\ }\href@noop {} {\bibfield  {journal} {\bibinfo
  {journal} {Appl. Phys. Lett.}\ }\textbf {\bibinfo {volume} {122}},\ \bibinfo
  {pages} {012103} (\bibinfo {year} {2023})}\BibitemShut {NoStop}%
\bibitem [{\citenamefont {Tajik}, \citenamefont {Haapamaki},\ and\
  \citenamefont {LaPierre}(2012)}]{Tajik12}%
  \BibitemOpen
  \bibfield  {author} {\bibinfo {author} {\bibfnamefont {N.}~\bibnamefont
  {Tajik}}, \bibinfo {author} {\bibfnamefont {C.~M.}\ \bibnamefont
  {Haapamaki}},\ and\ \bibinfo {author} {\bibfnamefont {R.~R.}\ \bibnamefont
  {LaPierre}},\ }\bibfield  {title} {\enquote {\bibinfo {title}
  {Photoluminescence model of sulfur passivated p-{InP} nanowires},}\
  }\href@noop {} {\bibfield  {journal} {\bibinfo  {journal} {Nanotechnology}\
  }\textbf {\bibinfo {volume} {23}},\ \bibinfo {pages} {315703} (\bibinfo
  {year} {2012})}\BibitemShut {NoStop}%
\bibitem [{\citenamefont {Lebedev}(2020)}]{LebedevMV20}%
  \BibitemOpen
  \bibfield  {author} {\bibinfo {author} {\bibfnamefont {M.~V.}\ \bibnamefont
  {Lebedev}},\ }\bibfield  {title} {\enquote {\bibinfo {title} {Modification of
  the atomic and electronic structure of {III}-{V} semiconductor surfaces at
  interfaces with electrolyte solutions},}\ }\href@noop {} {\bibfield
  {journal} {\bibinfo  {journal} {Semiconductors}\ }\textbf {\bibinfo {volume}
  {54}},\ \bibinfo {pages} {699} (\bibinfo {year} {2020})}\BibitemShut
  {NoStop}%
\bibitem [{\citenamefont {Bessolov}\ and\ \citenamefont
  {Lebedev}(1998)}]{Bessolov98}%
  \BibitemOpen
  \bibfield  {author} {\bibinfo {author} {\bibfnamefont {V.~N.}\ \bibnamefont
  {Bessolov}}\ and\ \bibinfo {author} {\bibfnamefont {M.~V.}\ \bibnamefont
  {Lebedev}},\ }\bibfield  {title} {\enquote {\bibinfo {title} {Chalcogenide
  passivation of {III}-{V} semiconductor surfaces},}\ }\href@noop {} {\bibfield
   {journal} {\bibinfo  {journal} {Semiconductors}\ }\textbf {\bibinfo {volume}
  {32}},\ \bibinfo {pages} {1141} (\bibinfo {year} {1998})}\BibitemShut
  {NoStop}%
\bibitem [{\citenamefont {Hazra}\ \emph {et~al.}(2010)\citenamefont {Hazra},
  \citenamefont {Pascal}, \citenamefont {Courtois},\ and\ \citenamefont
  {Gupta}}]{Hazra10}%
  \BibitemOpen
  \bibfield  {author} {\bibinfo {author} {\bibfnamefont {D.}~\bibnamefont
  {Hazra}}, \bibinfo {author} {\bibfnamefont {L.~M.~A.}\ \bibnamefont
  {Pascal}}, \bibinfo {author} {\bibfnamefont {H.}~\bibnamefont {Courtois}},\
  and\ \bibinfo {author} {\bibfnamefont {A.~K.}\ \bibnamefont {Gupta}},\
  }\bibfield  {title} {\enquote {\bibinfo {title} {Hysteresis in
  superconducting short weak links and {$\mu$}-{SQUIDs}},}\ }\href@noop {}
  {\bibfield  {journal} {\bibinfo  {journal} {Phys. Rev. B}\ }\textbf {\bibinfo
  {volume} {82}},\ \bibinfo {pages} {184530} (\bibinfo {year}
  {2010})}\BibitemShut {NoStop}%
\bibitem [{\citenamefont {Vodolazov}\ and\ \citenamefont
  {Peeters}(2011)}]{Vodolazov11}%
  \BibitemOpen
  \bibfield  {author} {\bibinfo {author} {\bibfnamefont {D.~Y.}\ \bibnamefont
  {Vodolazov}}\ and\ \bibinfo {author} {\bibfnamefont {F.~M.}\ \bibnamefont
  {Peeters}},\ }\bibfield  {title} {\enquote {\bibinfo {title} {Origin of the
  hysteresis of the current voltage characteristics of superconducting
  microbridges near the critical temperature},}\ }\href@noop {} {\bibfield
  {journal} {\bibinfo  {journal} {Phys. Rev. B}\ }\textbf {\bibinfo {volume}
  {84}},\ \bibinfo {pages} {094511} (\bibinfo {year} {2011})}\BibitemShut
  {NoStop}%
\bibitem [{\citenamefont {Nitta}\ \emph {et~al.}(1992)\citenamefont {Nitta},
  \citenamefont {Akazaki}, \citenamefont {Takayanagi},\ and\ \citenamefont
  {Arai}}]{Nitta92}%
  \BibitemOpen
  \bibfield  {author} {\bibinfo {author} {\bibfnamefont {J.}~\bibnamefont
  {Nitta}}, \bibinfo {author} {\bibfnamefont {T.}~\bibnamefont {Akazaki}},
  \bibinfo {author} {\bibfnamefont {H.}~\bibnamefont {Takayanagi}},\ and\
  \bibinfo {author} {\bibfnamefont {K.}~\bibnamefont {Arai}},\ }\bibfield
  {title} {\enquote {\bibinfo {title} {Transport properties in an
  {InAs}-inserted-channel
  {In{$_{0.52}$}Al{$_{0.48}$}As}/{In{$_{0.53}$}Ga{$_{0.47}$}As} heterostructure
  coupled superconducting junction},}\ }\href@noop {} {\bibfield  {journal}
  {\bibinfo  {journal} {Phys. Rev. B}\ }\textbf {\bibinfo {volume} {46}},\
  \bibinfo {pages} {14286(R)} (\bibinfo {year} {1992})}\BibitemShut {NoStop}%
\bibitem [{\citenamefont {Takayanagi}\ and\ \citenamefont
  {Akazaki}(1995)}]{Takayanagi95}%
  \BibitemOpen
  \bibfield  {author} {\bibinfo {author} {\bibfnamefont {H.}~\bibnamefont
  {Takayanagi}}\ and\ \bibinfo {author} {\bibfnamefont {T.}~\bibnamefont
  {Akazaki}},\ }\bibfield  {title} {\enquote {\bibinfo {title} {Temperature
  dependence of the critical current in a clean-limit
  superconductor-{2DEG}-superconductor junction},}\ }\href@noop {} {\bibfield
  {journal} {\bibinfo  {journal} {Solid State Commun.}\ }\textbf {\bibinfo
  {volume} {96}},\ \bibinfo {pages} {815} (\bibinfo {year} {1995})}\BibitemShut
  {NoStop}%
\bibitem [{\citenamefont {Heida}\ \emph {et~al.}(1998)\citenamefont {Heida},
  \citenamefont {van Wees}, \citenamefont {Klapwijk},\ and\ \citenamefont
  {Borghs}}]{Heida98}%
  \BibitemOpen
  \bibfield  {author} {\bibinfo {author} {\bibfnamefont {J.~P.}\ \bibnamefont
  {Heida}}, \bibinfo {author} {\bibfnamefont {B.~J.}\ \bibnamefont {van Wees}},
  \bibinfo {author} {\bibfnamefont {T.~M.}\ \bibnamefont {Klapwijk}},\ and\
  \bibinfo {author} {\bibfnamefont {G.}~\bibnamefont {Borghs}},\ }\bibfield
  {title} {\enquote {\bibinfo {title} {Nonlocal supercurrent in mesoscopic
  {Josephson} junctions},}\ }\href@noop {} {\bibfield  {journal} {\bibinfo
  {journal} {Phys. Rev. B}\ }\textbf {\bibinfo {volume} {57}},\ \bibinfo
  {pages} {R5618(R)} (\bibinfo {year} {1998})}\BibitemShut {NoStop}%
\bibitem [{\citenamefont {Giazotto}\ \emph {et~al.}(2004)\citenamefont
  {Giazotto}, \citenamefont {Grove-Rasmussen}, \citenamefont {Fazio},
  \citenamefont {Beltram}, \citenamefont {Linfield},\ and\ \citenamefont
  {Ritchie}}]{Giazotto04}%
  \BibitemOpen
  \bibfield  {author} {\bibinfo {author} {\bibfnamefont {F.}~\bibnamefont
  {Giazotto}}, \bibinfo {author} {\bibfnamefont {K.}~\bibnamefont
  {Grove-Rasmussen}}, \bibinfo {author} {\bibfnamefont {R.}~\bibnamefont
  {Fazio}}, \bibinfo {author} {\bibfnamefont {F.}~\bibnamefont {Beltram}},
  \bibinfo {author} {\bibfnamefont {E.~H.}\ \bibnamefont {Linfield}},\ and\
  \bibinfo {author} {\bibfnamefont {D.~A.}\ \bibnamefont {Ritchie}},\
  }\bibfield  {title} {\enquote {\bibinfo {title} {Josephson current in
  {Nb/InAs/Nb} highly transmissive ballistic junctions},}\ }\href@noop {}
  {\bibfield  {journal} {\bibinfo  {journal} {J. Supercond.}\ }\textbf
  {\bibinfo {volume} {17}},\ \bibinfo {pages} {317} (\bibinfo {year}
  {2004})}\BibitemShut {NoStop}%
\bibitem [{\citenamefont {Blonder}, \citenamefont {Tinkham},\ and\
  \citenamefont {Klapwijk}(1982)}]{Blonder82}%
  \BibitemOpen
  \bibfield  {author} {\bibinfo {author} {\bibfnamefont {G.~E.}\ \bibnamefont
  {Blonder}}, \bibinfo {author} {\bibfnamefont {M.}~\bibnamefont {Tinkham}},\
  and\ \bibinfo {author} {\bibfnamefont {T.~M.}\ \bibnamefont {Klapwijk}},\
  }\bibfield  {title} {\enquote {\bibinfo {title} {Transition from metallic to
  tunneling regimes in superconducting microconstrictions: {Excess} current,
  charge imbalance, and supercurrent conversion},}\ }\href@noop {} {\bibfield
  {journal} {\bibinfo  {journal} {Phys. Rev. B}\ }\textbf {\bibinfo {volume}
  {25}},\ \bibinfo {pages} {4515} (\bibinfo {year} {1982})}\BibitemShut
  {NoStop}%
\bibitem [{\citenamefont {Octavio}\ \emph {et~al.}(1983)\citenamefont
  {Octavio}, \citenamefont {Tinkham}, \citenamefont {Blonder},\ and\
  \citenamefont {Klapwijk}}]{Octavio83}%
  \BibitemOpen
  \bibfield  {author} {\bibinfo {author} {\bibfnamefont {M.}~\bibnamefont
  {Octavio}}, \bibinfo {author} {\bibfnamefont {M.}~\bibnamefont {Tinkham}},
  \bibinfo {author} {\bibfnamefont {G.~E.}\ \bibnamefont {Blonder}},\ and\
  \bibinfo {author} {\bibfnamefont {T.~M.}\ \bibnamefont {Klapwijk}},\
  }\bibfield  {title} {\enquote {\bibinfo {title} {Subharmonic energy-gap
  structure in superconducting constrictions},}\ }\href@noop {} {\bibfield
  {journal} {\bibinfo  {journal} {Phys. Rev. B}\ }\textbf {\bibinfo {volume}
  {27}},\ \bibinfo {pages} {6739} (\bibinfo {year} {1983})}\BibitemShut
  {NoStop}%
\bibitem [{\citenamefont {Flensberg}, \citenamefont {Hansen},\ and\
  \citenamefont {Octavio}(1988)}]{Flensberg88}%
  \BibitemOpen
  \bibfield  {author} {\bibinfo {author} {\bibfnamefont {K.}~\bibnamefont
  {Flensberg}}, \bibinfo {author} {\bibfnamefont {J.~H.~B.}\ \bibnamefont
  {Hansen}},\ and\ \bibinfo {author} {\bibfnamefont {M.}~\bibnamefont
  {Octavio}},\ }\bibfield  {title} {\enquote {\bibinfo {title} {Subharmonic
  energy-gap structure in superconducting weak links},}\ }\href@noop {}
  {\bibfield  {journal} {\bibinfo  {journal} {Phys. Rev. B}\ }\textbf {\bibinfo
  {volume} {38}},\ \bibinfo {pages} {8707} (\bibinfo {year}
  {1988})}\BibitemShut {NoStop}%
\bibitem [{\citenamefont {Cuevas}, \citenamefont {Martín-Rodero},\ and\
  \citenamefont {Yeyati}(1996)}]{Cuevas96}%
  \BibitemOpen
  \bibfield  {author} {\bibinfo {author} {\bibfnamefont {J.~C.}\ \bibnamefont
  {Cuevas}}, \bibinfo {author} {\bibfnamefont {A.}~\bibnamefont
  {Martín-Rodero}},\ and\ \bibinfo {author} {\bibfnamefont {A.~L.}\
  \bibnamefont {Yeyati}},\ }\bibfield  {title} {\enquote {\bibinfo {title}
  {Hamiltonian approach to the transport properties of superconducting quantum
  point contacts},}\ }\href@noop {} {\bibfield  {journal} {\bibinfo  {journal}
  {Phys. Rev. B}\ }\textbf {\bibinfo {volume} {54}},\ \bibinfo {pages} {7366}
  (\bibinfo {year} {1996})}\BibitemShut {NoStop}%
\bibitem [{\citenamefont {Niebler}, \citenamefont {Cuniberti},\ and\
  \citenamefont {Novotn{\`{y}}}(2009)}]{Niebler09}%
  \BibitemOpen
  \bibfield  {author} {\bibinfo {author} {\bibfnamefont {G.}~\bibnamefont
  {Niebler}}, \bibinfo {author} {\bibfnamefont {G.}~\bibnamefont {Cuniberti}},\
  and\ \bibinfo {author} {\bibfnamefont {T.}~\bibnamefont {Novotn{\`{y}}}},\
  }\bibfield  {title} {\enquote {\bibinfo {title} {Analytical calculation of
  the excess current in the {Octavio-Tinkham-Blonder-Klapwijk} theory},}\
  }\href@noop {} {\bibfield  {journal} {\bibinfo  {journal} {Supercond. Sci.
  Technol.}\ }\textbf {\bibinfo {volume} {22}},\ \bibinfo {pages} {085016}
  (\bibinfo {year} {2009})}\BibitemShut {NoStop}%
\bibitem [{\citenamefont {Kjaergaard}\ \emph {et~al.}(2017)\citenamefont
  {Kjaergaard}, \citenamefont {Suominen}, \citenamefont {Nowak}, \citenamefont
  {Akhmerov}, \citenamefont {Shabani}, \citenamefont {Palmstrom}, \citenamefont
  {Nichele},\ and\ \citenamefont {Marcus}}]{Kjaergaard17}%
  \BibitemOpen
  \bibfield  {author} {\bibinfo {author} {\bibfnamefont {M.}~\bibnamefont
  {Kjaergaard}}, \bibinfo {author} {\bibfnamefont {H.}~\bibnamefont
  {Suominen}}, \bibinfo {author} {\bibfnamefont {M.}~\bibnamefont {Nowak}},
  \bibinfo {author} {\bibfnamefont {A.}~\bibnamefont {Akhmerov}}, \bibinfo
  {author} {\bibfnamefont {J.}~\bibnamefont {Shabani}}, \bibinfo {author}
  {\bibfnamefont {C.}~\bibnamefont {Palmstrom}}, \bibinfo {author}
  {\bibfnamefont {F.}~\bibnamefont {Nichele}},\ and\ \bibinfo {author}
  {\bibfnamefont {C.}~\bibnamefont {Marcus}},\ }\bibfield  {title} {\enquote
  {\bibinfo {title} {Transparent semiconductor-superconductor interface and
  induced gap in an epitaxial heterostructure {Josephson} junction},}\
  }\href@noop {} {\bibfield  {journal} {\bibinfo  {journal} {Phys. Rev. Appl.}\
  }\textbf {\bibinfo {volume} {7}},\ \bibinfo {pages} {034029} (\bibinfo {year}
  {2017})}\BibitemShut {NoStop}%
\bibitem [{\citenamefont {Mayer}\ \emph {et~al.}(2020)\citenamefont {Mayer},
  \citenamefont {Schiela}, \citenamefont {Yuan}, \citenamefont {Hatefipour},
  \citenamefont {Sarney}, \citenamefont {Svensson}, \citenamefont {Leff},
  \citenamefont {Campos}, \citenamefont {Wickramasinghe}, \citenamefont
  {Dartiailh}, \citenamefont {Zutic},\ and\ \citenamefont
  {Shabani}}]{Mayer20-A}%
  \BibitemOpen
  \bibfield  {author} {\bibinfo {author} {\bibfnamefont {W.}~\bibnamefont
  {Mayer}}, \bibinfo {author} {\bibfnamefont {W.~F.}\ \bibnamefont {Schiela}},
  \bibinfo {author} {\bibfnamefont {J.}~\bibnamefont {Yuan}}, \bibinfo {author}
  {\bibfnamefont {M.}~\bibnamefont {Hatefipour}}, \bibinfo {author}
  {\bibfnamefont {W.~L.}\ \bibnamefont {Sarney}}, \bibinfo {author}
  {\bibfnamefont {S.~P.}\ \bibnamefont {Svensson}}, \bibinfo {author}
  {\bibfnamefont {A.~C.}\ \bibnamefont {Leff}}, \bibinfo {author}
  {\bibfnamefont {T.}~\bibnamefont {Campos}}, \bibinfo {author} {\bibfnamefont
  {K.~S.}\ \bibnamefont {Wickramasinghe}}, \bibinfo {author} {\bibfnamefont
  {M.~C.}\ \bibnamefont {Dartiailh}}, \bibinfo {author} {\bibfnamefont
  {I.}~\bibnamefont {Zutic}},\ and\ \bibinfo {author} {\bibfnamefont
  {J.}~\bibnamefont {Shabani}},\ }\bibfield  {title} {\enquote {\bibinfo
  {title} {Superconducting proximity effect in {InAsSb} surface quantum wells
  with in-situ {Al} contact},}\ }\href@noop {} {\bibfield  {journal} {\bibinfo
  {journal} {ACS Appl. Electron. Mat.}\ }\textbf {\bibinfo {volume} {2}},\
  \bibinfo {pages} {2351} (\bibinfo {year} {2020})}\BibitemShut {NoStop}%
\bibitem [{\citenamefont {Hertel}\ \emph {et~al.}(2021)\citenamefont {Hertel},
  \citenamefont {Andersen}, \citenamefont {van Zanten}, \citenamefont
  {Eichinger}, \citenamefont {Scarlino}, \citenamefont {Yadav}, \citenamefont
  {Karthik}, \citenamefont {Gronin}, \citenamefont {Gardner}, \citenamefont
  {Manfra}, \citenamefont {Marcus},\ and\ \citenamefont
  {Petersson}}]{Hertel21}%
  \BibitemOpen
  \bibfield  {author} {\bibinfo {author} {\bibfnamefont {A.}~\bibnamefont
  {Hertel}}, \bibinfo {author} {\bibfnamefont {L.~O.}\ \bibnamefont
  {Andersen}}, \bibinfo {author} {\bibfnamefont {D.~M.~T.}\ \bibnamefont {van
  Zanten}}, \bibinfo {author} {\bibfnamefont {M.}~\bibnamefont {Eichinger}},
  \bibinfo {author} {\bibfnamefont {P.}~\bibnamefont {Scarlino}}, \bibinfo
  {author} {\bibfnamefont {S.}~\bibnamefont {Yadav}}, \bibinfo {author}
  {\bibfnamefont {J.}~\bibnamefont {Karthik}}, \bibinfo {author} {\bibfnamefont
  {S.}~\bibnamefont {Gronin}}, \bibinfo {author} {\bibfnamefont {G.~C.}\
  \bibnamefont {Gardner}}, \bibinfo {author} {\bibfnamefont {M.~J.}\
  \bibnamefont {Manfra}}, \bibinfo {author} {\bibfnamefont {C.~M.}\
  \bibnamefont {Marcus}},\ and\ \bibinfo {author} {\bibfnamefont {K.~D.}\
  \bibnamefont {Petersson}},\ }\bibfield  {title} {\enquote {\bibinfo {title}
  {Electrical properties of selective-area-grown superconductor-semiconductor
  hybrid structures on {Silicon}},}\ }\href@noop {} {\bibfield  {journal}
  {\bibinfo  {journal} {Phys. Rev. Appl.}\ }\textbf {\bibinfo {volume} {16}},\
  \bibinfo {pages} {044015} (\bibinfo {year} {2021})}\BibitemShut {NoStop}%
\bibitem [{\citenamefont {Haberkorn}, \citenamefont {Knauer},\ and\
  \citenamefont {Richter}(1978)}]{Haberkorn78}%
  \BibitemOpen
  \bibfield  {author} {\bibinfo {author} {\bibfnamefont {W.}~\bibnamefont
  {Haberkorn}}, \bibinfo {author} {\bibfnamefont {H.}~\bibnamefont {Knauer}},\
  and\ \bibinfo {author} {\bibfnamefont {J.}~\bibnamefont {Richter}},\
  }\bibfield  {title} {\enquote {\bibinfo {title} {A theoretical study of the
  current-phase relation in {Josephson} contacts},}\ }\href@noop {} {\bibfield
  {journal} {\bibinfo  {journal} {Phys. Stat. Sol. (a)}\ }\textbf {\bibinfo
  {volume} {47}},\ \bibinfo {pages} {K161} (\bibinfo {year}
  {1978})}\BibitemShut {NoStop}%
\bibitem [{\citenamefont {Mayer}\ \emph {et~al.}(2019)\citenamefont {Mayer},
  \citenamefont {Yuan}, \citenamefont {Wickramasinghe}, \citenamefont {Nguyen},
  \citenamefont {Dartiailh},\ and\ \citenamefont {Shabani}}]{Mayer19}%
  \BibitemOpen
  \bibfield  {author} {\bibinfo {author} {\bibfnamefont {W.}~\bibnamefont
  {Mayer}}, \bibinfo {author} {\bibfnamefont {J.}~\bibnamefont {Yuan}},
  \bibinfo {author} {\bibfnamefont {K.~S.}\ \bibnamefont {Wickramasinghe}},
  \bibinfo {author} {\bibfnamefont {T.}~\bibnamefont {Nguyen}}, \bibinfo
  {author} {\bibfnamefont {M.~C.}\ \bibnamefont {Dartiailh}},\ and\ \bibinfo
  {author} {\bibfnamefont {J.}~\bibnamefont {Shabani}},\ }\bibfield  {title}
  {\enquote {\bibinfo {title} {Superconducting proximity effect in epitaxial
  {Al-InAs} heterostructures},}\ }\href@noop {} {\bibfield  {journal} {\bibinfo
   {journal} {Appl. Phys. Lett.}\ }\textbf {\bibinfo {volume} {114}},\ \bibinfo
  {pages} {103104} (\bibinfo {year} {2019})}\BibitemShut {NoStop}%
\bibitem [{\citenamefont {Li}\ \emph {et~al.}(2018)\citenamefont {Li},
  \citenamefont {Gallop}, \citenamefont {Hao},\ and\ \citenamefont
  {Romans}}]{LiT18}%
  \BibitemOpen
  \bibfield  {author} {\bibinfo {author} {\bibfnamefont {T.}~\bibnamefont
  {Li}}, \bibinfo {author} {\bibfnamefont {J.}~\bibnamefont {Gallop}}, \bibinfo
  {author} {\bibfnamefont {L.}~\bibnamefont {Hao}},\ and\ \bibinfo {author}
  {\bibfnamefont {E.}~\bibnamefont {Romans}},\ }\bibfield  {title} {\enquote
  {\bibinfo {title} {Ballistic josephson junctions based on cvd graphene},}\
  }\href@noop {} {\bibfield  {journal} {\bibinfo  {journal} {Supercond. Sci.
  Technol.}\ }\textbf {\bibinfo {volume} {31}},\ \bibinfo {pages} {045004}
  (\bibinfo {year} {2018})}\BibitemShut {NoStop}%
\bibitem [{\citenamefont {Lee}\ \emph {et~al.}(2015)\citenamefont {Lee},
  \citenamefont {Kim}, \citenamefont {Jhi},\ and\ \citenamefont
  {Lee}}]{LeeGH15}%
  \BibitemOpen
  \bibfield  {author} {\bibinfo {author} {\bibfnamefont {G.-H.}\ \bibnamefont
  {Lee}}, \bibinfo {author} {\bibfnamefont {S.}~\bibnamefont {Kim}}, \bibinfo
  {author} {\bibfnamefont {S.-H.}\ \bibnamefont {Jhi}},\ and\ \bibinfo {author}
  {\bibfnamefont {H.-J.}\ \bibnamefont {Lee}},\ }\bibfield  {title} {\enquote
  {\bibinfo {title} {Ultimately short ballistic vertical graphene {Josephson}
  junctions},}\ }\href@noop {} {\bibfield  {journal} {\bibinfo  {journal} {Nat.
  Commun.}\ }\textbf {\bibinfo {volume} {6}},\ \bibinfo {pages} {6181}
  (\bibinfo {year} {2015})}\BibitemShut {NoStop}%
\bibitem [{\citenamefont {Borzenets}\ \emph {et~al.}(2016)\citenamefont
  {Borzenets}, \citenamefont {Amet}, \citenamefont {Ke}, \citenamefont
  {Draelos}, \citenamefont {Wei}, \citenamefont {Seredinski}, \citenamefont
  {Watanabe}, \citenamefont {Taniguchi}, \citenamefont {Bomze}, \citenamefont
  {Yamamoto}, \citenamefont {Tarucha},\ and\ \citenamefont
  {Finkelstein}}]{Borzenets16}%
  \BibitemOpen
  \bibfield  {author} {\bibinfo {author} {\bibfnamefont {I.}~\bibnamefont
  {Borzenets}}, \bibinfo {author} {\bibfnamefont {F.}~\bibnamefont {Amet}},
  \bibinfo {author} {\bibfnamefont {C.}~\bibnamefont {Ke}}, \bibinfo {author}
  {\bibfnamefont {A.}~\bibnamefont {Draelos}}, \bibinfo {author} {\bibfnamefont
  {M.}~\bibnamefont {Wei}}, \bibinfo {author} {\bibfnamefont {A.}~\bibnamefont
  {Seredinski}}, \bibinfo {author} {\bibfnamefont {K.}~\bibnamefont
  {Watanabe}}, \bibinfo {author} {\bibfnamefont {T.}~\bibnamefont {Taniguchi}},
  \bibinfo {author} {\bibfnamefont {Y.}~\bibnamefont {Bomze}}, \bibinfo
  {author} {\bibfnamefont {M.}~\bibnamefont {Yamamoto}}, \bibinfo {author}
  {\bibfnamefont {S.}~\bibnamefont {Tarucha}},\ and\ \bibinfo {author}
  {\bibfnamefont {G.}~\bibnamefont {Finkelstein}},\ }\bibfield  {title}
  {\enquote {\bibinfo {title} {Ballistic graphene {Josephson} junctions from
  the short to the long junction regimes},}\ }\href@noop {} {\bibfield
  {journal} {\bibinfo  {journal} {Phys. Rev. Lett.}\ }\textbf {\bibinfo
  {volume} {117}},\ \bibinfo {pages} {237002} (\bibinfo {year}
  {2016})}\BibitemShut {NoStop}%
\end{thebibliography}
\end{document}


\title{\LARGE{SUPPLEMENTARY MATERIAL}\\~\\
\Large{High transparency induced superconductivity in field effect two-dimensional electron gases in undoped InAs/AlGaSb surface quantum wells}}

\author{~\\ E. Annelise Bergeron, F. Sfigakis, A. Elbaroudy, A. W. M. Jordan, F. Thompson,\\ George Nichols, Y. Shi, M. C. Tam, Z. R. Wasilewski, and J. Baugh}

\affiliation{University of Waterloo, Waterloo N2L 3G1, Canada}

\begin{abstract}
~\\~\\ \noindent \textbf{Table of Contents:}\\~\vspace{-5mm}\\
\indent\qquad Section \ref{sec:MBE}: MBE growth \\
\indent\qquad Section \ref{sec:bandstructure}: Bandstructure profiles \\
\indent\qquad Section \ref{sec:fabrication}: Sample fabrication \\
\indent\qquad Section \ref{sec:magnetotransport}: Characterization of Hall bars \\
\indent\qquad Section \ref{sec:SNS}: Characterization of SNS junctions \\
\end{abstract}

\maketitle

\section{MBE growth}
\label{sec:MBE}

All samples in this study were grown on undoped p-type GaSb (100) substrates. G743 was grown on a quarter 2-inch wafer, while G782 was grown on a full 2-inch wafer. However, the growth procedure remained consistent across both wafers. The substrates were initially outgassed at 200~$^\circ$C for two hours in the system's load lock, followed by an additional hour of outgassing at 300~$^\circ$C in the preparation module. They were then transferred to the growth module to start the epitaxial growth process. In the growth chamber, the wafer was radiatively heated by the substrate manipulator, controlled by a proportional-integral-derivative (PID) controller. The substrate's temperature was monitored using an integrated spectral pyrometry (ISP) technique developed by our Molecular Beam Epitaxy (MBE) group [\onlinecite{AlanTam17}]. This technique was necessary because the opaque nature of the GaSb substrate makes its absorption edge undetectable.

The native oxide layer was removed by heating the substrate to 530\,$^\circ$C under an Sb flux overpressure, while maintaining the cracker temperature at 900\,$^\circ$C, to prevent the desorption of Ga. Oxide desorption was monitored using Reflective High Energy Electron Diffraction (RHEED), during which a reconstruction from (1$\times$5) to (1$\times$3) was observed at approximately 520\,$^\circ$C [\onlinecite{Bracker00}]. Subsequently, the substrate was annealed for 5 minutes at 530\,$^\circ$C. The substrate temperature was subsequently lowered to 500\,$^\circ$C to deposit a 25 nm smoothing layer of GaSb. This was followed by the growth of the quaternary buffer layer Al$_{0.8}$Ga$_{0.2}$Sb$_{0.93}$As$_{0.07}$ at the same temperature, following the procedure outlined in Ref.\,[\onlinecite{ThomasC18}]. To achieve the targeted 7\% As concentration in the quaternary buffer and to prevent dislocation propagation into the InAs quantum well (QW), the Sb and As fluxes were adjusted before the oxide desorption process. It was determined that the substrate temperature significantly influences the incorporation of As and Sb into the quaternary buffer, thereby affecting their respective ratios. Through a series of experiments that varied the As and Sb fluxes during buffer growth and subsequent analysis of these ratios using ex-situ X-ray diffraction (XRD), it was established that maintaining consistent flux ratios leads to an As concentration of 6-7\% in the quaternary buffer when the substrate temperature is held between 480\,$^\circ$C and 500\,$^\circ$C. Furthermore, this ratio range does not adversely impact the mobility or carrier density in the InAs quantum well, as evidenced by results from ungated hall bar measurements.

The bottom barrier, consisting of Al$_{0.8}$Ga$_{0.2}$Sb with a thickness of 20~nm, was then grown. Possessing a higher band gap, this layer serves to confine the two-dimensional electron gas within the InAs quantum well. During the transition from the buffer to the barrier layer, an Sb flux was supplied to create an antimony-rich surface, which facilitated the growth of the barrier. Subsequently, the substrate temperature was maintained at 490$-$500\,$^\circ$C for the deposition of a 24~nm thick InAs quantum well. At this stage, critical interface-like decisions were made, and following the methodology of Tuttle et al. [\onlinecite{Tuttle90}], the shutters for As, Sb, and In were operated strategically to form either InSb-like or AlAs-like interfaces. The growth process was concluded with the deposition of a 6~nm In$_{0.75}$Ga$_{0.25}$As cap layer, with all layers being grown at a uniform rate of 2 {\AA}/s.

\begin{figure}[b]
  \includegraphics[width=1.0\columnwidth]{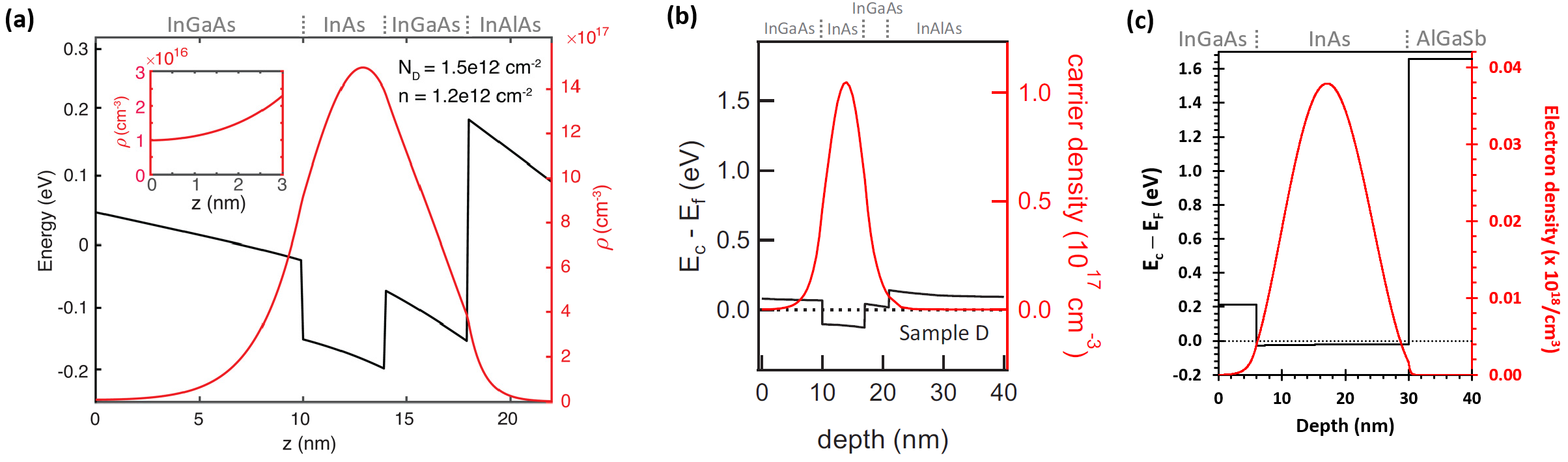}
  \caption{Calculated bandstructure profiles in the MBE growth direction of ungated wafers consisting of: (a) an In(Ga)As/InAlAs surface quantum wells [modified from Fig.~1d in Ref.\,\onlinecite{YuanJ21}] with a 4~nm InAs core, (b) an In(Ga)As/InAlAs surface quantum wells [modified from Fig.~1c in Ref.\,\onlinecite{LeeJS19}] with a 7~nm InAs core, and (c) an InAs/AlGaSb surface quantum well, the focus of this paper's main text. ``0 nm'' corresponds to the wafer surface. The 2DEG wavefunction $\psi(z)$ is represented by a solid red line, the conduction band edge by a solid black line, and the Fermi level by a dashed grey line.}
  \label{fig:bandstructure}
\end{figure}

\section{Bandstructure profiles}
\label{sec:bandstructure}

Figure \ref{fig:bandstructure} shows bandstructure profiles calculated from nextnano\texttrademark ~self-consistent simulations, solving both the Poisson and Schr\"{o}dinger equations [\onlinecite{NextNano-A,NextNano-B,NextNano-C}]. Figure~\ref{fig:bandstructure}a and \ref{fig:bandstructure}b are modified from literature, and Figure~\ref{fig:bandstructure}c is our own calculations.

The width of the InAs quantum well (typ.~4$-$7 nm) in an In(Ga)As/InAlAs heterostructure must be equal or less than its small critical thickness in that material system. As a direct consequence, the 2DEG wavefunction spills out of the InAs quantum well and spreads into the surrounding InGaAs and even into In$_{0.8}$Al$_{0.2}$As barrier layer. This is shown in Fig.~\ref{fig:bandstructure}a and Fig.~\ref{fig:bandstructure}b, where a significant fraction (25\%$-$50\%) of the 2DEG electrons are outside the InAs QW.

On the other hand, the InAs quantum well in a InAs/AlGaAs heterostructure can be very wide: 24 nm in our heterostructures. Because the QW is wider and the AlGaSb barrier is larger than InAlAs, the 2DEG wavefunction is mostly confined within the InAs quantum well. This is shown in Fig.~\ref{fig:bandstructure}c.

\section{Sample fabrication}
\label{sec:fabrication}

The fabrication steps of Hall bars and SNS junctions discussed here and in the main text are presented in the following two sub-sections.

\subsection{Hall bars}

Samples are cleaned prior to lithography by sonication in acetone and subsequently propanol for 5 minutes each before a final blow dry with nitrogen. Mesa regions are defined with optical lithography using Shipley S1811 photoresist. The resist is spun at 5000 rpm for 60 seconds and baked at 120\,$^\circ$C for 90 seconds. Following exposure, the photoresist is developed in MF319 developer for one minute. In order to ensure no unintentional thin film of photoresist remains in the exposed regions, samples are ashed in an oxygen plasma at 50 W for twenty seconds prior to wet etching to remove any residual photoresist in the exposed (off-mesa) regions. Wet etching proceeds with a ten second dip in buffered oxide etch (BOE) (1:10) to remove any native oxide on the surface of the sample caused by ashing and exposure to air. The mesa is etched with a solution of H$_2$O$_2$:H$_3$PO$_4$:C$_6$H$_8$O$_7$:H$_2$O mixed 3:4:9:44 by volume for approximately 30 seconds or until an etch depth of at least 100 nm has been reached. After etching, the photoresist etch mask is removed by sonication in acetone and isopropanol.

Optical lithography for definition of Ohmic contacts uses a bilayer resist recipe of MMA/Shipley. First the MMA (methyl methacrylate) is spun at 5000 rpm for 60 seconds and baked at 150\,$^\circ$C for 5 minutes. Next the Shipley is spun in the same manner with a bake at 120\,$^\circ$C for 90 seconds. Optical exposure and development of the sample in MF319 succesfully removes Shipley in regions where Ohmic contacts are to be formed. This exposure and development does not remove the MMA which protects the surface from being etched by the MF319 developer. MMA is subsequently removed by a fifteen minute UV exposure, and developed in a solution of isopropanol:H$_2$O at a 7:3 concentration. An angled 45$^\circ$ deposition of 20/60 nm of Ti/Au is performed in a thermal evaporator.

The 60 nm thick gate dielectric layer (HfO$_2$ or SiO$_2$) which isolates the top gate from the quantum well and Ohmic contacts in a gated Hallbar is deposited using atomic layer deposition at 150\,$^\circ$C; the dielectric breakdown field of HfO$_2$ (SiO$_2$) is $\sim$1.5 MV/cm ($>$3.3 MV/cm) at $T=1.6$ K. Following deposition, optical lithography with Shipley is used to define vias above the Ohmic contacts. The oxide in the exposed vias is etched in BOE at a concentration of 1:10. Following etching, via resist is removed and processing proceeds with optical lithography of the top-gate and bond pads to metallic contacts. A bilayer of MMA/Shipley as discussed for the Ohmic contacts is again used and the Ti/Au (20/60 nm) top-gate and bond pads are similarly deposited in a thermal evaporator at an angle of 45\,$^\circ$.

\subsection{SNS junctions}

\begin{figure}
  \includegraphics[width=0.75\columnwidth]{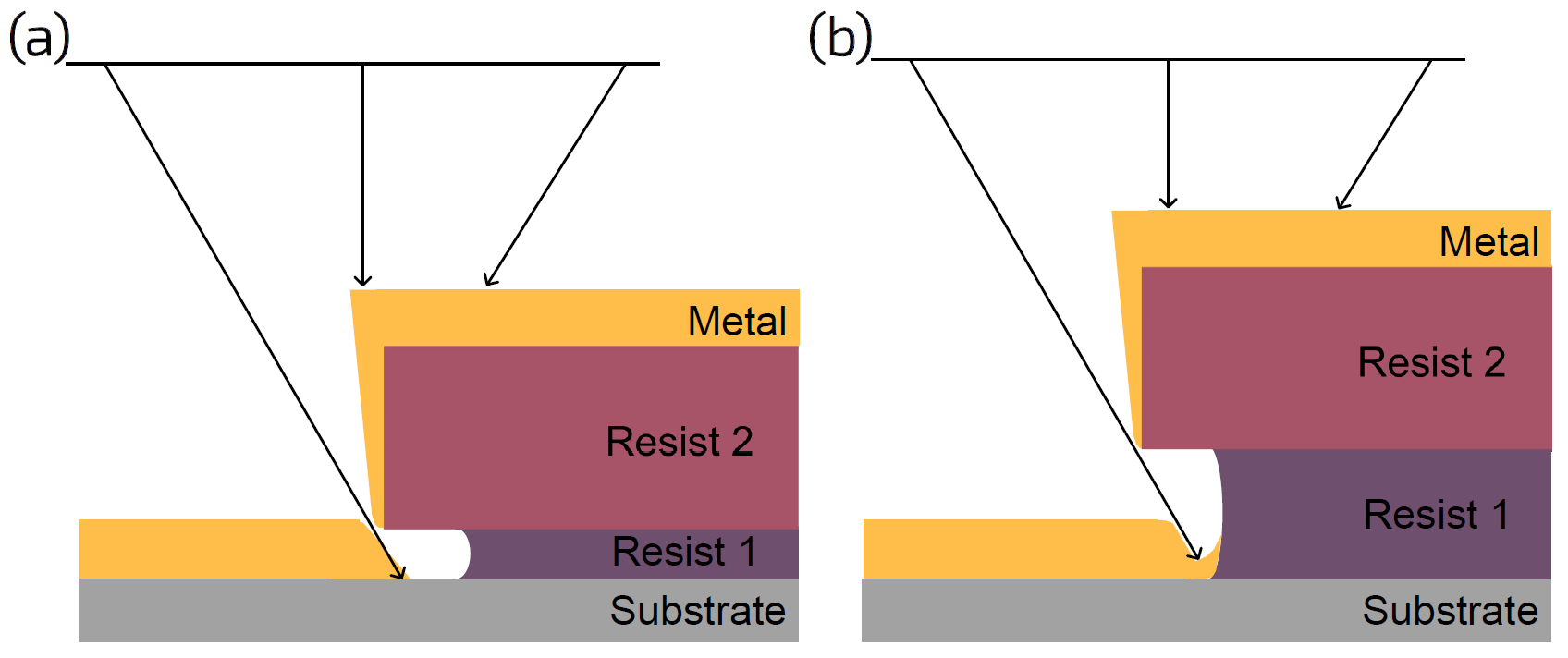}
  \caption{Bilayer PMMA profiles for Nb sputter deposition, with 495K PMMA being ``resist 1'' and 950K PMMA being ``resist 2''. (a) Ideal profile. (b) Non-ideal profile resulting in ``lily pads'', because resist 1 is too thick.}
  \label{fig:Nb-EBL}
\end{figure}

Sample preparation methods of the previous section applies to fabrication of SNS junction apart from details of the superconducting leads which is discussed here. A bilayer resist profile of PMMA(495K)/PMMA(950K) is used to define a suitable bilayer undercut profile for liftoff of sputtered Nb, where 495K/950K refers to the PMMA molecular weight. Deposition by sputtering occurs at a relatively wide range of deposition angles (e.g., Fig.~\ref{fig:Nb-EBL}), unlike thermal or electron beam evaporation. Figure~\ref{fig:Nb-EBL}a shows the optimum profile of a PMMA bilayer specifically for sputtering. The bottom layer is PMMA 495K and thin, while the top layer is PMMA 950K. The latter is mostly responsible for determining feature size. If the bottom 495K layer is too thick, metal is deposited on the sidewall of the bottom layer [see Fig.~\ref{fig:Nb-EBL}b], resulting in ``lily pads'': tall, sharp edges that can cause shorts between two metal layers separated by a gate dielectric. A 100~kV acceleration voltage is used for e-beam lithography to define the superconducting leads using a JEOL JBX-6300FS direct-write system with proximity effect correction. The beam current for the write is $\sim$1 nA with a beam size of $\sim$17 nm. Following exposure, the sample is developed in IPA:DI water 7:3 by vol. for 30 seconds, rinsed in DI water, and checked under a microscope.

The sulfur passivation process is introduced after the development of a lithographically patterned sample and prior to loading the sample into the sputter chamber. The passivation process begins by removing the native oxide of the semiconductor with buffered oxide etchant (diluted 1:10 in DI water) in areas where the surface is clear of resist. The sample is then rinsed in DI water and transferred to a polysulide solution for 5-20 minutes of passivation under illumination and at room temperature. The polysulfide solution is prepared first by mixing a 3.125~M solution consisting of sulfur powder dissolved in (NH$_4$)$_2$S (20\%), which is then diluted 1:500 in DI water. Upon completion of the sulfur passivation, the sample is rinsed in DI water and quickly transferred (less than 30 s) in air to the loadlock of an ATC Orion sputter system from AJA International. The sample is transferred to the deposition chamber and ion milled for 6.5 minutes at 50 W. The ion milling recipe was calibrated to remove 2-3 nm of native oxide and sulfur. Finally, we sputter 2 nm of Ti at 200 W corresponding to a deposition rate of 8.92 nm/min and subsequently 80 nm of Nb at 200 W corresponding to a deposition rate of 9.36 nm/min.

Lift-off is performed overnight in PG remover followed by 20 minutes in heated PG remover (no stirrer) and 20 minutes in heated PG remover with stirrer at 350 rpm. The sample is removed from heat and pipetted for 5 minutes before being transferred to a plastic beaker filled with acetone and sonicated for 20 seconds. The sample is then rinsed in acetone and propanol and blown dry with nitrogen.

\section{Characterization of Hall bars}
\label{sec:magnetotransport}

\begin{figure}[h]
  \includegraphics[width=1.0\columnwidth]{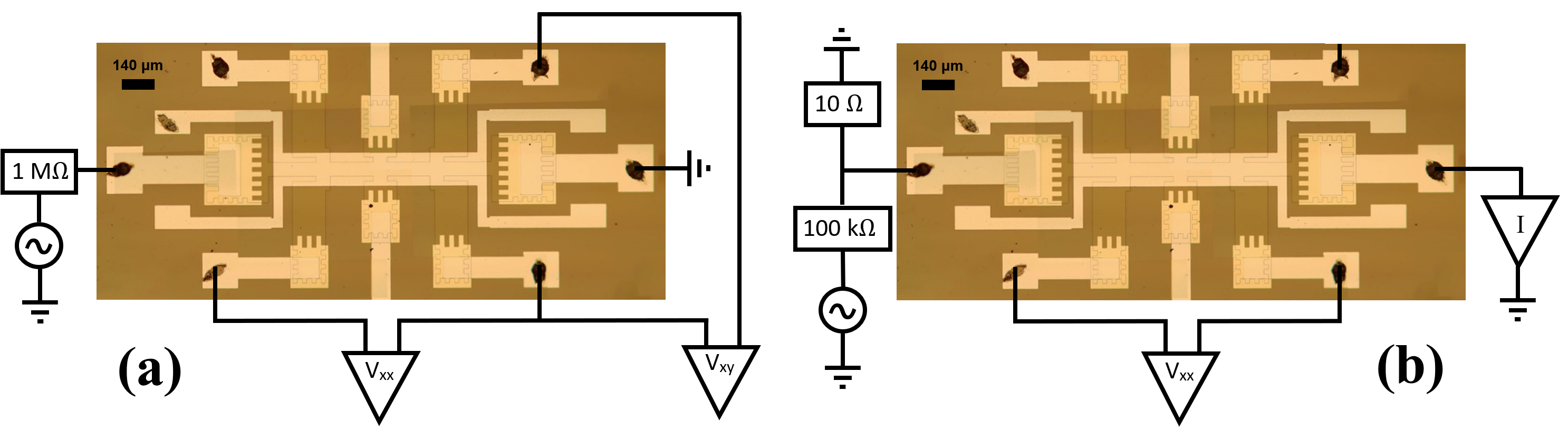}
  \caption{Optical images of a Hall bar with SiO$_2$ gate dielectric. (a) Constant current (100 nA) 4-terminal circuit setup for measuring Hall density and mobility at $T=1.6$~K. (b) Constant voltage (100 $\mu$V) 4-terminal circuit setup for measuring pinch-off characteristics. Triangles represent current/voltage pre-amplifiers, whose voltage output is fed into a SR-830 lock-in (not shown).}
  \label{fig:circuits-hallbars}
\end{figure}

\begin{figure}
  \includegraphics[width=1.0\columnwidth]{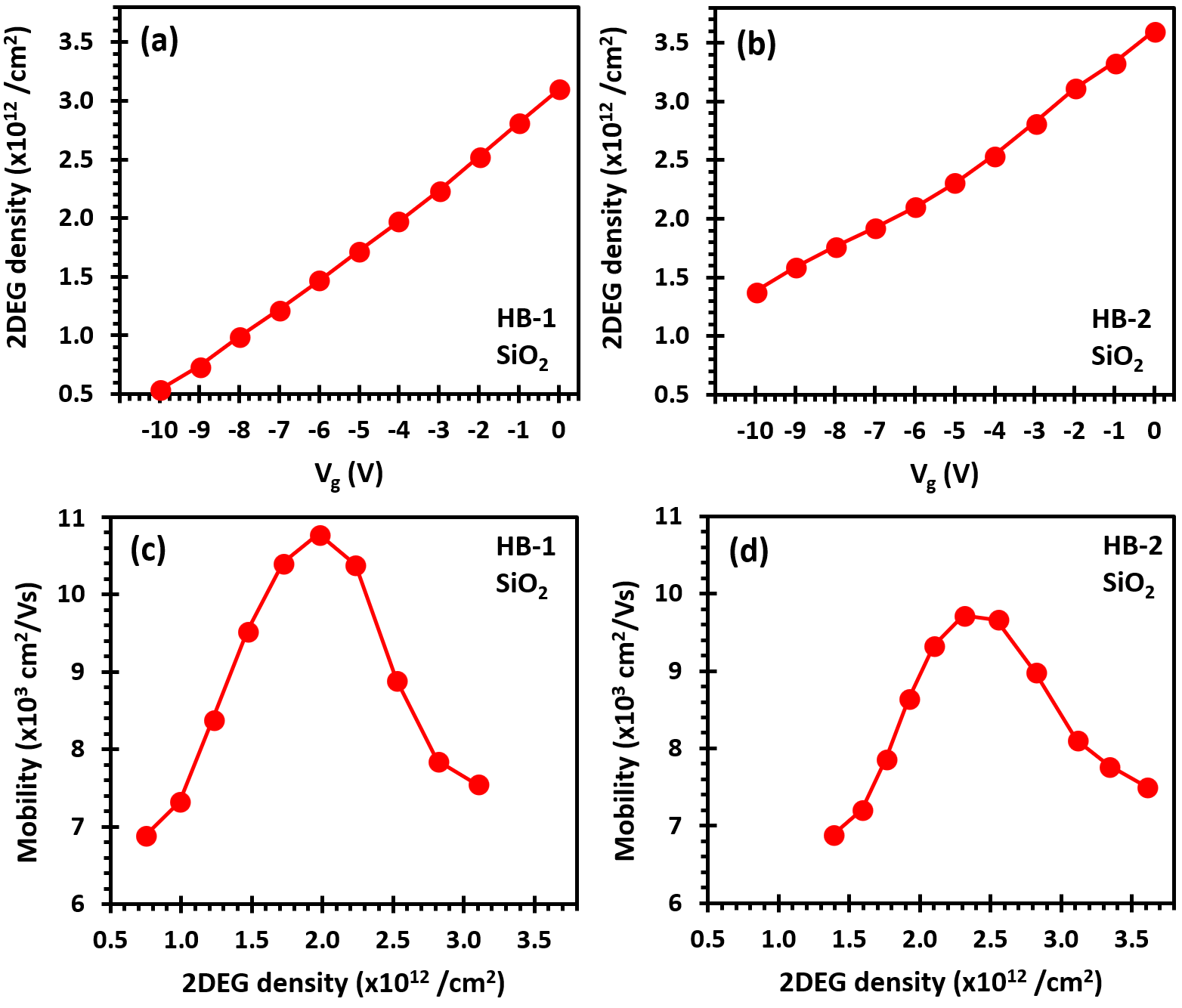}
  \caption{(a)$-$(b) Hall density versus top-gate voltage of Hall bars with \textbf{SiO}$_2$ gate dielectric. (c)$-$(d) Mobility at $T=1.6$~K in the same devices. Lines are a guide to the eye.}
  \label{fig:hallbars-SiO2}
\end{figure}

\begin{figure}
  \includegraphics[width=1.0\columnwidth]{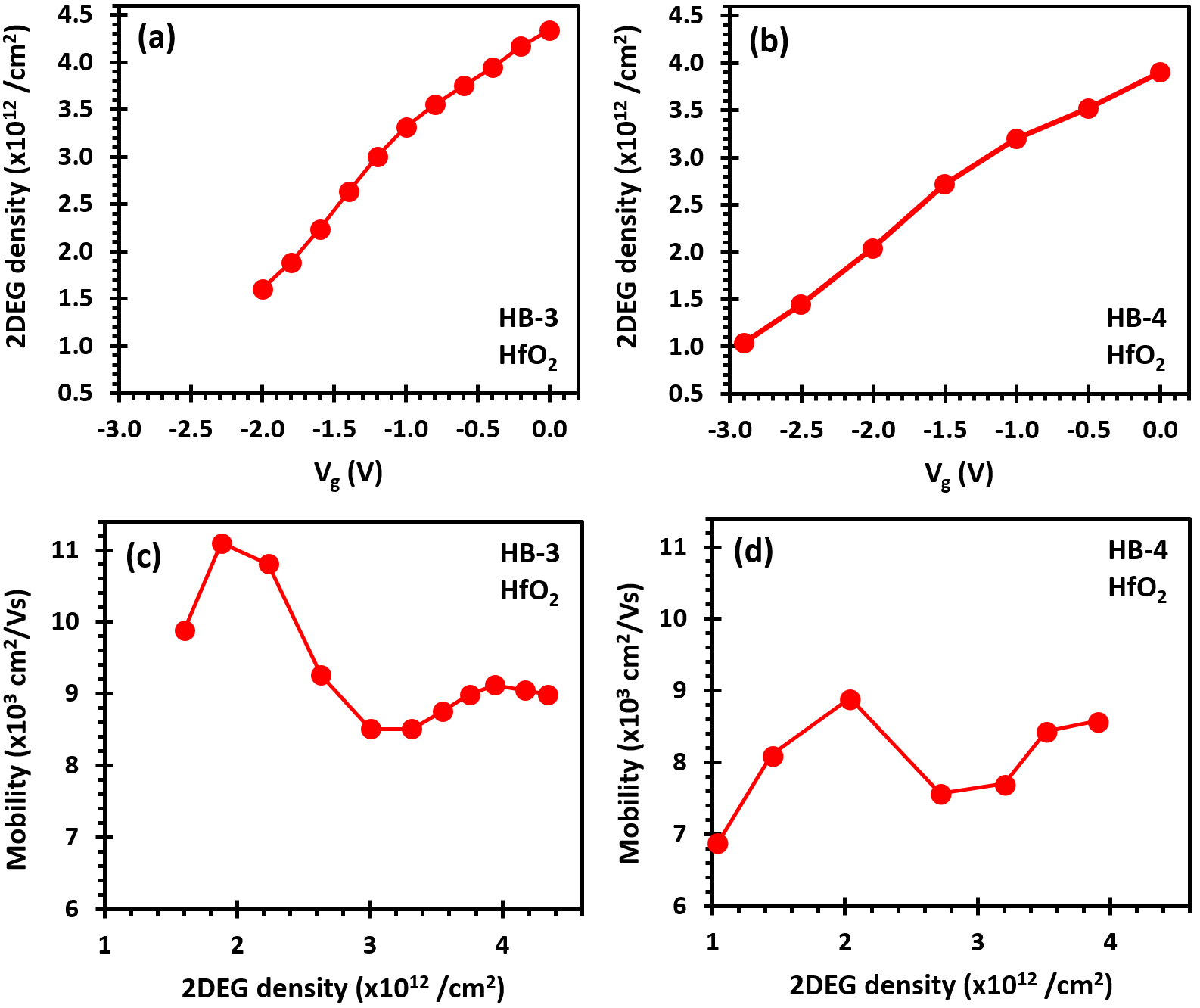}
  \caption{(a)$-$(b) Hall density versus top-gate voltage of Hall bars with \textbf{HfO}$_2$ gate dielectric. (c)$-$(d) Mobility at $T=1.6$~K in the same devices. Lines are a guide to the eye.}
  \label{fig:hallbars-HfO2}
\end{figure}

Figure~\ref{fig:circuits-hallbars} shows a composite photograph of a typical Hall bar and the electrical circuits used for measuring Hall density, mobility, and pinch-off characteristics. Table \ref{tab:Hallbars} lists Hall bars reported here. Figures~\ref{fig:hallbars-SiO2} and \ref{fig:hallbars-HfO2} shows their magnetotransport data.\\

\begin{table}[h]
    \begin{ruledtabular}
    \begin{tabular}{ccccc}
    Hall bar ID & Gate dielectric & Peak mobility & ~ & 2DEG density \vspace{0.5 mm} \\ \hline
    HB-1 & SiO$_2$ & 10,800~cm$^2$/Vs & at & 2.0$\times$10$^{12}$~/cm$^2$ \\
    HB-2 & SiO$_2$ & 9,700~cm$^2$/Vs & at & 2.3$\times$10$^{12}$~/cm$^2$ \\ 
    HB-3 & HfO$_2$ & 11,100~cm$^2$/Vs & at & 1.9$\times$10$^{12}$~/cm$^2$ \\ 
    HB-4 & HfO$_2$ & 8,900~cm$^2$/Vs & at & 2.0$\times$10$^{12}$~/cm$^2$ 
    \end{tabular}
    \end{ruledtabular}
    \caption{List of Hall bar devices, with their peak mobility and corresponding 2DEG density.}
    \label{tab:Hallbars}
\end{table}

\clearpage
\newpage

\clearpage
\newpage

\clearpage
\newpage

\begin{figure}
  \includegraphics[width=1.0\columnwidth]{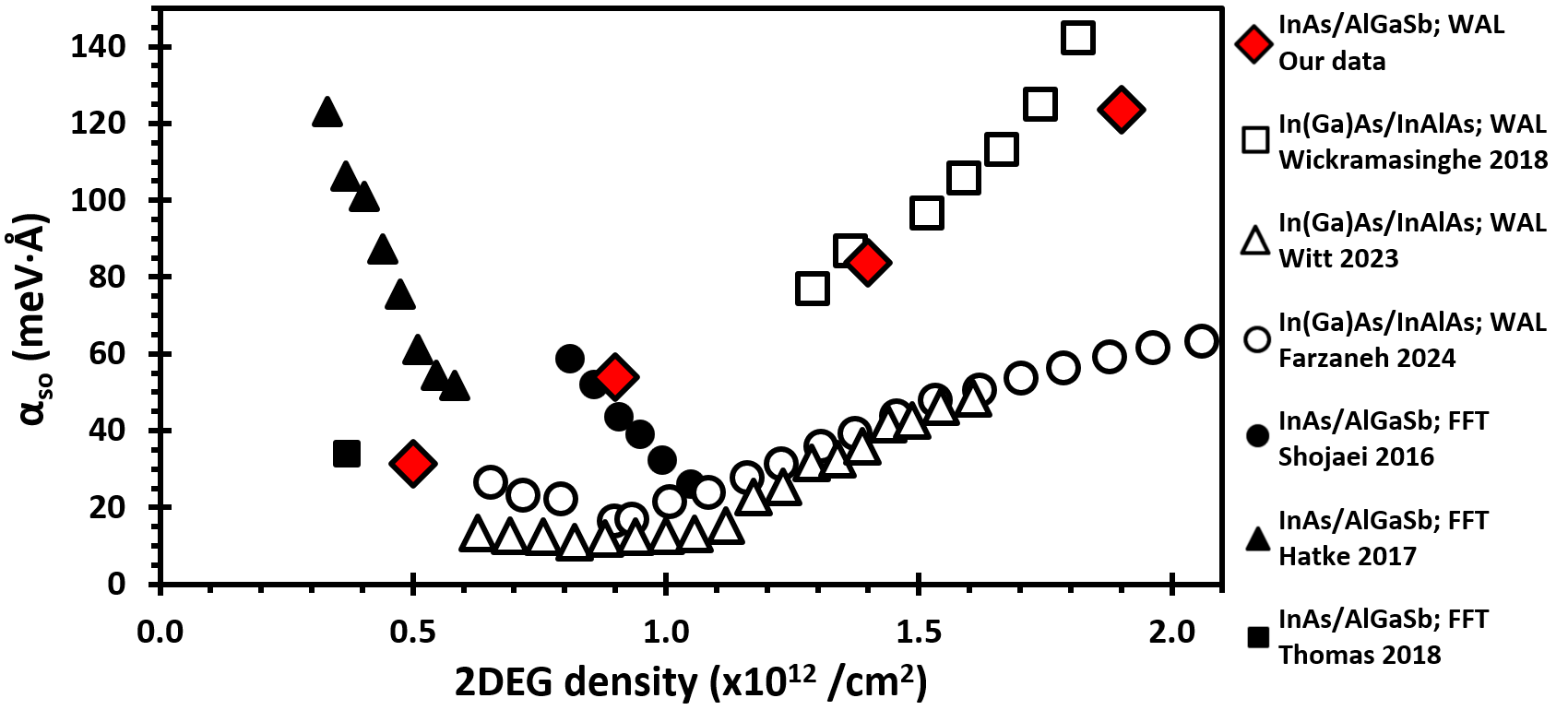}
  \caption{Comparison between our experimentally-measured Rashba coefficients $\alpha_{so}$ (red diamonds) and those from the literature (squares, circles, and triangles). Filled/closed symbols represent $\alpha_{so}$ from InAs/AlGaSb quantum wells,\cite{Shojaei16-A,Hatke17,ThomasC18} and empty/open symbols represent $\alpha_{so}$ from In(Ga)As/InAlAs quantum wells.\cite{Wickramasinghe18,Witt23,Farzaneh24} In the legend, data labeled ``WAL'' were measured from fits of the WAL peak to the ILP model (see main text). Data labeled ``FFT'' were measured from the Fourier transform of Shubnikov-de-Haas oscillations. At first sight, it would appear that $\alpha_{so}$ decreases with 2DEG density in InAs/AlGaSb quantum wells,\cite{Shojaei16-A,Hatke17} whereas it increases with 2DEG density in In(Ga)As/InAlAs quantum wells.\cite{Wickramasinghe18,Witt23,Farzaneh24} However, our data goes against that trend. In fact, the behaviour of the function $\alpha_{so}(n_{\text{2D}})$ is mostly determined by the degree of structural inversion asymmetry (SIA) present in the 2DEG, i.e. from the amplitude and sign of the electric field $E_z$ across the quantum well in the MBE growth direction. When the quantum well is symmetric (i.e. $E_z$=0), $\alpha_{so}(n_{\text{2D}})$ should have a minimum. This is exactly what happens in the experimental data from Ref.~\onlinecite{Farzaneh24} (empty circles in Fig.~\ref{fig:WAL} above, with a minimum near $n_{\text{2D}}=0.9\times 10^{12}$) and Ref.~\onlinecite{ZhangT23} (not shown; see Fig.~9c in Ref.~\onlinecite{ZhangT23}). From the behavior of  $\alpha_{so}(n_{\text{2D}})$ with a front gate and/or back gate, one can thus experimentally determine the direction (sign) of $E_z$, and relate it to bandstructure simulations (e.g., Fig.~\ref{fig:bandstructure}).}
  \label{fig:WAL}
\end{figure}

\clearpage
\newpage

\section{Characterization of SNS junctions}
\label{sec:SNS}

\begin{figure}[h]
  \includegraphics[width=0.5\columnwidth]{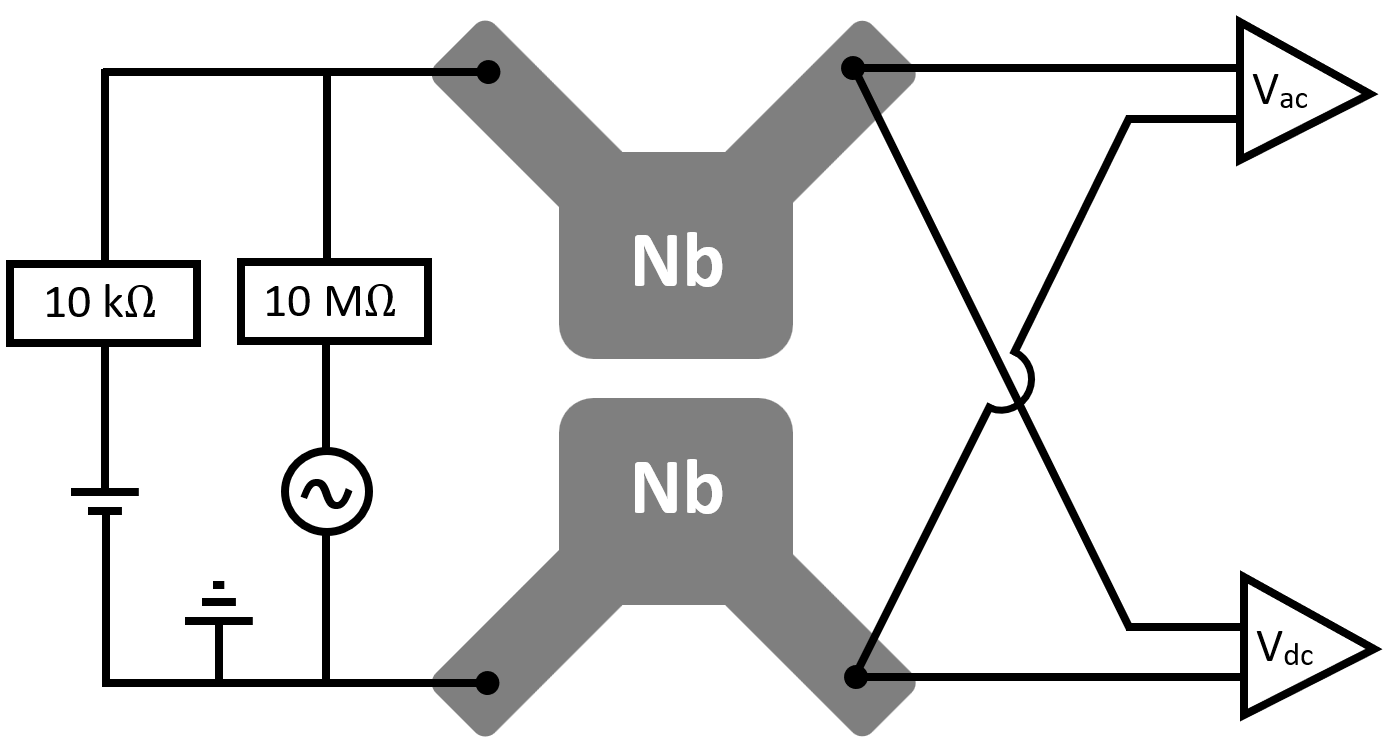}
  \caption{Constant current four-terminal ac and dc circuits used to measure SNS devices.}
  \label{fig:circuits-SNS}
\end{figure}

\begin{figure}
  \includegraphics[width=0.85\columnwidth]{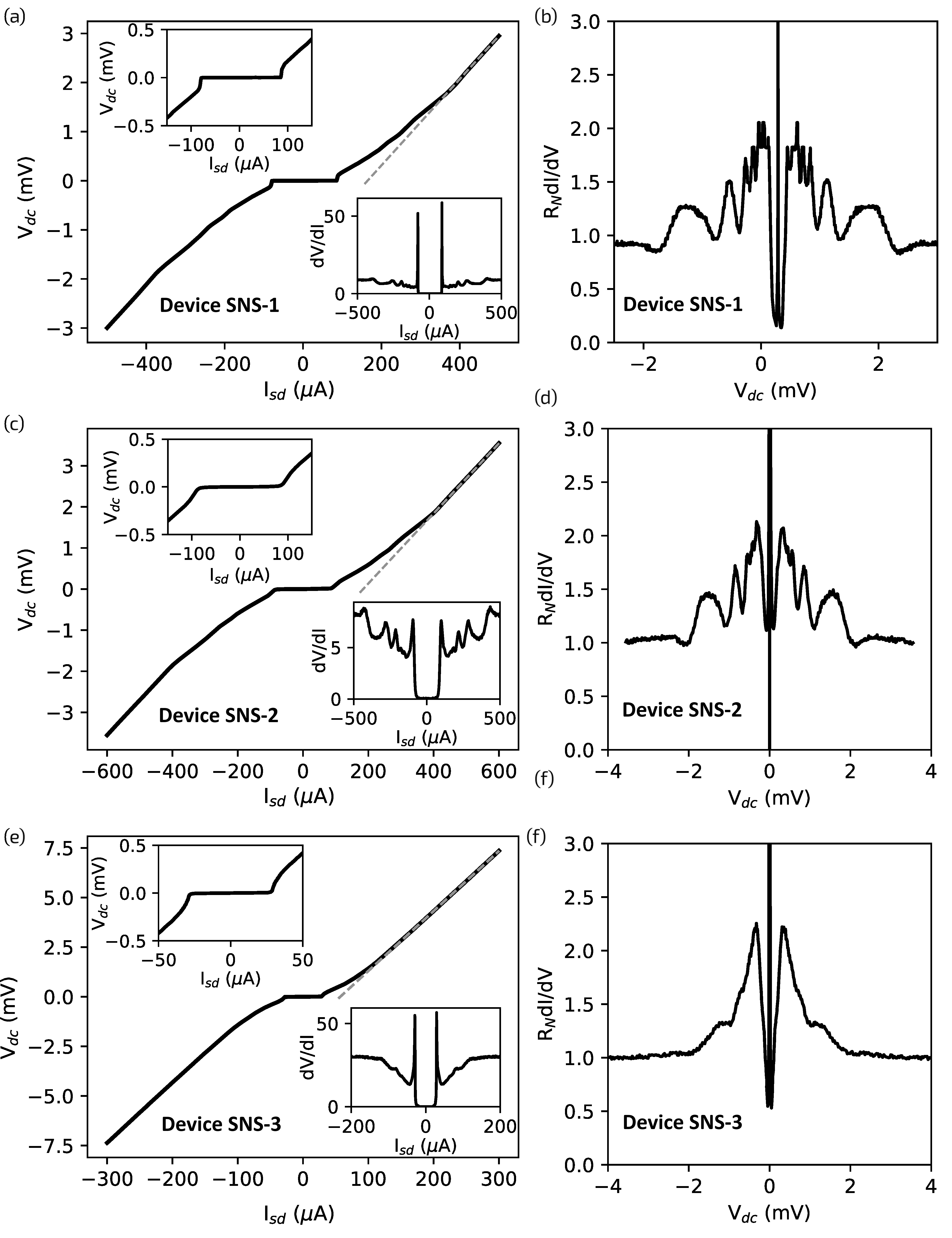}
  \caption{The panels show data for devices: (a)$-$(b) SNS-1, (c)$-$(d) SNS-2, and (e)$-$(f) SNS-3. All panels on the left show dc four-terminal I-V traces of SNS junctions; identical in format to Figure~3a in the main text. All panels on the right show normalized ac four-terminal differential conductance $dI/dV$ and MAR peaks; identical in format to Figure~3b in the main text.}
  \label{fig:SNS-123}
\end{figure}

\begin{figure}
  \includegraphics[width=0.85\columnwidth]{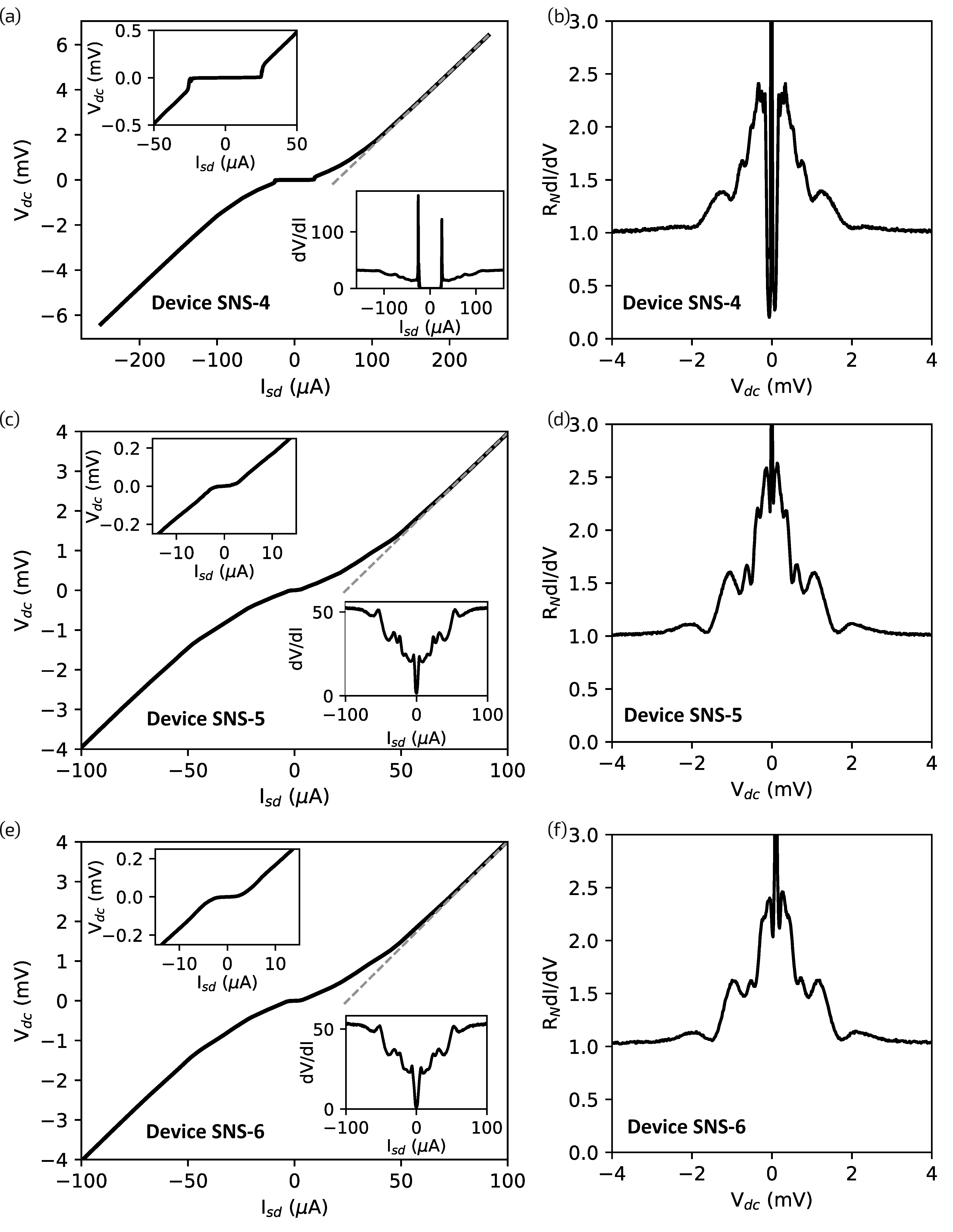}
  \caption{The panels show data for devices: (a)$-$(b) SNS-4, (c)$-$(d) SNS-5, and (e)$-$(f) SNS-6. All panels on the left show dc four-terminal I-V traces of SNS junctions; identical in format to Figure~3a in the main text. All panels on the right show normalized ac four-terminal differential conductance $dI/dV$ and MAR peaks; identical in format to Figure~3b in the main text.}
  \label{fig:SNS-456}
\end{figure}

Figure \ref{fig:circuits-SNS} shows the electrical circuit used for measuring the dc I-V traces and ac MAR peaks in SNS devices. A constant 100 nA ac current was used, whereas dc $I_{sd}$ was swept from 0 to up to 600 $\mu$A. Identical in format to Figures 3a and 3b from the main text, Figures \ref{fig:SNS-123} and \ref{fig:SNS-456} show the dc four-terminal I-V traces and ac normalized four-terminal differential conductances traces for all SNS devices.

\begin{figure}
  \includegraphics[width=0.85\columnwidth]{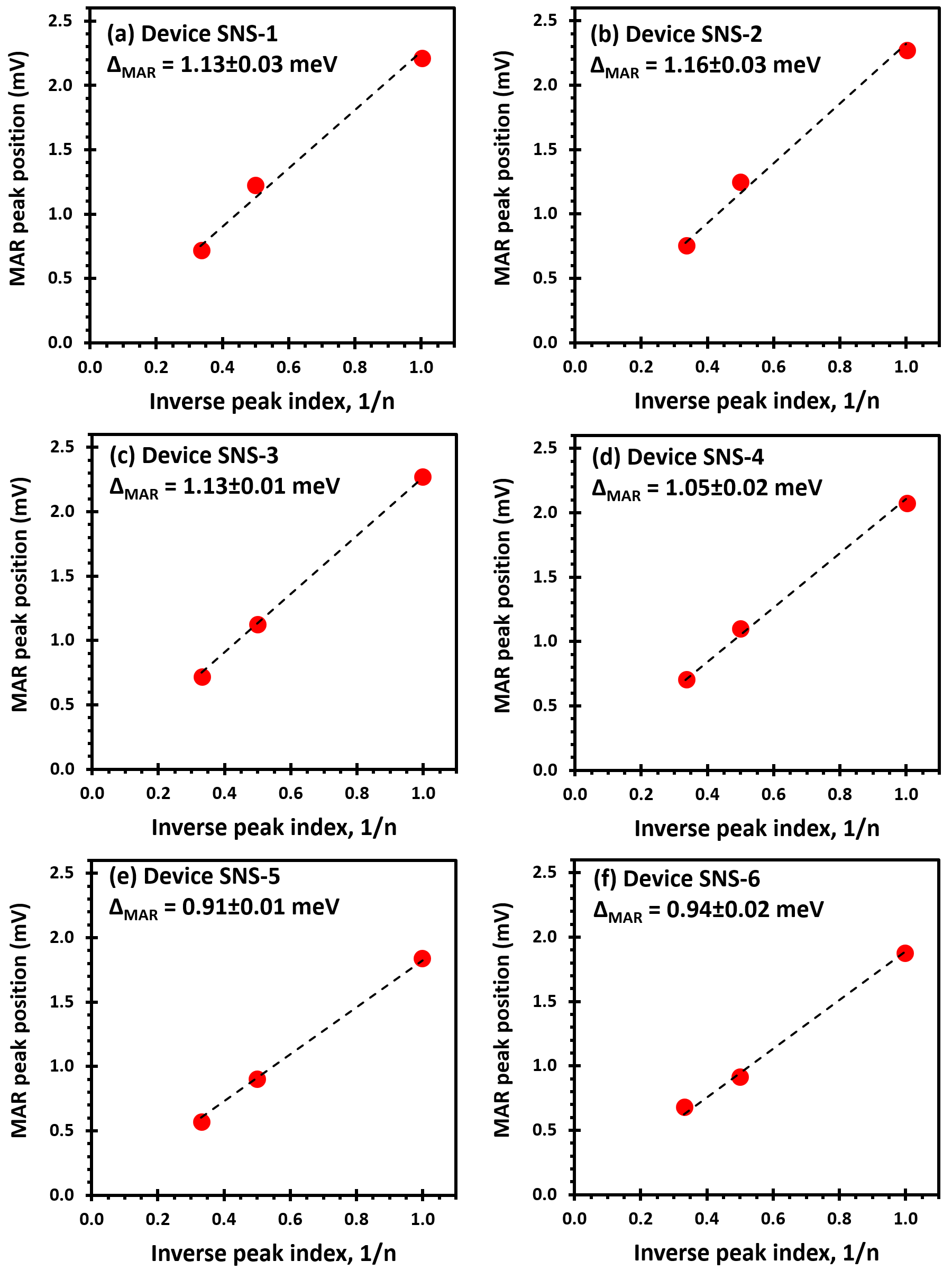}
  \caption{MAR peak positions (red circles) to $V_n = 2\Delta_{\textsc{mar}}/en$, with individual fits (dashed lines) shown for device: (a) SNS-1, (b) SNS-2, (c) SNS-3, (d) SNS-4, (e) SNS-5, and (f) SNS-6. From the fit analysis, the respective $\Delta_{\textsc{mar}}$ are stated. The $R^2$ correlation coefficient is 0.999 in all fits above.}
  \label{fig:MARfits}
\end{figure}

Figure \ref{fig:MARfits} shows the linear fit of the MAR peak positions for each SNS device, from which the value of $\Delta_{\textsc{mar}}$ is obtained. Depending on the value of $Z$ and $T/T_c$, theory [\onlinecite{Flensberg88}] predicts the precise positioning of the MAR peaks with respect to the equation $eV_n = 2\Delta_{\textsc{mar}}/n$ can vary from the conductance peak maximum to the inflection point on the side nearest $V=0$ [see $n=2$ MAR peaks in Fig.~2b and Fig.~3b of Ref.~\onlinecite{Flensberg88}]. In our experiments, since we observe $Z\approx 0.5$ and measure at $T/T_c \approx 0.01$, we thus selected $V_n$ for $n=1,2,3$ in our fits to be the inflection point on the side nearest $V_{dc}=0$ of the conductance peak.

\begin{figure}
  \includegraphics[width=0.5\columnwidth]{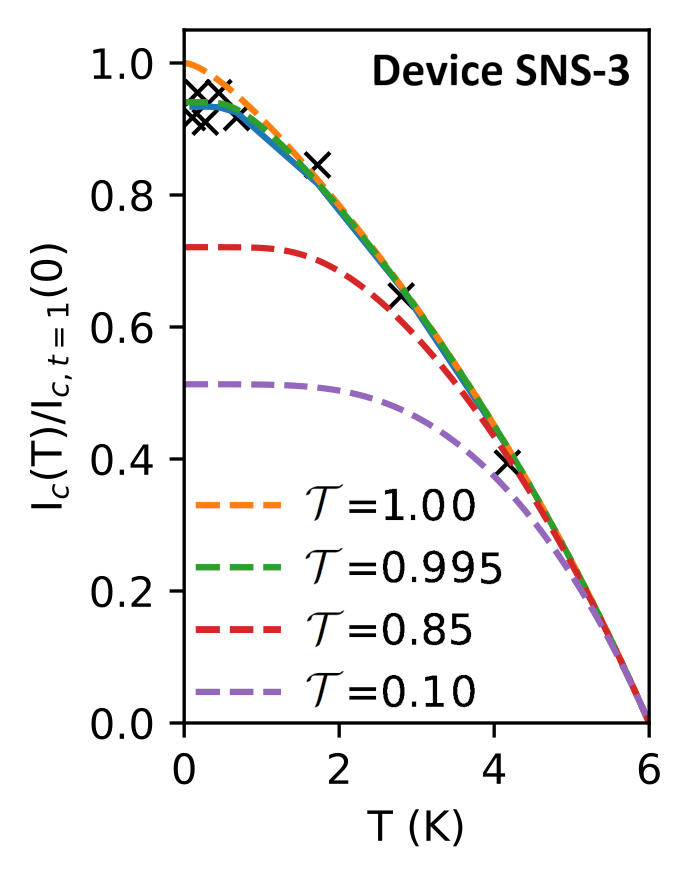}
  \caption{Temperature dependence of $I_c$ of SNS device, where crosses are experimental data and dashed lines are fits to Eqn.~(3) in the main text for $T_c=6$~K and different values of $\mathcal{T}$.}
  \label{fig:Tdep}
\end{figure}

Finally, Figure~\ref{fig:Tdep} shows the temperature dependence of $I_c$ in device SNS-3, to compare measured transparencies between Eqn.~(1) and Eqn.~(3) from the main text.

\clearpage
\newpage

\clearpage
\newpage

\clearpage
\newpage

\clearpage
\newpage

